%% file: B2G-25-009_temp.tex
\documentclass[11pt,twoside,a4paper,cmspaper,final,collab]{cms-tdr}
\def\svnVersion{4f15d4f-D}\def\svnDate{2026/03/24}\def\cmsCernNoTag{CERN-EP-2026-049}\def\cmsCernDate{\today}\def\cmsMessage{Submitted to the Journal of High Energy Physics}
\begin{document}\cmsNoteHeader{B2G-25-009}

\providecommand{\cmsTable}[1]{\resizebox{\textwidth}{!}{#1}}
\newlength\cmsTabSkip\setlength\cmsTabSkip{1ex}

\newcommand{\pp}{\ensuremath{\Pp\Pp}\xspace}
\newcommand{\mtt}{\ensuremath{m_{\ttbar}}\xspace}
\newcommand{\mtop}{\ensuremath{m_{\PQt}}\xspace}
\newcommand{\mH}{\ensuremath{m_{\PH}}\xspace}
\newcommand{\costhetastar}{\ensuremath{\cos(\theta^\ast)}\xspace}
\newcommand{\BR}{\ensuremath{\mathcal{B}}\xspace}

\newcommand{\akfour}{\text{AK4}\xspace}
\newcommand{\akeight}{\text{AK8}\xspace}
\newcommand{\softdropmass}{\ensuremath{m_{\text{SD}}}\xspace}
\newcommand{\deepak}{\textsc{DeepAK8}\xspace}

\newcommand{\PA}{\PSA}
\newcommand{\Zprime}{\PZpr}
\newcommand{\ZprimeDM}{{\HepParticle{\PZ}{\text{DM}}{\prime}}\xspace}
\newcommand{\mDM}{\ensuremath{m_{\text{DM}}}\xspace}
\newcommand{\gKK}{{\HepParticle{g}{\text{KK}}{}}\xspace}
\newcommand{\gDM}{\ensuremath{g_{\text{DM}}}\xspace}
\newcommand{\gq}{\ensuremath{g_{\PQq}}\xspace}
\newcommand{\gell}{\ensuremath{g_{\Pell}}\xspace}

\newcommand{\ptrel}{\ensuremath{p_{\text{T,rel}}}\xspace}

\newcommand{\btagged}{\ensuremath{\PQb\text{ tagged}}\xspace}
\newcommand{\btagging}{\ensuremath{\PQb\text{ tagging}}\xspace}
\newcommand{\ttagged}{\ensuremath{\PQt\text{ tagged}}\xspace}
\newcommand{\ttagging}{\ensuremath{\PQt\text{ tagging}}\xspace}

\newcommand{\zttag}{0 \PQt \text{tag}\xspace}
\newcommand{\ottag}{1 \PQt \text{tag}\xspace}

\newcommand{\gptt}{\ensuremath{g_{\PGF\ttbar}}\xspace}
\newcommand{\Gammaptt}{\ensuremath{\Gamma_{\PGF\ttbar}}\xspace}
\newcommand{\Gammap}{\ensuremath{\Gamma_{\PGF}}\xspace}

\newcommand{\mlep}{\ensuremath{M_{\text{lep}}}\xspace}
\newcommand{\mhad}{\ensuremath{M_{\text{had}}}\xspace}
\newcommand{\mlepexp}{\ensuremath{\mlep^{\text{exp}}}\xspace}
\newcommand{\mhadexp}{\ensuremath{\mhad^{\text{exp}}}\xspace}
\newcommand{\sigmalep}{\ensuremath{\sigma_{\text{lep}}}\xspace}
\newcommand{\sigmahad}{\ensuremath{\sigma_{\text{had}}}\xspace}

\newcommand{\Stot}{\ensuremath{S_{i}^{\text{comb.}}}\xspace}
\newcommand{\Sres}{\ensuremath{S_{i}^{\text{res.}}}\xspace}
\newcommand{\Sint}{\ensuremath{S_{i}^{\text{int.}}}\xspace}

\newcommand{\st}{\ensuremath{S_{\mathrm{T}}}\xspace}
\newcommand{\DRsum}{\ensuremath{\DR_{\text{sum}}}\xspace}
\newcommand{\Vjets}{\ensuremath{\PV\text{+jets}}\xspace}
\newcommand{\Wjets}{\ensuremath{\PW\text{+jets}}\xspace}
\newcommand{\Zjets}{\ensuremath{\PZ\text{+jets}}\xspace}
\newcommand{\PQj}{{\HepParticle{j}{}{}}\xspace}

\newcommand{\ee}{\ensuremath{\Pe\Pe}\xspace}
\newcommand{\emu}{\ensuremath{\Pe\PGm}\xspace}
\newcommand{\mumu}{\ensuremath{\PGm\PGm}\xspace}
\newcommand{\muR}{\ensuremath{\mu_{\mathrm{R}}}\xspace}
\newcommand{\muF}{\ensuremath{\mu_{\mathrm{F}}}\xspace}

\cmsNoteHeader{B2G-25-009}
\title{Search for new particles decaying into top quark-antiquark pairs in proton-proton collisions at \texorpdfstring{$\sqrt{s}=13\TeV$}{sqrt(s)=13 TeV}}

\date{\today}

\abstract{A search for new particles decaying to top quark-antiquark pairs is performed using proton-proton collision data at a centre-of-mass energy of 13\TeV. The data set recorded with the CMS detector between 2016 and 2018 is used, corresponding to an integrated luminosity of 138\fbinv. Final states with 0, 1, and 2 leptons are analyzed, covering all decay modes of the top quark-antiquark pairs. Heavy \Zprime bosons with relative widths of 1, 10, and 30\% are excluded for masses in the ranges 0.4--4.8, 0.4--6.2, and 0.4--7.4\TeV, respectively. A Kaluza--Klein gluon in the Randall--Sundrum model and a dark-matter mediator are excluded for masses between 0.5--5.5 and 1.0--4.2\TeV, respectively. These results set the most stringent limits to date for the considered models in the \ttbar final state. In addition, in the two-Higgs-doublet models, upper limits are set on the coupling strength modifier for scalar and pseudoscalar Higgs bosons with relative widths of 2.5, 10, and 25\% in the mass range of 0.5--1.0\TeV.}

\hypersetup{
pdfauthor={CMS Collaboration},
pdftitle={Search for new particles decaying into top quark-antiquark pairs in proton-proton collisions at sqrt(s)=13 TeV},
pdfsubject={CMS},
pdfkeywords={CMS, B2G}}

\maketitle

\section{Introduction}

The top quark is the heaviest known fermion. Its large coupling to the Higgs field
could be an indication of a special role of the top quark in any model that
explains the difference between the electroweak (EW) and Planck scales.
The high production rate of top quarks at the CERN LHC provides a unique opportunity
to seek out small top quark couplings in searches for anomalous production
of physics beyond the standard model (BSM).
Particularly, in addition to standard model (SM) processes, BSM particles decaying into a top quark-antiquark pair (\ttbar) could manifest themselves
as a resonant or non-resonant contribution to the \ttbar invariant mass (\mtt) or to other kinematic observables.

Examples of BSM production of \ttbar resonances include
models with massive color-singlet \PZ-like bosons in extended gauge theories~\cite{zprime_Rosner,zprime_Lynch,zprime_Carena},
colorons~\cite{Hill1991419,Hill:1993hs,Hill:1994hp,Jain11124928} or axigluons~\cite{axigluon,Choudhury:2007ux,Godbole:2008qw}. In addition, anomalous production of \ttbar can occur via Kaluza--Klein (KK) excitations of gluons~\cite{Agashe:2006hk}
or gravitons~\cite{Davoudiasl:1999jd} in various extensions of the Randall--Sundrum model of extra dimensions~\cite{Randall:1999ee,PhysRevLett.83.4690}.

Searches for such resonant signatures have been performed by the ATLAS~\cite{ATLAS_run2} and CMS~\cite{Sirunyan2019} Collaborations,
focusing primarily on resonant BSM effects.
In contrast, the present search is sensitive to both resonant and non-resonant BSM signatures.
This analysis is performed using proton-proton (\pp) collision data at a centre-of-mass energy of 13\TeV
recorded by the CMS experiment between 2016 and 2018,
corresponding to a total integrated luminosity of 138\fbinv.

{\tolerance=800
Recent results~\cite{CMS:2025kzt,CMS:2025dzq} from the CMS Collaboration demonstrate an
excess of \ttbar events, above perturbative quantum chromodynamics (QCD) predictions near the kinematic production threshold (2\mtop),
that may be consistent with a quasi-bound toponium state.
This analysis, on the other hand, is optimized for a higher mass regime.
Its sensitivity to such low-mass signals is limited, and the mass region below 0.4\TeV is
excluded from the interpretations.
\par}

All decay channels of the top quark-antiquark pair are considered. Thus, events may contain 0, 1, or 2 charged leptons
(``0\Pell'' or ``all-hadronic'', ``1\Pell'' or ``single-lepton'', ``2\Pell'' or ``dilepton''). In this analysis, only electrons and muons are considered.
As \PGt leptons are not explicitly vetoed in the analysis, the \ttbar decays with \PGt leptons in the final state may contribute to the search channels, and are counted in the \ttbar background (for SM production) or the signals (for BSM production). The event overlap between the channels is found to be negligible. The data from all channels are combined into a single maximum-likelihood fit constraining the product of the BSM signal production cross section and the branching fraction.
Results are presented for several signal hypotheses, providing the most stringent limits
to date on spin-1 \ttbar resonances. Compared to previous CMS measurements~\cite{Sirunyan2019}, substantive improvements have been made in all channels.

{\tolerance=850
The event selection in the 0\Pell channel utilizes a deep neural network (DNN), named the \deepak algorithm~\cite{DeepAK8}, to identify hadronically decaying top quarks with large Lorentz boosts (known as ``\ttagging''),
as well as a new background estimation technique based on a two-dimensional (2D) fit to the \ttagged jet mass (\mtop) and \mtt spectrum.
The analysis in Ref.~\cite{Sirunyan2019} separated the signal region (SR) into
bottom quark (\PQb) tagging categories, as well as the rapidity difference between the two top quark candidates,
whereas here the \btagging has been subsumed into the \deepak algorithm.
The categorization based on the rapidity difference between the top quark candidates is retained,
and provides discrimination between SM \ttbar production and BSM \ttbar production.
\par}

The event selection in the 1\Pell channel has been optimized with respect to the analysis in Ref.~\cite{Sirunyan2019} by lowering the thresholds on the lepton's transverse momentum (\pt) and the missing transverse momentum, and uses the same \deepak \ttagging algorithm employed by the 0\Pell channel to identify hadronic decays of \PQt quarks.
Another DNN is employed to enhance event categorization by providing improved discrimination between \ttbar and non-\ttbar backgrounds.
This approach also facilitates the construction of background-dominated control regions (CRs),
which are used to constrain the normalization of the non-\ttbar backgrounds in the SRs.
Additionally, a spin-sensitive variable is used to further enhance the separation between BSM signals
and the SM \ttbar background in the SRs, exploiting differences in their angular distributions.

The event selection in the 2\Pell channel has also been improved with respect to the analysis in Ref.~\cite{Sirunyan2019}.
An event categorization based on the angular distance between the two leptons and the jets nearest to them
has been developed. Events are separated into regions based on the variable \DRsum, defined as the sum of the \DR values between each lepton and the closest jet, where $\DR=\sqrt{\smash[b]{(\Delta\eta)^2+(\Delta\phi)^2}}$. As previously done in Ref.~\cite{Sirunyan2019},
the \st variable (the scalar sum of the transverse momenta of the jets, leptons, and missing transverse momentum) is used instead of \mtt, since the 2\Pell channel is affected by the presence of two neutrinos that complicates the reconstruction of \mtt.

Several BSM benchmark scenarios are considered in this search.
These include the previously investigated benchmarks of \Zprime bosons in Ref.~\cite{Sirunyan2019} with relative widths of 1, 10, and 30\% in the mass range of 0.4--9\TeV,
as well as KK gluons (\gKK) in the Randall--Sundrum model in the mass range of 0.5--6\TeV.
A new model of a dark-matter (DM) mediator in the mass range of 1--5\TeV has been added.
Finally, for the 1\Pell channel, scalar and pseudoscalar Higgs bosons predicted
in two-Higgs-doublet models are considered, with relative widths of 2.5, 10, and 25\%, in the mass range of 0.5--1\TeV. The two-Higgs-doublet interpretation is restricted to this channel, as it is the only one providing sensitivity in this mass range.

This paper is organized as follows.
Section~\ref{sec:signals} introduces the signal models under investigation.
Section~\ref{sec:cms_detector} provides a brief description of the CMS detector and event reconstruction.
Section~\ref{sec:data_and_simulated_samples} details the trigger selection, and the simulation of signal and background events.
Section~\ref{sec:event_selection_and_categorization} describes the event selection, categorization strategies,
and the methods used for background estimation,
while Section~\ref{sec:systematic_uncertainties} discusses the systematic uncertainties.
The results are presented in Section~\ref{sec:results}, and Section~\ref{sec:summary} provides a summary of the paper. Numerical results of the analysis are available in HEPData~\cite{hepdata}.

\section{Signal models}
\label{sec:signals}

Several signal models are considered in this search. All models predict a signal in the process $\pp\to\ttbar$ that can be observed either as a
resonant or non-resonant modification of the SM \ttbar production.
These signals change the shape of the \mtt and angular distributions, as well as the distributions of other similar kinematic variables. The modified distributions may contain resonant or non-resonant enhancements or deficits (the latter in the case of spin-0 resonances that destructively interfere with the SM production).

\subsection{Spin-1 resonant production}

\begin{figure}[!tp]
\centering
\includegraphics[width=0.49\textwidth]{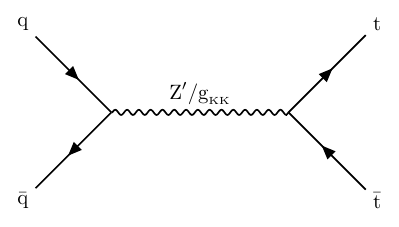}
\hfill
\includegraphics[width=0.49\textwidth]{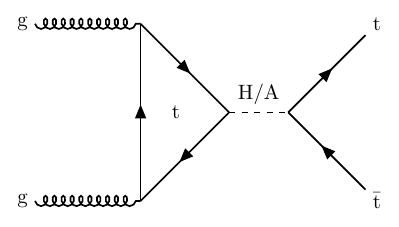}
\caption{Example Feynman diagrams at leading order for the production and decay of a spin-1 \Zprime/\gKK boson (left)
and a scalar \PH or pseudoscalar \PA resonance (right).}
\label{fig:Feynman_diagrams}
\end{figure}

Multiple BSM theories predict the existence of a massive spin-1 particle that strongly couples
to top quarks. An example is the Randall--Sundrum model of extra dimensions~\cite{Randall:1999ee,PhysRevLett.83.4690},
where the KK gluon \gKK can have a large branching fraction (\BR) into \ttbar.
Another example is an extension of the SM gauge groups~\cite{Hill:1994hp, Barger:1980ix, Contino:2011np, Bellazzini:2014yua},
where leptophobic and top-philic \Zprime bosons can appear. These resemble topcolor models~\cite{Jain11124928}
if the \Zprime boson couples strongly only to the first and third generations of quarks with
no significant couplings to the leptons.
The existence of \Zprime bosons as mediators in the interaction between ordinary matter and DM
can also be parametrized with simplified models~\cite{Albert:2017}.
In these models, a Dirac fermion \PGc is proposed as the DM candidate, with a heavy particle serving as the mediator.
Two scenarios are considered: an axial-vector mediator and a vector mediator.
These mediators produce the same effects in the observed final states selected in this analysis.
Consequently, both scenarios are treated as equivalent for the purposes of this study.
An example Feynman diagram for the production of a \Zprime or \gKK boson and its decay to \ttbar are shown in
Fig.~\ref{fig:Feynman_diagrams} (left).
We consider only the production of massive spin-1 particles through \qqbar annihilation and neglect
any interference with SM \ttbar, which predominantly occurs via gluon-gluon fusion at the LHC~\cite{Bonciani:2015hgv}.

\subsection{Spin-0 resonant production}

Several BSM models with extended Higgs sectors predict massive spin-0 particles that couple to top quarks.
For instance, two-Higgs-doublet models (2HDM)~\cite{Lee:1973iz,Branco:2011iw,Haber:2015pua,Kling:2016opi}
contain a massive neutral scalar boson (\PH) and a massive neutral pseudoscalar boson (\PA).
In several scenarios, these have suppressed couplings to the SM vector bosons, such that decays
to \ttbar are enhanced~\cite{pseudohiggs}.
The production mechanism of the \PH and \PA bosons is dominantly through gluon-gluon fusion,
such that there are interference effects with SM \ttbar production.
An example Feynman diagram is shown in Fig.~\ref{fig:Feynman_diagrams} (right).

\section{The CMS detector and event reconstruction}
\label{sec:cms_detector}

The CMS detector~\cite{Chatrchyan:2008aa, Hayrapetyan_2024} is a multipurpose, nearly hermetic apparatus,
designed to identify electrons, muons, photons,
and charged and neutral hadrons~\cite{CMS:2020uim,CMS:2018rym,Chatrchyan:2014fea}.
A global particle-flow (PF) algorithm~\cite{Sirunyan:2017ulk} aims to reconstruct all individual
particles in an event, combining information provided by the all-silicon inner tracker
and by the crystal electromagnetic (ECAL) and brass-scintillator hadron calorimeters,
operating inside a 3.8\unit{T} superconducting solenoid, with data from the gas-ionization muon detectors
interleaved with the layers of the steel flux-return outside the solenoid. The reconstructed particles are used
to build leptons, jets, and missing transverse momentum~\cite{Khachatryan:2016kdb,CMS:2018jrd,CMS:2019ctu}.

Proton bunches collide every 25\unit{ns}, producing a large number of events, not all of which are of interest for physics analyses.
Events of interest are selected using a two-tiered trigger system. The first level (L1),
composed of custom hardware processors, uses information from the calorimeters and muon detectors
to select events at a rate of around 100\unit{kHz} within a fixed latency of 4\mus~\cite{CMS:2020cmk}.
The second level, known as the high-level trigger, consists of a farm of processors
running a version of the full event reconstruction software optimized for fast processing,
and reduces the event rate to a few kHz before data storage~\cite{Khachatryan:2016bia,CMS:2024aqx}.
The primary vertex is taken to be the vertex corresponding to the hardest scattering in the event, evaluated using tracking information alone, as described in Section 9.4.1 of Ref.~\cite{CMS-TDR-15-02}.
Electrons are reconstructed by combining the momentum measured in the tracker, the energy
of the associated ECAL cluster, and the energy of bremsstrahlung photons spatially compatible with
the electron trajectory~\cite{CMS:2020uim}.
This combination corrects for energy losses due to radiation in the tracker material and improves
the resolution and accuracy of the reconstructed electron four-momentum.
Muons are reconstructed by combining tracks in the silicon tracker with hits in the muon chambers~\cite{CMS:2018rym}.
For very high-\pt muons, the reconstruction is initiated in the muon system and subsequently fitted to tracks in the pixel and strip tracker,
to mitigate inefficiencies observed in data while improving the momentum resolution and ensuring correct vertex association~\cite{Chatrchyan:2014fea}.
This two-step approach provides efficient and precise muon reconstruction across the full \pt spectrum.

Hadronic jets are reconstructed from PF candidates, which include charged and neutral particles originating from the primary interaction,
as well as additional contributions from other \pp interactions occurring in the same or nearby bunch crossings (pileup).
To mitigate this effect, charged particles identified to be originating from pileup vertices are discarded and an offset correction
is applied to correct for remaining contributions.
Jets reconstructed excluding charged particles associated with pileup vertices, also known as the
``charged hadron subtraction'' (CHS) technique, are referred to as ``CHS jets'', and are used by the 2\Pell analysis.
The pileup-per-particle identification (PUPPI) algorithm~\cite{Bertolini:2014bba,Sirunyan:2020foa}
is used to mitigate the effect of pileup at the reconstructed-particle level, making use of
local shape information, event pileup properties, and tracking information. A local shape variable
is defined, which distinguishes between collinear and soft diffuse distributions of other
particles surrounding the particle under consideration. The former is attributed to particles
originating from the hard scatter and the latter to particles originating from pileup interactions.
Charged particles identified to be originating from pileup vertices are discarded. For each neutral
particle, a local shape variable is computed using the surrounding charged particles compatible with
the primary vertex within the tracker acceptance ($\abs{\eta}<2.5$), and using both charged and
neutral particles in the region outside of the tracker coverage. The momenta of the neutral particles
are then rescaled according to their probability to originate from the primary interaction vertex
deduced from the local shape variable, superseding the need for jet-based pileup
corrections~\cite{Sirunyan:2020foa}.
Jets reconstructed with constituents weighted by the PUPPI algorithm are referred to as "PUPPI jets" and are used by the 0\Pell and 1\Pell analyses.

Following pileup mitigation at the particle level with the PUPPI or CHS algorithms, jets are clustered from PF candidates
with the \FASTJET package~\cite{Cacciari:2011ma} using the anti-\kt algorithm~\cite{Cacciari:2008gp}
with distance parameters of $R=0.4$ and $R=0.8$, and are referred to as \akfour and \akeight jets, respectively. Jet momenta are computed as the vector sum of the constituent particle momenta.

The \akfour jets are required to have $\pt>30\GeV$ and $\abs{\eta}<2.5$.
The \textsc{DeepJet} DNN algorithm~\cite{Sirunyan:2017ezt} is used to identify jets originating from the decay of \PQb hadrons, and relies on information from the calorimeters and the tracking detector~\cite{Bols:2020bkb}. The 1\Pell (2\Pell) analysis uses a working point (WP) with a misidentification rate of 1 (10)\% and an efficiency of 70--80 (85--90)\%.

Jets from the \akeight algorithm are required to have $\pt>400\GeV$ and $\abs{\eta}<2.5$ to be considered for the \ttagging algorithm. This ensures that the decay products from the top quark are fully contained in a single, large-radius jet. For these \akeight jets, jet grooming is applied to remove soft, wide-angle radiation that contributes to the mass and substructure observables.
The soft-drop algorithm~\cite{Larkoski:2014wba},
which is a generalization of the modified mass drop tagger algorithm~\cite{Dasgupta:2013ihk},
is used to groom the jets and identify up to two subjets inside the \akeight jets.
This algorithm, with an angular exponent $\beta=0$ and a soft-cutoff threshold $z_{\text{cut}}<0.1$,
is applied to \akeight jets reclustered using the Cambridge--Aachen algorithm~\cite{Dokshitzer:1997in,Wobisch:1998wt},
and removes soft, wide-angle radiation from the jet. We refer to the resulting jet mass as the soft-drop mass (\softdropmass).
Jets originating from the hadronic decay of top quarks are identified (\ttagged) using a machine-learning technique
that relies on jet substructure variables, jet constituents, and secondary vertices~\cite{DeepAK8}.
This algorithm is referred to as \deepak.

Jet momenta are corrected using jet energy corrections (JEC) to match the average energy of particle-level jets.
The JEC sequence begins with an offset correction that removes the residual energy contribution from pileup.
Subsequent corrections are derived from simulation-based calibrations and refined with in situ
measurements of the momentum balance in dijet, {\PGg}+jet, {\PZ}+jet, and multijet events to account for any residual
differences between data and simulation~\cite{Khachatryan:2016kdb}.
Additional selection criteria are applied to each jet to remove jets potentially dominated by instrumental effects or reconstruction failures. The JEC values are also propagated to the jet mass by correcting the subjets comprising the groomed \akeight jets from the soft drop algorithm.

The missing transverse momentum vector \ptvecmiss is computed as the negative vector sum of
the transverse momenta of all PF candidates in the event,
and its magnitude is denoted as \ptmiss~\cite{CMS:2019ctu}.
To reduce pileup dependence, the PUPPI algorithm is applied at the particle level, and
\ptvecmiss is computed from PF candidates weighted by their probability to originate from the primary interaction vertex.
The \ptvecmiss is modified to account for corrections to the energy scale of the reconstructed jets in the event.

Quality control criteria are also applied to all channels to remove detector noise and correct other technical issues.

\section{Data and simulated samples}
\label{sec:data_and_simulated_samples}

{\tolerance=800
Data events are collected with the CMS detector in \pp collisions recorded between 2016 and 2018 at a centre-of-mass energy of 13\TeV, corresponding to an integrated luminosity of 138\fbinv.
The trigger paths used to select events are determined by the final-state particles in each decay channel.
The all-hadronic channel uses a trigger requiring that the scalar sum of the transverse momenta of the \akfour jets (\HT) be larger than a specified value, and the offline selection requires $\HT>1.3\TeV$ to ensure that the online \HT trigger is fully efficient for the selected events.
In the single-lepton channel, events are recorded by triggers requiring the presence of a single muon, with or without an isolation requirement~\cite{CMS:2018rym}, or using a combination of isolated and non-isolated electron and photon triggers~\cite{CMS:2020uim} to achieve optimal efficiency across the full electron energy range.
For the dilepton channel, events are selected using triggers that require either a single muon or two electrons, without any isolation requirement.
\par}

{\tolerance=1200
Top quark-antiquark pair and single top quark electroweak production are simulated at next-to-leading order (NLO) with \POWHEG~v2~\cite{Nason:2004rx,Frixione:2007nw,Frixione:2007vw,Alioli:2010xd,Re:2010bp},
with the \ttbar cross section normalized to next-to-next-to-leading order (NNLO) precision in perturbative QCD,
using a next-to-next-to-leading-logarithmic soft-gluon approximation from \textsc{top++}~2.0~\cite{Czakon:2011xx}.
Single top, \PZ, and \PW boson production (\Vjets) are generated at leading order (LO) with \MGvATNLO~2.6.5~\cite{Alwall:2007fs, Alwall:2014hca},
with \Vjets events reweighted for NLO QCD and electroweak effects~\cite{Lindert:2017olm} as a function of the vector boson transverse momentum.
The QCD multijet and diboson processes are simulated with \PYTHIA~8.240~\cite{Sjostrand:2014zea} at LO.
\par}

The \ttbar transverse momentum spectrum is known to be mismodeled by NLO generators, partially because of the destructive interference with higher-order terms in the perturbative expansion~\cite{Kidonakis:2012rm}. The dilepton analysis corrects for this effect explicitly, whereas the all-hadronic and single-lepton analyses account for it with uncertainties that cover the variations.

Signal samples for \gKK are generated at LO with \PYTHIA for masses within the range 0.5--6\TeV in 0.5\TeV steps,
with the \gKK coupling to right-handed top quarks set to 5, yielding a branching fraction to \ttbar of about 94\%.
Cross sections are corrected to NLO QCD~\cite{Gao:2010bb}.
Leptophobic \Zprime boson signals are generated at LO with \MGvATNLO for masses in the range 0.4--9\TeV, with widths of 1, 10, and 30\% of the resonance mass, decaying exclusively to \ttbar.
Dark-matter mediator \ZprimeDM boson signals are generated at LO with \MGvATNLO for masses in the range 1--5\TeV.
The model parameters are set according to the V1 and A1 benchmark models described in Ref.~\cite{Albert:2017}:
the DM mass is $\mDM=10\GeV$, the DM-DM-mediator coupling is $\gDM=1$,
the mediator-quark coupling is $\gq=0.25$, and the mediator-lepton coupling is $\gell=0$.
As mentioned above, interference with SM \ttbar is neglected for all spin-1 models, including \Zprime, \ZprimeDM, and \gKK.

Scalar (\PH) and pseudoscalar (\PA) Higgs bosons in the 2HDM model are generated at LO with \MGvATNLO with masses between 0.5 and 1\TeV, with widths of 2.5, 10, and 25\% of the resonance mass.
Additionally, the \PH and \PA bosons are forced to decay to \ttbar, producing the final states with one lepton and jets.
Cross sections are corrected to NNLO using the $k$-factors from Ref.~\cite{Hespel:2016qaf}.

For all samples, parton showering and hadronization are simulated with \PYTHIA, using the CP5 tune~\cite{Sirunyan:2019dfx} for backgrounds and most signals, except for \Zprime samples generated with CP2~\cite{Sirunyan:2019dfx}.
The NNPDF~3.1 NNLO parton distribution function (PDF) set is used~\cite{Ball_2017}.
All simulated samples are processed through a \GEANTfour-based~\cite{GEANT4} simulation of the CMS detector.
To simulate the effect of pileup collisions, additional inelastic events are generated using \PYTHIA
with a total inelastic cross section of 69.2\unit{mb}~\cite{Sirunyan:2018nqx}
and superimposed on the hard-scattering events.
The simulation is corrected to reproduce the distribution of the number of pileup interactions observed in the data.

\section{Event selection, categorization, and background estimates}
\label{sec:event_selection_and_categorization}

Events are selected that are consistent with the production of a \ttbar pair.
All final states of \ttbar are analyzed. The 0\Pell (all-hadronic) channel assumes a fully merged decay topology and uses two well-separated jets identified with a \ttagging algorithm.
The 1\Pell (single-lepton) channel has exactly one lepton, and is separated into categories based on lepton flavors and whether the quarks from the sequential top quark decays correspond to individual jets (resolved topology) or have multiple quarks within a single jet (merged topology). The 2\Pell (dilepton) channel requires exactly two leptons (\ee, \mumu, or \emu), at least two jets, and missing transverse momentum, and has separate categories for resolved and merged topologies, as well as lepton flavor.

\subsection{All-hadronic channel}

This channel focuses on events in which both top quarks decay hadronically ($\PQt\to\PQb\PW\to\PQb\PQq\PAQq^\prime$).

Events must have at least two \akeight jets within the tracker acceptance, \ie, a rapidity $\abs{y}<2.4$, and with $\pt>400\GeV$. The two jets with the highest \pt are used to construct the \ttbar candidate mass, and they are sorted according to their \ttagging \deepak score. The leading \ttagging score jet is required to pass the ``medium'' WP criterion of the \deepak algorithm, which corresponds to a 0.5\% mistag rate for light quarks and gluons and to a signal efficiency of 50\%. In the SR, both jets must pass the identification criteria of the \deepak algorithm at the medium WP. Additionally, both jets are required to have \softdropmass in a $[105,210]\GeV$ window around the top quark mass.

\begin{figure}[!ht]
\centering
\includegraphics[width=0.6\textwidth]{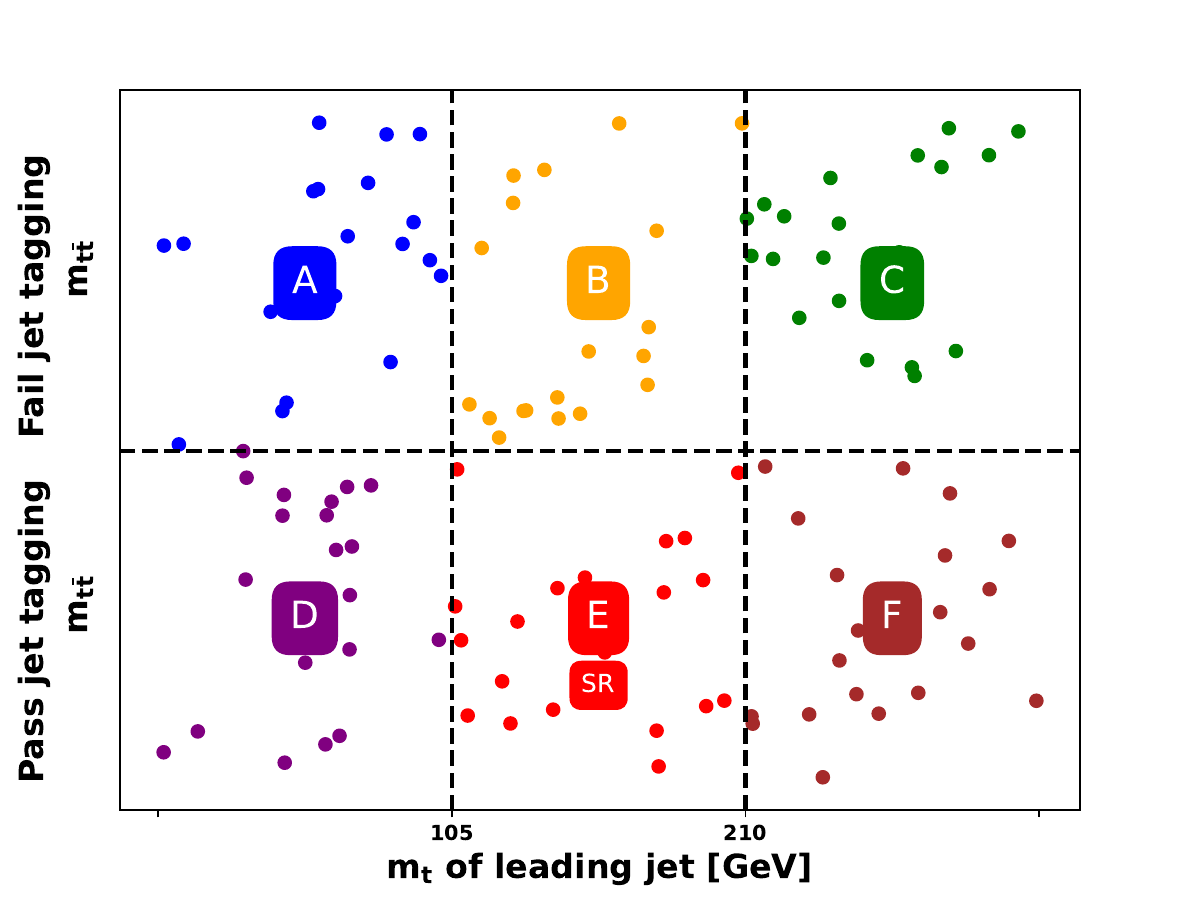}
\caption{Illustration of the background estimation method. The data set is binned in the leading jet mass \mtop and in the reconstructed
\ttbar invariant mass \mtt. Disjoint regions are defined according to whether \mtop lies inside or outside the top quark mass window and whether the subleading jet passes or fails the \ttagging requirement. A method based on control samples in data is used to estimate the QCD background in the signal region E from regions A, B, C, D, and F. The colored dotted points are shown for illustrative purposes only.}
\label{fig:cartoon2DABCD}
\end{figure}

The background estimate for this SR is derived from a CR defined in sidebands of the \ttagging score and \mtop distributions in a grid, as shown in Fig.~\ref{fig:cartoon2DABCD}. Since SM processes, such as QCD and SM \ttbar production, tend to be more forward while the resonant signals tend to be more central, the events are further separated into ``central'' and ``forward'' categories based on the leading jet's rapidity ($\abs{\Delta y}<1$ and $\abs{\Delta y}>1$, respectively). These criteria are used to separate the events into 12 categories simultaneously, with background model parameters that are allowed to vary in the likelihood fit. The SR is in the ``Pass jet tagging'' region, with the requirement $105<\mtop<210\GeV$ in both central and forward regions in order to select the majority of the signal events. Events in the ``Fail jet tagging'' region have a subleading jet that fails the \ttagging selection but passes the ``loose'' WP of the \deepak tagger, corresponding to a misidentification rate of 1\% for light quarks and gluons and to a signal efficiency of 55\%. These regions have varying signal purities and efficiencies. Overall, the efficiency of the selection in SM \ttbar events is around 25--40\% depending on the \pt of the jets.

The central and forward regions are fitted separately with different functional forms for the QCD background, but the SM \ttbar component is fitted jointly. This is discussed in detail in Section~\ref{sec:background_estimation_0l}.

At low \mtt ($\mtt<1.5\TeV$), the main loss of signal events originates from the trigger selection. Additional losses arise from the top quark mass and the \ttagging requirements on the jets, reducing the signal efficiency to approximately 5 to 20\%, depending on the mass point. This inefficiency is primarily  due to cases where, within a jet cone of $\DR=0.8$, only the \PW boson decay is reconstructed instead of the full top quark decay, as well as the limited efficiency of the \deepak algorithm.

\begin{figure}[!b]
\centering
\includegraphics[width=0.45\textwidth]{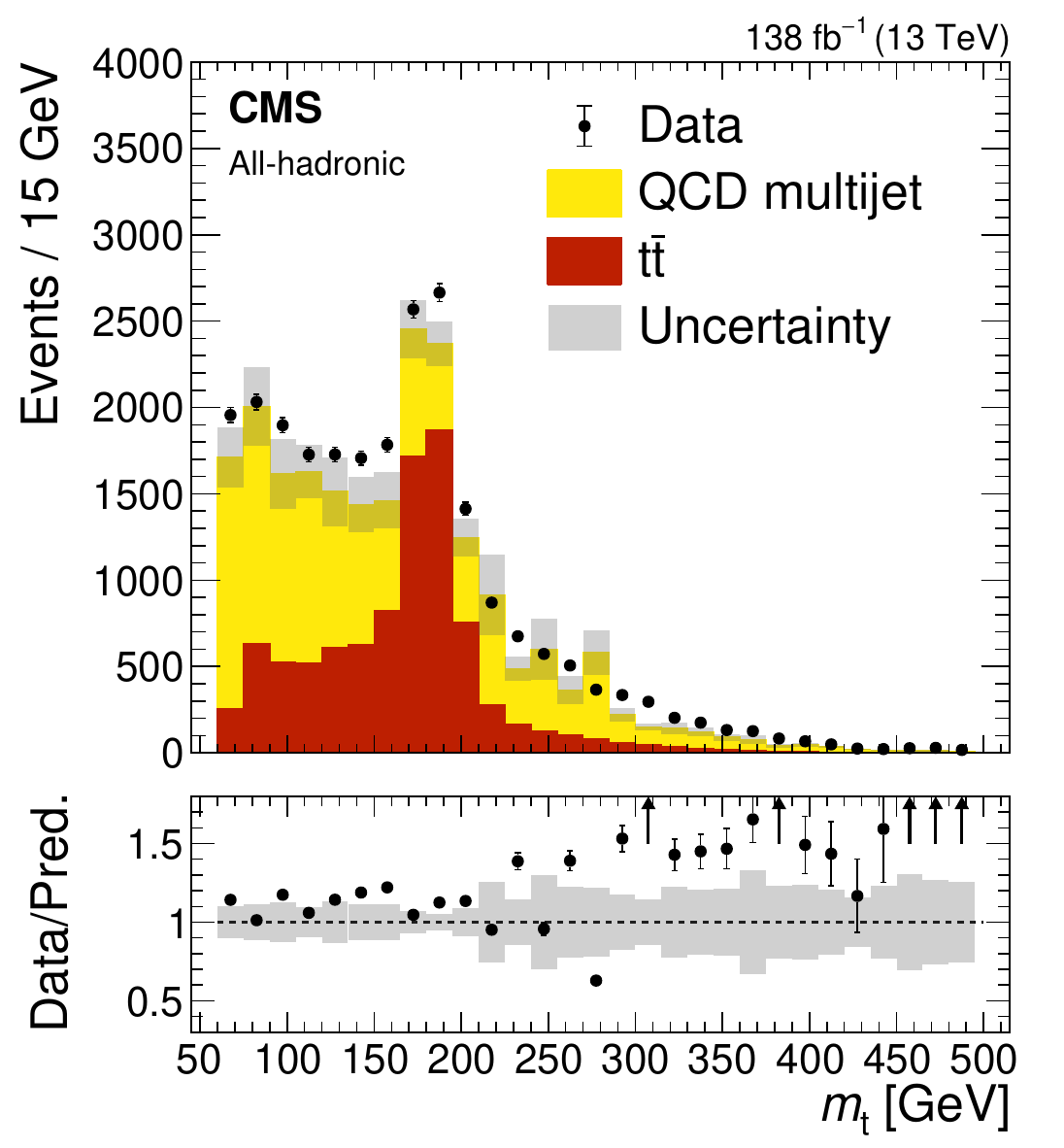}
\hfill
\includegraphics[width=0.45\textwidth]{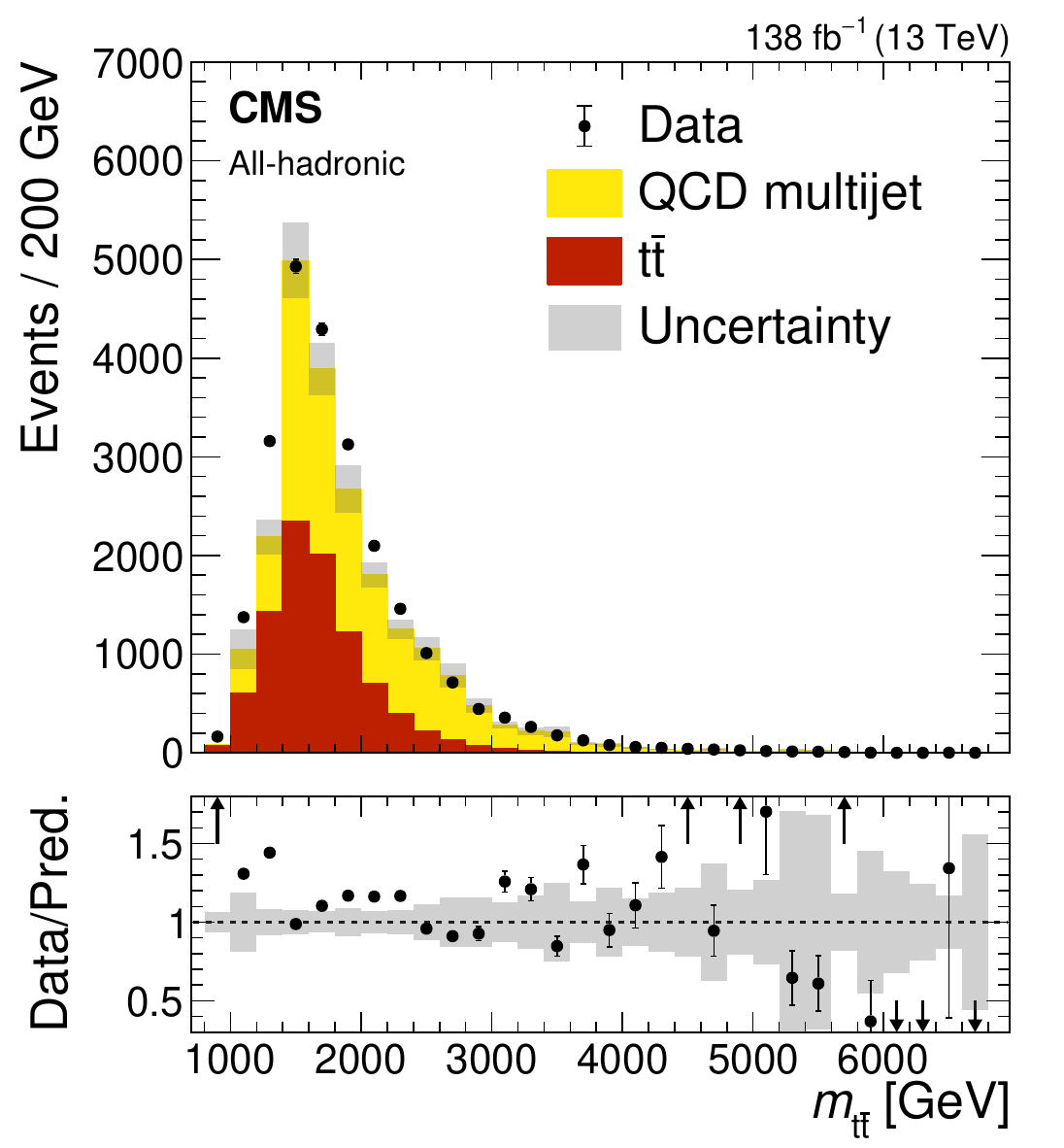}
\caption{Prefit data-to-simulation comparison of distributions in the all-hadronic channel for the mass of the leading \ttagged jet (left) and the reconstructed \ttbar mass (right) in the central and forward categories combined, where both jets pass the \ttagging requirement. The QCD background is taken from simulation for comparison, whereas in the analysis it is estimated from data. No cut on the \mtop variable is applied.}
\label{fig:msd_mtt}
\end{figure}

\begin{figure}[!b]
\centering
\includegraphics[width=0.45\textwidth]{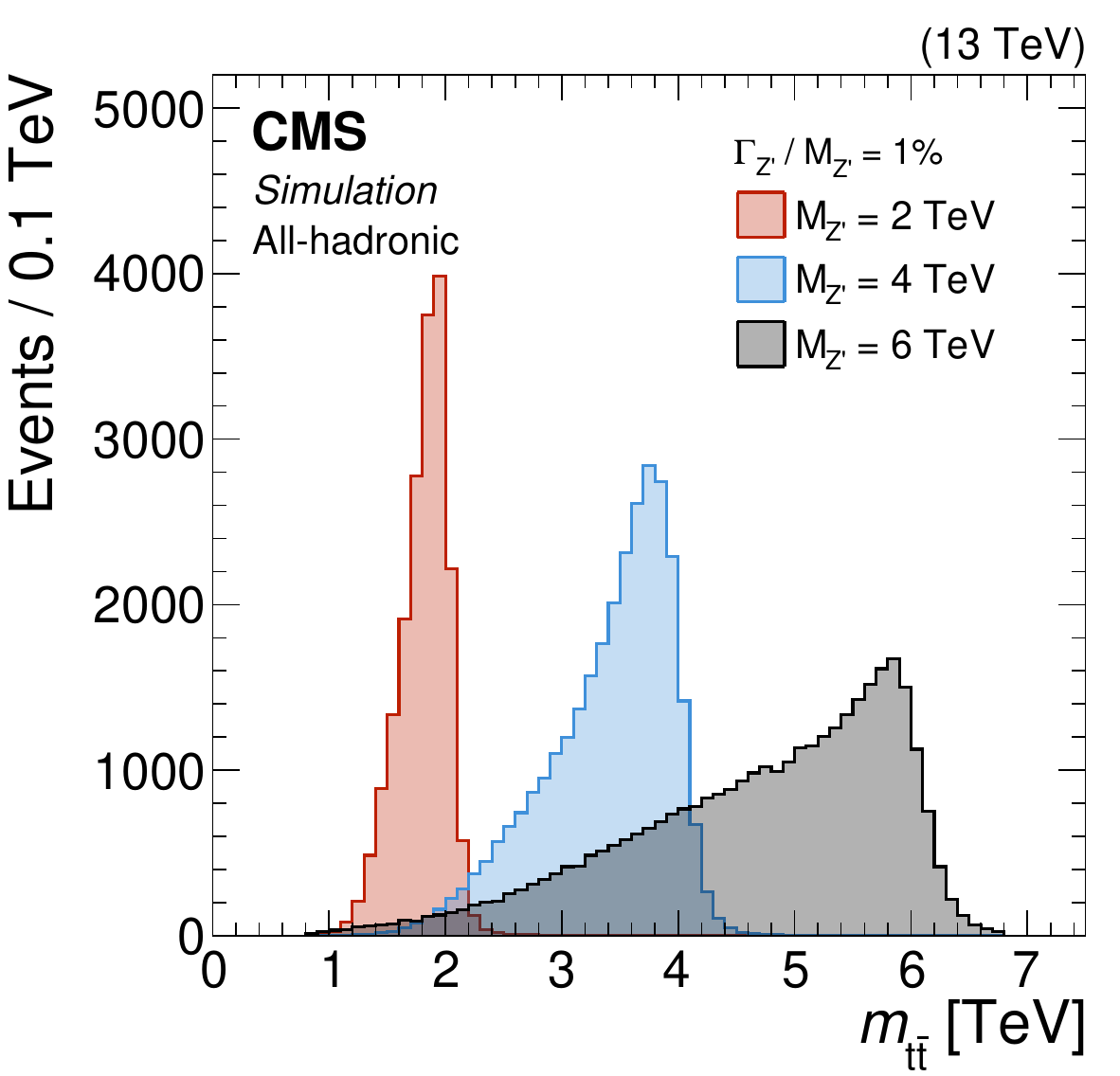}
\hfill
\includegraphics[width=0.45\textwidth]{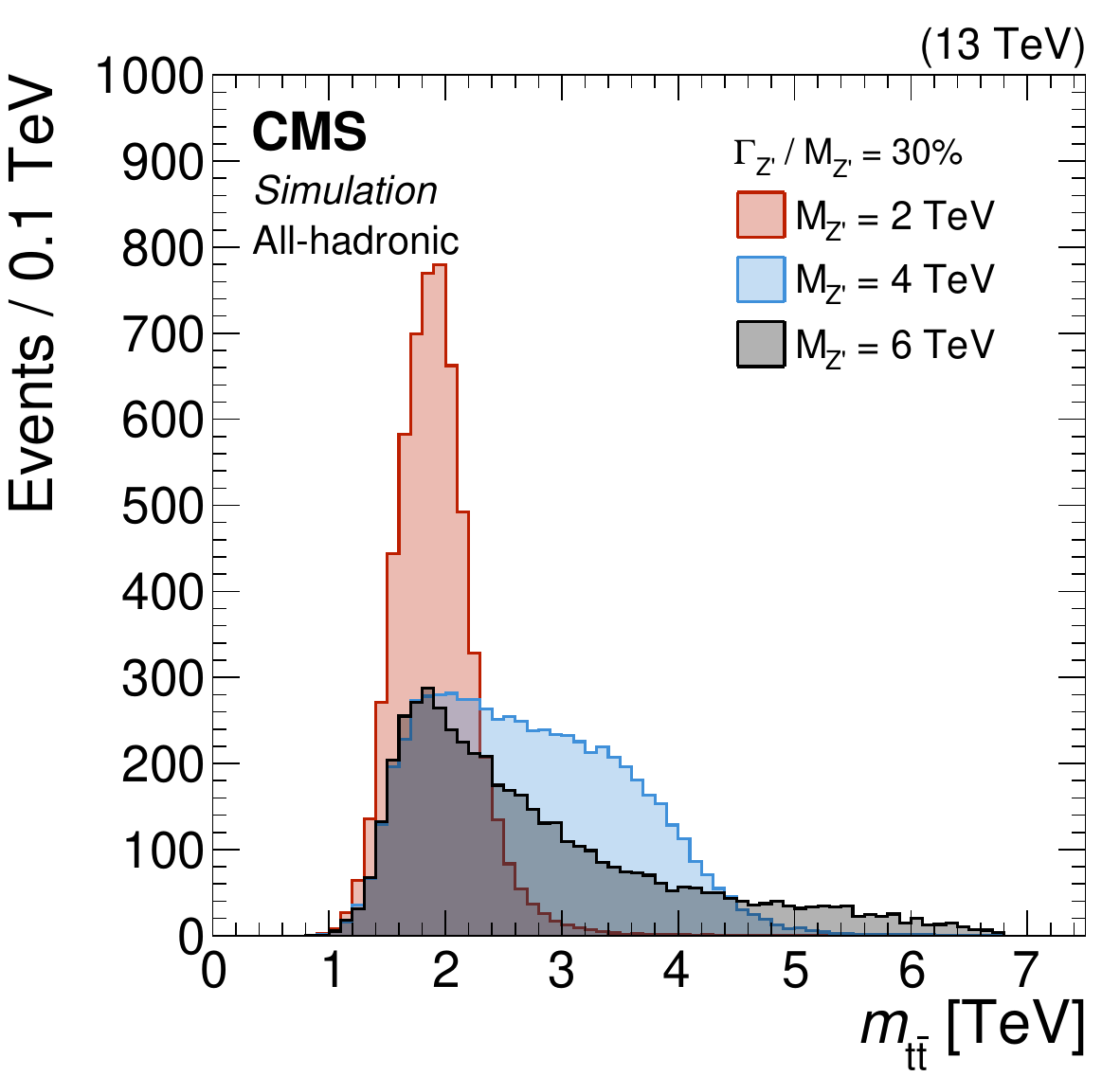}
\caption{Reconstructed \mtt distributions in simulation in the all-hadronic channel for \Zprime bosons with 1 and 30\% relative widths, shown in the left and right panels, respectively. The signals correspond to \Zprime boson masses of 2, 4, and 6\TeV, where both jets pass the \ttagging requirement. Signals are normalized to a cross section of 1\unit{pb} and an integrated luminosity of 138\fbinv. No cut on the \mtop variable is applied.}
\label{fig:mttbar_signal}
\end{figure}

The distributions of the data-to-simulation comparison for the leading jet mass \mtop and the \ttbar resonance mass (\mtt) are shown in Fig.~\ref{fig:msd_mtt} for the central and forward regions combined. The QCD background in this figure is estimated from simulation. The reconstructed \mtt distributions for two signal scenarios, \Zprime with 1 and 30\% relative widths, are shown in Fig.~\ref{fig:mttbar_signal}. The convolution of the available parton luminosity and the resonance width results in a significant fraction of events being produced off-shell, at masses below the nominal resonance mass.
The simulated QCD background is not used in the analysis, and instead that background is estimated from data using a 2D likelihood fit to the joint \mtt and \mtop distribution, which we describe in the next section.

\subsubsection{Background estimate}
\label{sec:background_estimation_0l}

Events from the SM \ttbar process and QCD multijet production constitute the primary sources of background in the SR. The SM \ttbar background is estimated using Monte Carlo simulations. In contrast, the QCD background is estimated from control samples in data, specifically via a modified version of the traditional sideband extrapolation (``ABCD'') method. This choice is motivated by the inability of simulations to accurately capture the complex QCD processes and jet multiplicities that characterize QCD events, resulting in discrepancies between predicted and observed data.

The background estimation uses four observables in a multidimensional distribution: the rapidity of the leading jet, the mass of the leading jet, the \deepak score of the subleading jet, and the reconstructed invariant mass of the top quark-antiquark pair \mtt. The rapidity of the leading jet is used to separate events into ``central'' and ``forward'' regions. The \deepak score of the subleading jet is used to separate events into ``pass'' and ``fail'' categories. Finally the fit uses a multidimensional polynomial in 2D (\mtt and \mtop), connecting the relevant parameters in the central/forward and pass/fail regions.

Figure~\ref{fig:cartoon2DABCD} illustrates this methodology, designating region E as the SR, characterized by both jets satisfying $\mtop\in[105,205]\GeV$ and passing the \deepak tagger requirements (``Pass jet tagging'').

The regions A, C, D, and F are CRs corresponding to the \mtop distribution sidebands. Region B also acts as a CR. It corresponds to events where both jets pass the \mtop requirement, but the subleading jet fails the tagger selection (``Fail jet tagging'').

The total number of events in the ``pass'' (P) and ``fail'' (F) jet tagging regions are given by :
\begin{equation}\begin{aligned}
    n_{\mathrm{F}}(i, \vec{\theta}) = n_{\mathrm{F}}^{\text{QCD}}(i, \vec{\theta}) + n_{\mathrm{F}}^{\ttbar} (i, \vec{\theta}) + n_{\mathrm{F}}^{\text{signal}} (i, \vec{\theta}),  \\
    n_{\mathrm{P}}(i, \vec{\theta}) = n_{\mathrm{P}}^{\text{QCD}}(i, \vec{\theta}) + n_{\mathrm{P}}^{\ttbar} (i, \vec{\theta}) + n_{\mathrm{P}}^{\text{signal}} (i, \vec{\theta}),
\end{aligned}\end{equation}
where the index $i$ is a given bin in the \mtop, \mtt plane and $\theta$ are the nuisance parameters considered for the analysis, as described in Section~\ref{sec:systematic_uncertainties}.

The multijet background contribution in the ``Pass jet tagging'' region is then determined from the CR, ``Fail jet tagging'',  by subtracting the simulated \ttbar background from the observed data. This remaining contribution is then scaled using a 2D transfer function (TF) in the (\mtop, \mtt) plane, denoted $R_{\mathrm{P/F}}$ , to estimate the number of QCD events in the ``Pass jet tagging'' region:
\begin{equation}
    n^{\text{QCD}}_{\mathrm{P}}(i) = n^{\text{QCD}}_{\mathrm{F}}(i) R_{\mathrm{P/F}} (\mtop, \mtt).
\end{equation}
In particular, the multijet background contribution in the SR (E) is determined as follows:
\begin{equation}
    n^{\text{QCD}}_{\mathrm{E}}(i) = n^{\text{QCD}}_{\mathrm{B}}(i) R_{\mathrm{P/F}} (\mtop, \mtt).
\end{equation}
Since events are categorized as either central or forward, different TFs are used for each category within each data-taking year. The TFs are obtained by fitting the \mtop and \mtt spectrum in two dimensions as a binned histogram.

The optimal TF parametrizations are chosen using Fisher tests~\cite{ftest}, which evaluate whether additional parameters significantly improve the quality of the fit. If no substantial improvement is observed, the simpler TF model is selected to avoid overfitting. In all cases, the chosen functional form is a polynomial of degree at most one.

The TF parameters are determined through a simultaneous maximum likelihood fit to data in both the ``Fail jet tagging'' and ``Pass jet tagging'' regions, in the CR and SR. This approach enhances both the accuracy of the background estimate and the analysis sensitivity, as the TF parameters and signal extraction are derived in parallel. The likelihood fit constructs the total background from the sum of individual contributions in each bin of the (\mtop, \mtt)  distribution using a Poisson model. To avoid potential bias, signal contamination in the ``Fail jet tagging'' region and mass sidebands is explicitly accounted for in the fit.

This Fisher test is performed for different signal scenarios and widths. The results confirm the absence of bias even for wide-width, high-mass signal scenarios, where the signal is expected to be highly off-shell, such as a \Zprime boson with a 30\% relative width and a mass of 6\TeV, as shown in Fig.~\ref{fig:mttbar_signal}.

\begin{figure}[!htp]
\centering
\includegraphics[width=0.4\textwidth]{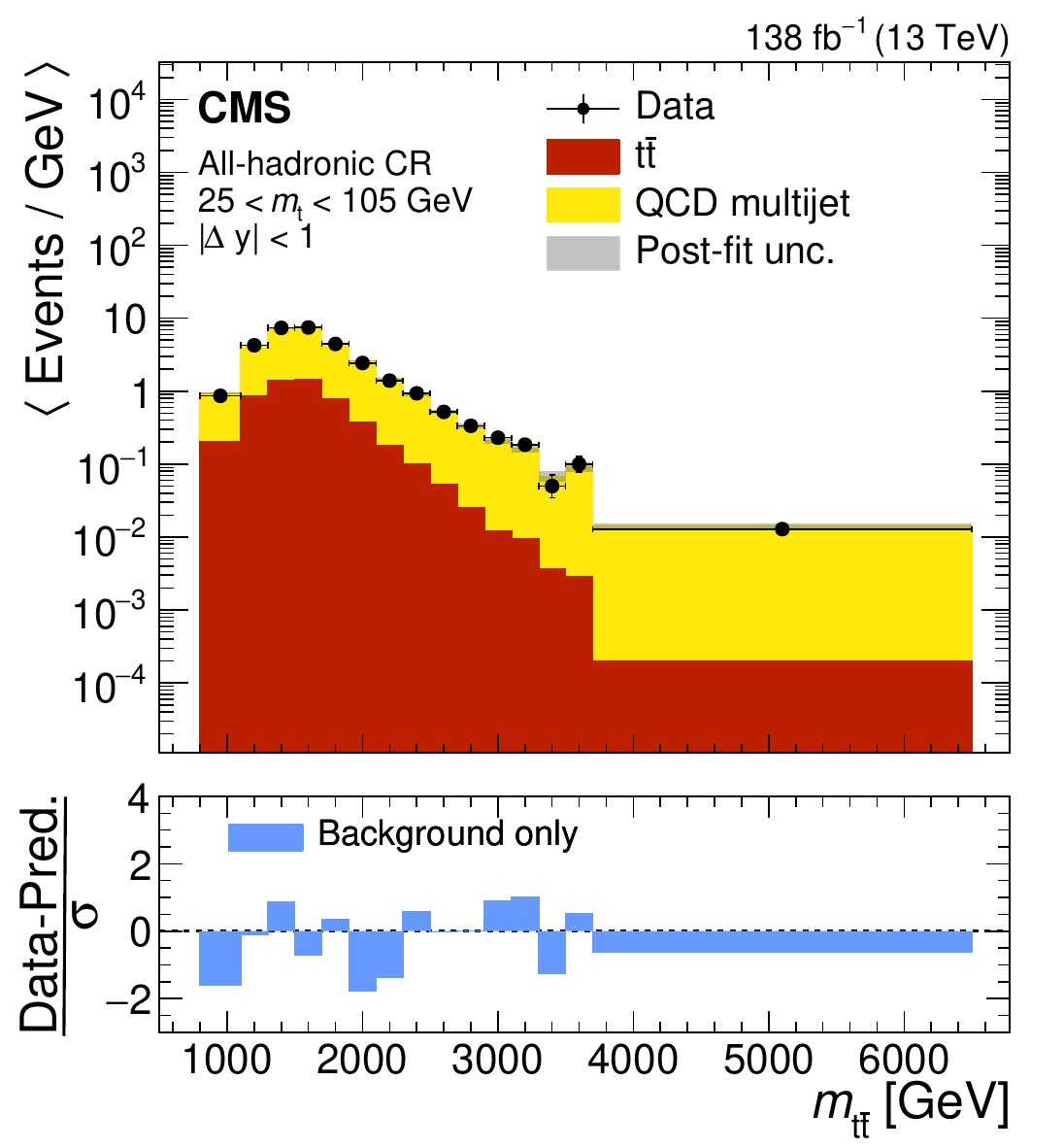}
\hspace{0.1\textwidth}
\includegraphics[width=0.4\textwidth]{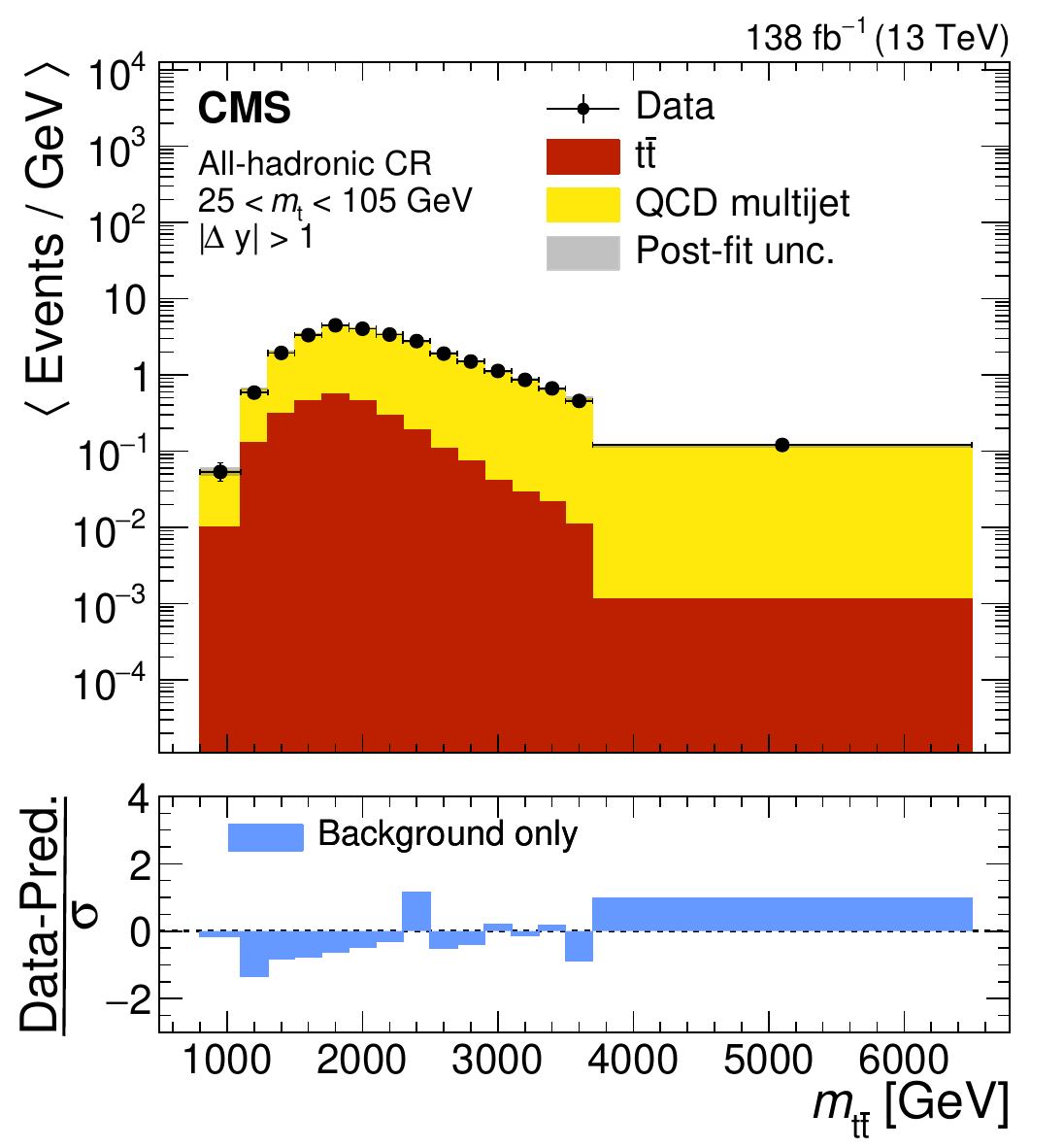} \\
\includegraphics[width=0.4\textwidth]{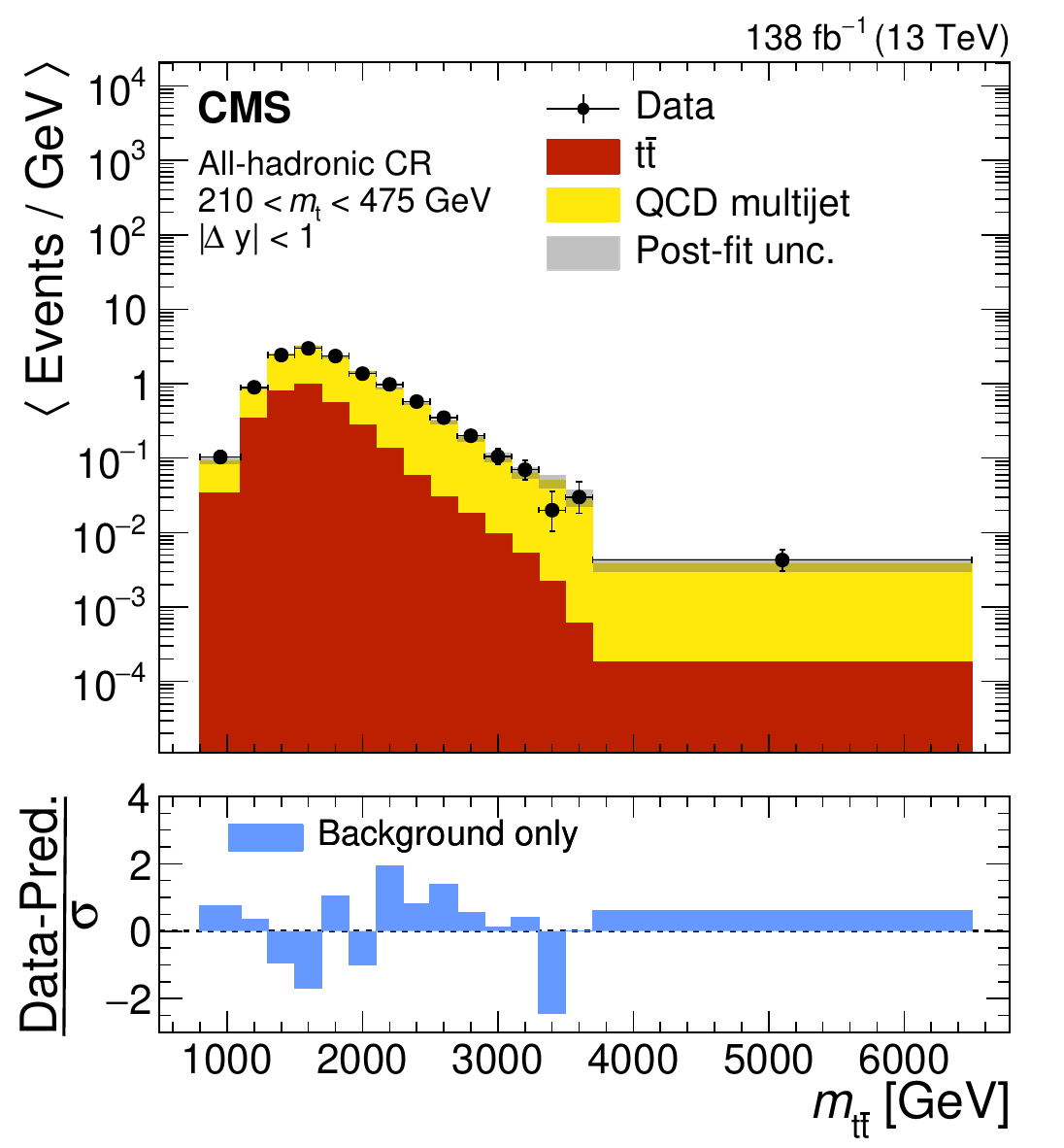}
\hspace{0.1\textwidth}
\includegraphics[width=0.4\textwidth]{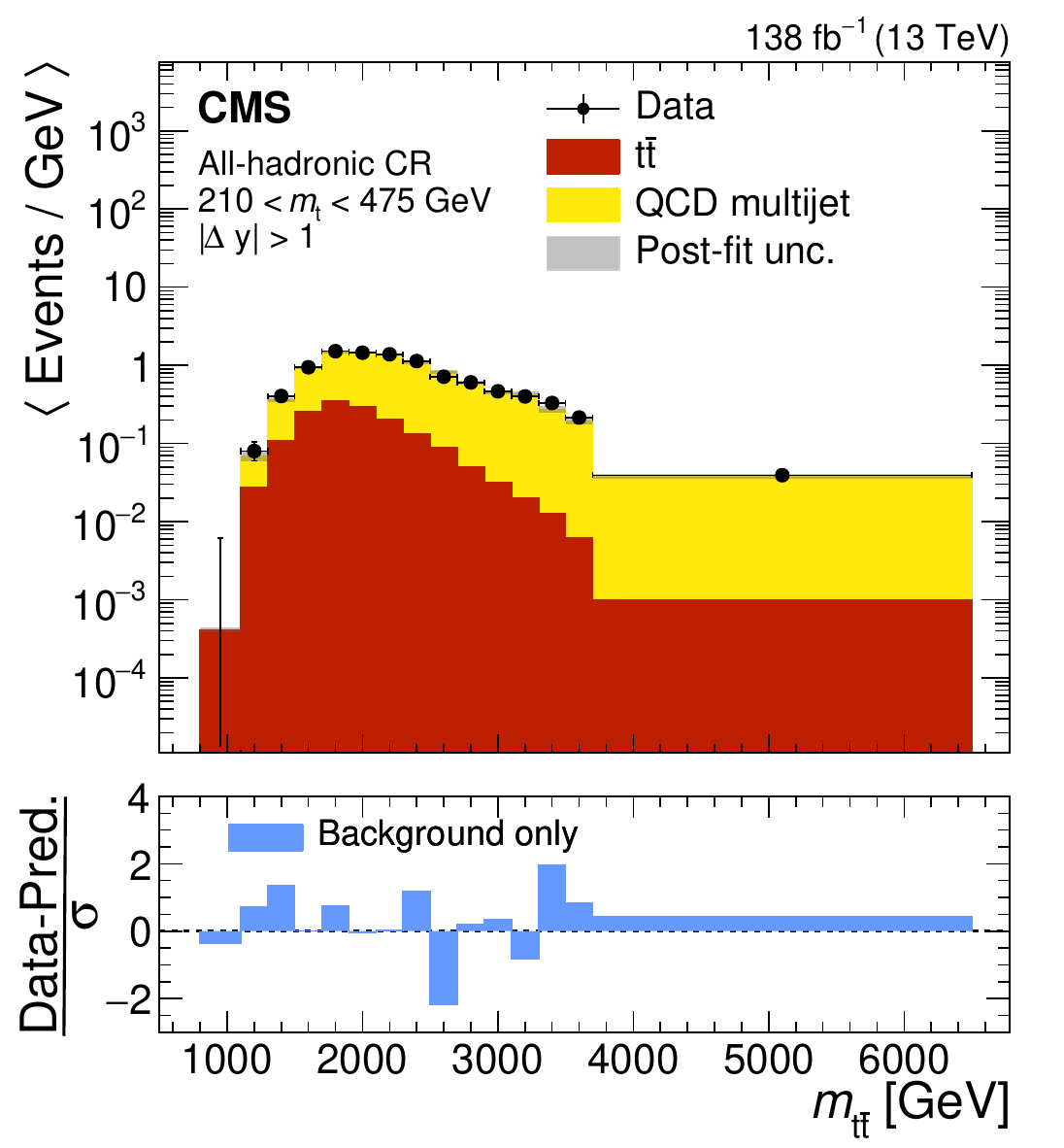} \\
\includegraphics[width=0.4\textwidth]{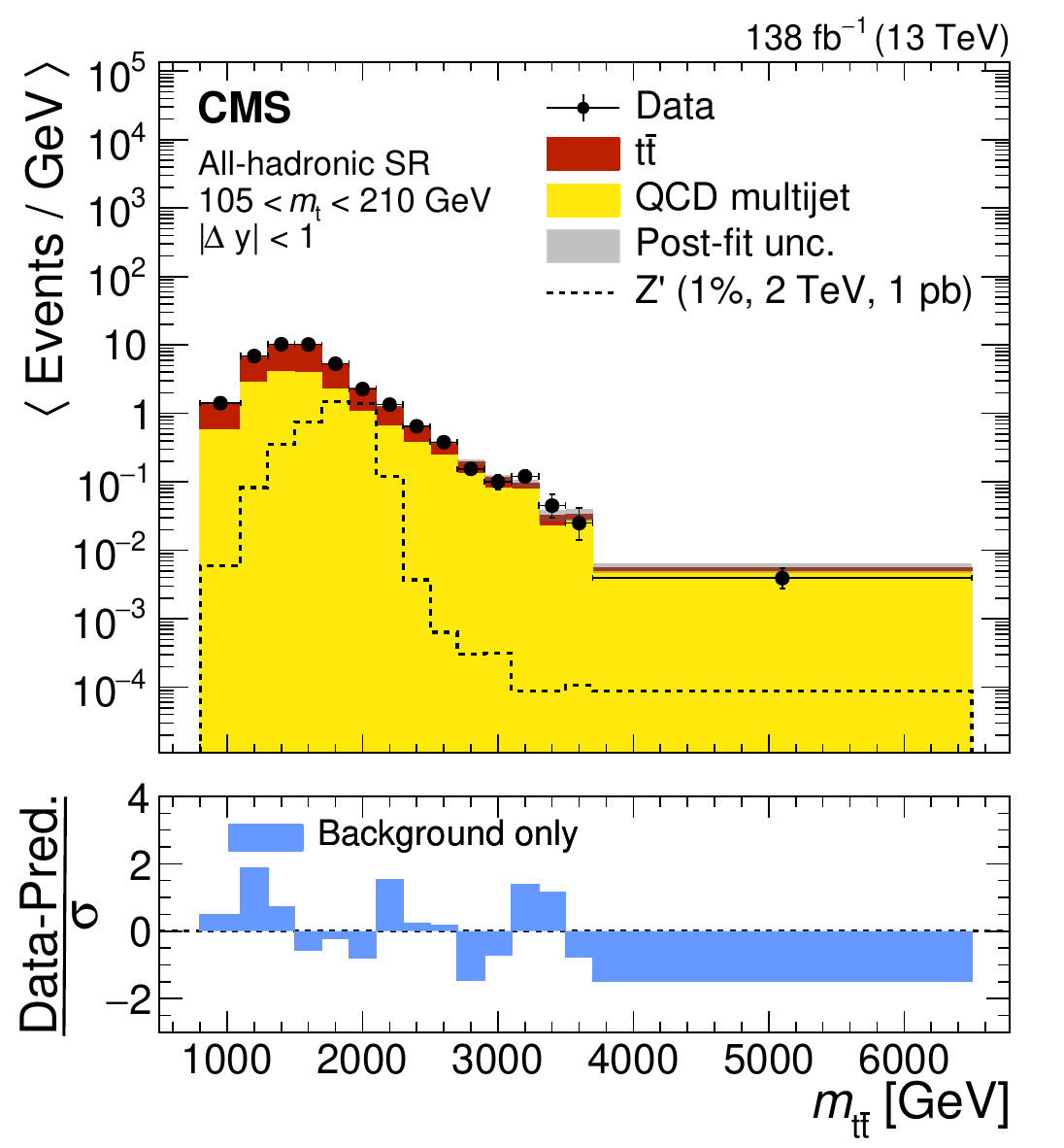}
\hspace{0.1\textwidth}
\includegraphics[width=0.4\textwidth]{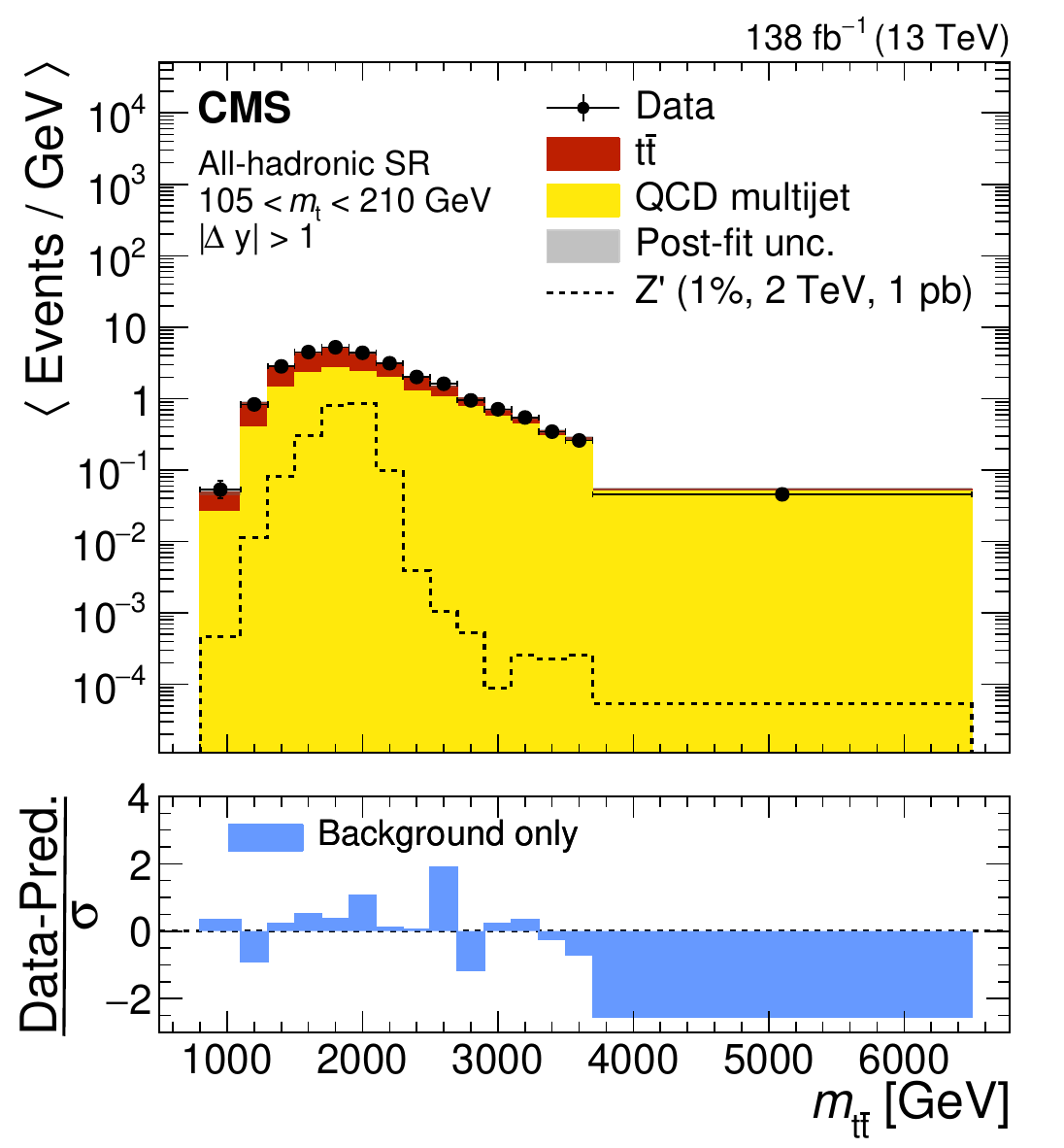}
\caption{Postfit distributions in \mtt for data and simulation for the central (left) and forward (right) categories for the all-hadronic channel, under the background-only hypothesis.
Distributions are shown for the low-\mtop (upper) and high-\mtop (middle) sidebands, as well as the SR (lower). The horizontal bars on the data points indicate the bin width. For illustrative purposes, the \Zprime boson signal with a relative width of 1\% and a mass of 2\TeV is normalized to a cross section of 1\unit{pb} and overlaid to the backgrounds in the signal regions.
The lower panels show the pulls, defined as $(\text{Data}-\text{Prediction})/\sigma$, where $\sigma$ denotes the total postfit uncertainty in each bin, relative to the SM prediction.}
\label{fig:bkgest_run2}
\end{figure}

The fit results in the \mtop sidebands and in the \mtop SR, obtained under the background-only hypothesis as described in Section~\ref{sec:results}, are presented in Fig.~\ref{fig:bkgest_run2}. The binning choice is a trade-off between optimizing signal sensitivity and ensuring fit stability by reducing statistical fluctuations. The impact of binning on narrow-width signals has been tested through a signal injection study and was found to be minimal, even when the signal falls within only a few bins. A good agreement between data and the total background prediction is observed.

\subsection{Single-lepton channel}

In the single-lepton channel, one top quark decays hadronically ($\PQt\to\PQb\PW\to\PQb\PQq\PAQq^\prime$) and the other top quark decays semileptonically ($\PQt\to\PQb\PW\to\PQb\Pell\PGn_{\Pell}$).

The selected events are placed into mutually exclusive categories based on their topology and the flavour of the reconstructed charged lepton.
In the resolved category, the decay products of the \PQt quarks are well-separated,
resulting in isolated leptons and \akfour jets.
In the merged category, the hadronic decay products of the \PQt quark
are collimated and reconstructed as a single \akeight jet,
while the semileptonic decay of the other \PQt quark produces a lepton that is usually non-isolated, requiring dedicated selections to separate it from the overlapping jet.

Events in the muon channel must have exactly one muon with $\pt>30\GeV$ and $\abs{\eta}<2.4$.
In the resolved categories, only isolated muons with $\pt<55\GeV$ are considered,
while in the merged category muons must have $\pt>55\GeV$, where the choice of \pt thresholds is determined by the trigger requirements.
The isolation methods and algorithms for the charged leptons used in the resolved categories
are detailed further in Refs.~\cite{CMS:2020uim, CMS:2018rym}.  In the merged categories, high-\pt muons must satisfy a two-dimensional isolation requirement
defined as:
\begin{equation}\label{eq:2dcut}
    \DR(\Pell,\text{jet})>0.4
    \quad\text{or}\quad
    \ptrel(\Pell,\text{jet})>25\GeV,
\end{equation}
where \DR is the separation
of the muon candidate from any \akfour jet with $\pt>15\GeV$, and \ptrel is
the muon momentum component perpendicular to the axis of the closest \akfour jet.
Events in the electron channel must have exactly one electron with $\abs{\eta}<2.5$
and $\pt>35/38/35\GeV$ for the years 2016/2017/2018, respectively.
The offline thresholds are dictated by the online trigger requirement of each data-taking period.
In the resolved category, isolated electrons with $\pt<120\GeV$ are considered,
as well as electrons that do not satisfy the isolation requirement defined in Eq.~\ref{eq:2dcut},
while in the merged category electrons must have $\pt>120\GeV$ and
satisfy the isolation requirement.
The higher \pt threshold used to distinguish between resolved and merged categories in the electron channel,
compared to the muon channel, is driven by the stricter electron \pt requirement ($\pt>115\GeV$)
for triggers that do not impose isolation criteria on electron candidates.

Events containing additional charged leptons with $\pt>25\GeV$
and $\abs{\eta}<2.4$ or missing transverse momentum $\ptmiss<70$ (60)\GeV are rejected in the muon (electron) category.
Additionally, the leading and subleading \akfour jets are required to have
$\pt>50\GeV$ and 50 (40)\GeV in the muon (electron) channels, respectively.
In both muon and electron channels, at least one \akfour jet has to be \btagged,
passing the medium working point of the \textsc{DeepJet} algorithm corresponding to
a 1\% misidentification rate for jets originating from light quarks or gluons,
and an efficiency of 70--80\% in selecting \PQb quark jets.
The $\Delta\eta$ between the two leading \akfour jets must be less than
3, to further reduce the QCD multijet background contribution.

\subsubsection{Reconstruction of the \texorpdfstring{\ttbar}{ttbar} system}

The reconstruction of the \ttbar system proceeds as follows.
First, the charged lepton and \ptmiss are assigned to the semileptonically decaying top quark.
It is assumed that there is an on-shell \PW boson and that the entire \ptmiss can be interpreted as the transverse momentum of the neutrino, allowing a quadratic equation to be used to derive the longitudinal component of the neutrino's momentum, resulting in 0, 1, or 2 real solutions~\cite{CMS:2022ged}.
In the absence of a real solution, the real part of the complex solutions is used.
In the case of two real solutions, both hypotheses are further tested for that event.

Events are divided into categories based on the presence or absence of one \ttagged jet.
The working point of the algorithm is chosen to correspond to a 1\% misidentification rate and to a signal efficiency of 55\%,
and only \akeight jets with \softdropmass in the range of 105--210\GeV are considered.
Events containing more than one \ttagged \akeight jet are rejected.
In the merged category, exactly one \akeight jet must be \ttagged.
The selected \ttagged \akeight jet is assigned to the hadronic leg of the decay.
Each \akfour jet in the event is assigned to the semileptonic top quark decay and used to build
a different hypothesis for the \ttbar system.
In the resolved category, all possible combinations of \akfour jets
are used to reconstruct both the semileptonic and hadronic decays of the \ttbar system.
For events with more than ten jets, only the leading ten are considered.

Finally, a single \ttbar hypothesis is selected for each event.
The chosen hypothesis is the one with the smallest $\chi^2$, defined as:
\begin{equation}
    \chi^2 = \left[\frac{\mlep-\mlepexp}{\sigmalep}\right]^2  + \left[\frac{\mhad-\mhadexp}{\sigmahad}\right]^2,
\end{equation}
{\tolerance=800
where \mlep and \mhad are the invariant masses of the reconstructed semileptonically and hadronically decaying top quarks, respectively.
The parameters \mlepexp, \mhadexp, \sigmalep, and \sigmahad are obtained from simulation
by matching the reconstructed objects to generator-level particles from the \ttbar decay.
The efficiency for correctly reconstructing the \ttbar system ranges from 50\% when no \ttagged jet is found, up to 70\% when a \ttagged jet is present in the event.
\par}

To remove events with misreconstructed top quarks and to reduce the background contribution,
a maximum value of 30 is requested for the $\chi^2$ value of each hypothesis.
This criterion is applied to the events in the SR.

\begin{figure}[!ht]
\centering
\includegraphics[width=0.45\textwidth]{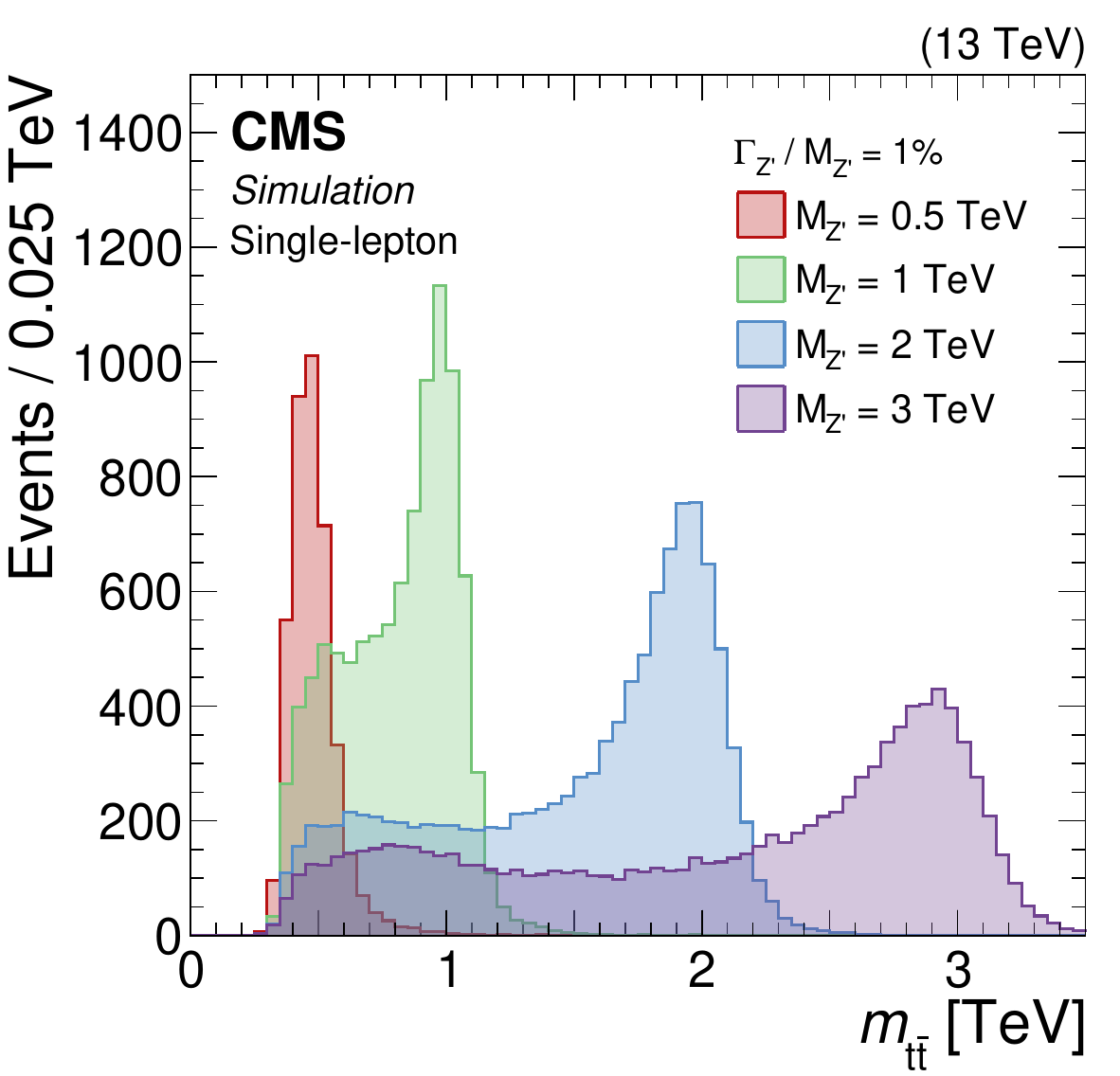}
\caption{Reconstructed invariant mass distribution in simulation in the single-lepton channel for \Zprime bosons with 1\% relative width, for different mass hypotheses.
Each distribution corresponds to a production cross section of 1\unit{pb}.}
\label{fig:1L_signals_Zprime}
\end{figure}

\begin{figure}[!ht]
\centering
\includegraphics[width=0.45\textwidth]{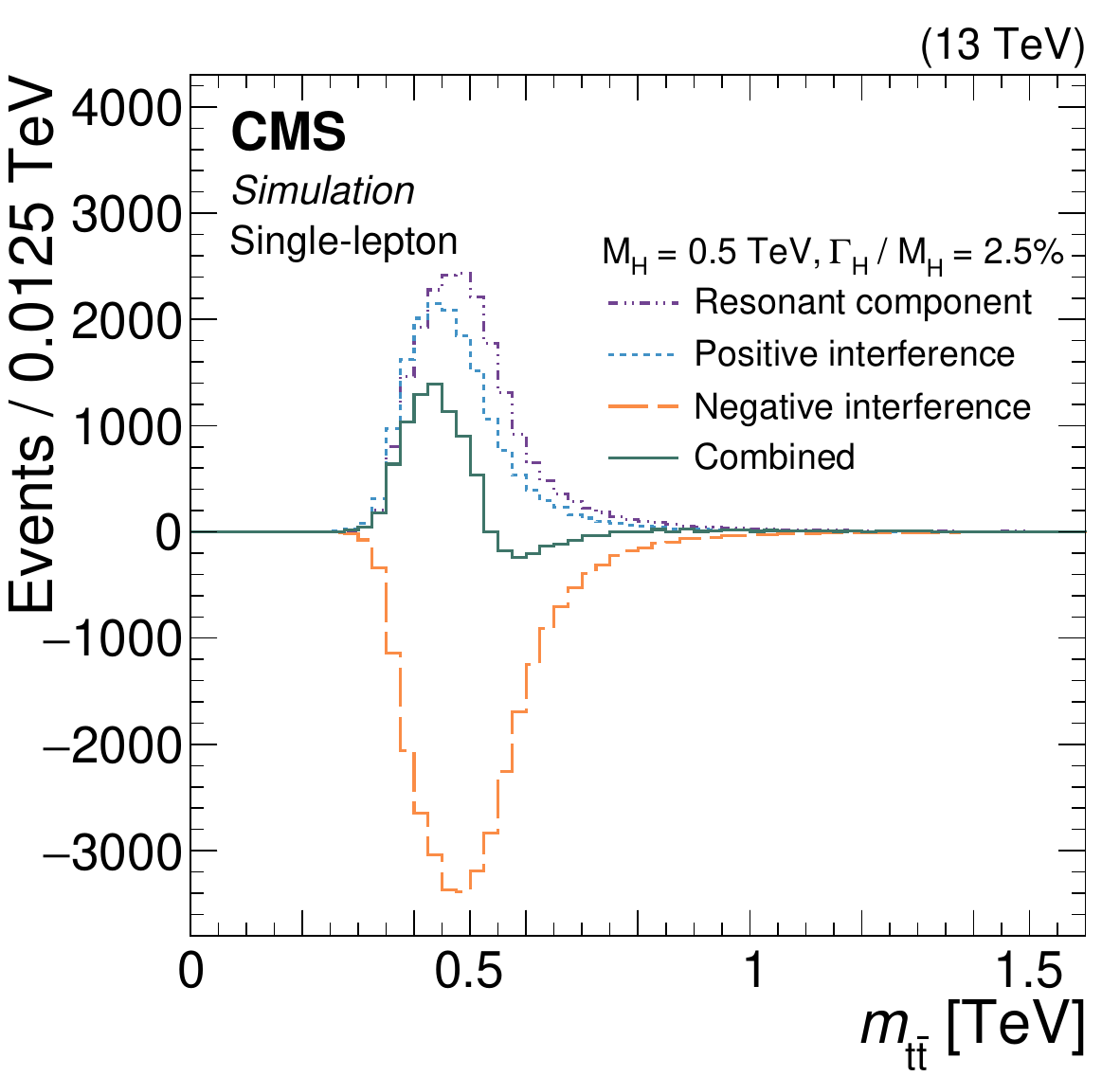}
\hfill
\includegraphics[width=0.45\textwidth]{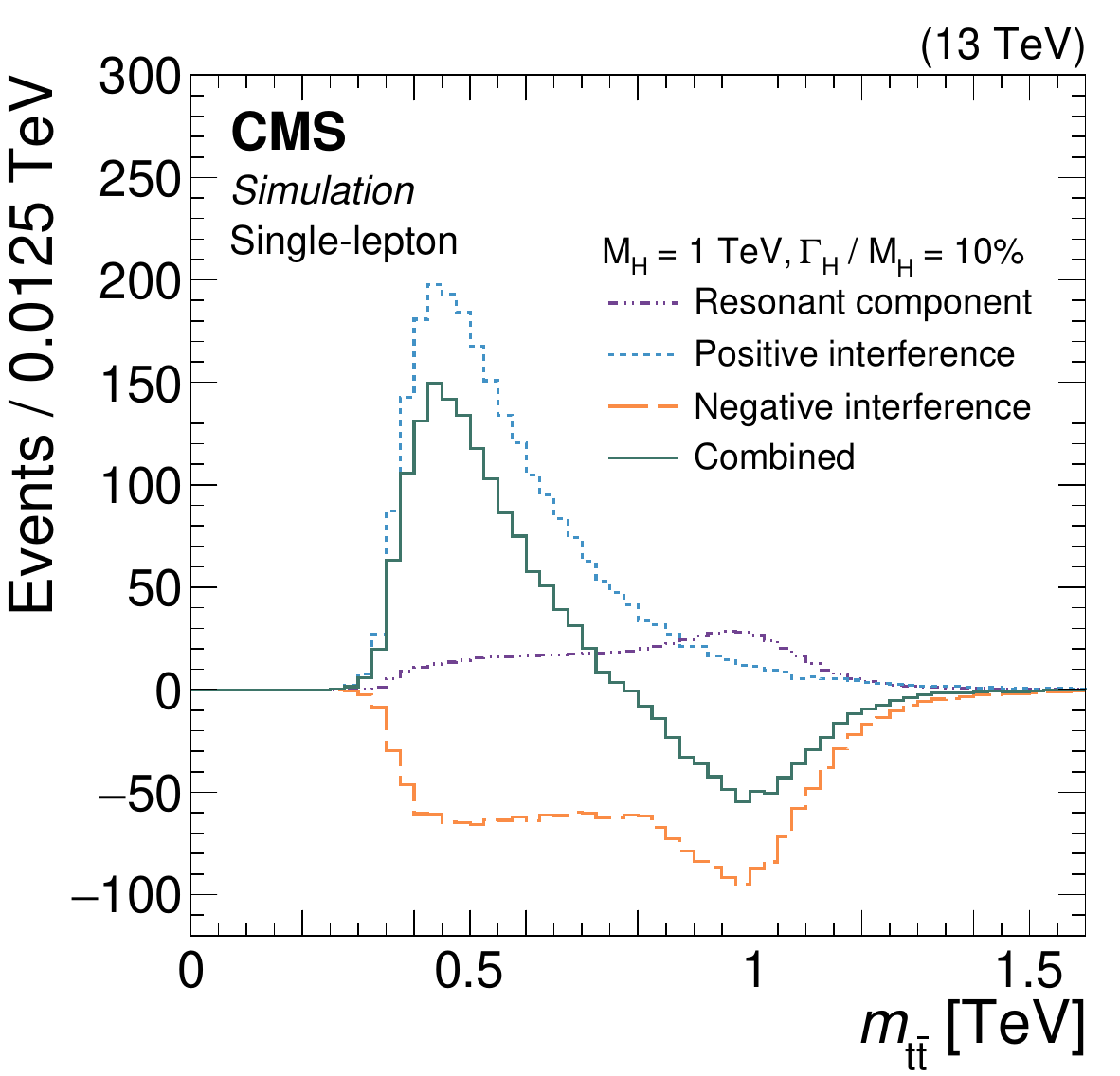}
\caption{Different contributions to the \mtt distribution in simulation in the single-lepton channel for scalar Higgs bosons
with masses of 0.5 (left) and 1\TeV (right), and corresponding relative widths of 2.5 and 10\%, respectively.
Each distribution is normalized to the corresponding production cross section.}
\label{fig:1L_signals_Higgs}
\end{figure}

Example distributions of reconstructed \mtt distribution
for different signal hypotheses are shown in Figs.~\ref{fig:1L_signals_Zprime} and~\ref{fig:1L_signals_Higgs}
for the \Zprime bosons with a 1\% relative width and for scalar \PH signals, respectively. A resonant structure can be
observed for the \Zprime boson signal with mass of 0.5\TeV, while at higher masses the off-shell contribution becomes
increasingly pronounced, resulting in a non-resonant component of the signals.
In the case of \PH/\PA bosons, the interference produces a peak-dip structure in the \ttbar invariant mass distribution.
The size of the interference depends on the mass, the width, and the $CP$-structure of the particle~\cite{Carena:2016,Djouadi:2019}.

\subsubsection{DNN for event classification}

To enhance the sensitivity of this search, we utilize a DNN
to distinguish events originating from various processes,
namely \ttbar, single \PQt, and \Vjets.

The DNN architecture comprises a fully connected feed-forward neural network, developed using \textsc{keras}~\cite{chollet2015keras},
featuring two hidden layers, each containing 512 nodes.
The DNN has three output nodes, corresponding to the three SM background processes.
To ensure that the DNN is model-agnostic, we exclusively consider SM processes during training.
Due to the different event selection criteria in the muon and electron categories,
two separate DNNs are trained to optimize the sensitivity.
The input variables are:
\ptmiss and its azimuthal angle; the four-momenta of the selected lepton, \akfour jets, and \akeight jets;
the \btagging scores and masses of the selected \akfour jets;
the top quark candidate mass \mtop, and the $\tau_{21}$ and $\tau_{32}$ $N$-subjettiness values of the selected \akeight jets;
and the multiplicity of the \akfour and \akeight jets.
The $N$-subjettiness~\cite{Thaler:2010tr,Thaler:2011gf} observables are calculated
using all PF candidates in the \akeight jet. Each corresponds to a \pt-weighted minimum
distance from one of $N$ hypothesized subjet axes, defined by the one-pass minimization
procedure. These observables are used to quantify the consistency of jet constituents
with an $N$-prong decay topology.
Up to five \akfour jets and three \akeight jets, sorted by \pt, are considered for each event,
resulting in a total of up to 59 input variables.

\begin{figure}[!bh]
\centering
\includegraphics[width=0.45\textwidth]{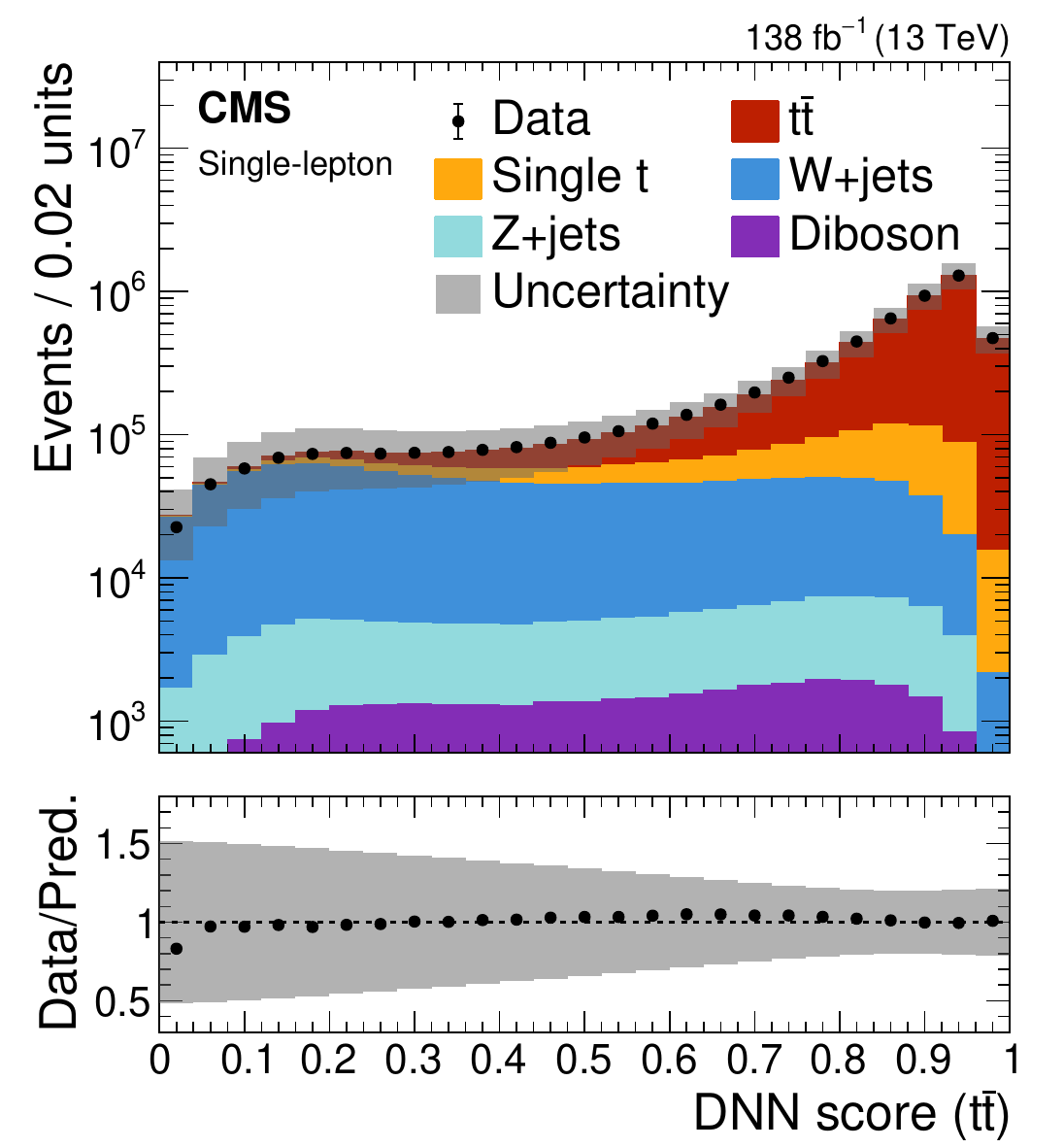}
\hfill
\includegraphics[width=0.45\textwidth]{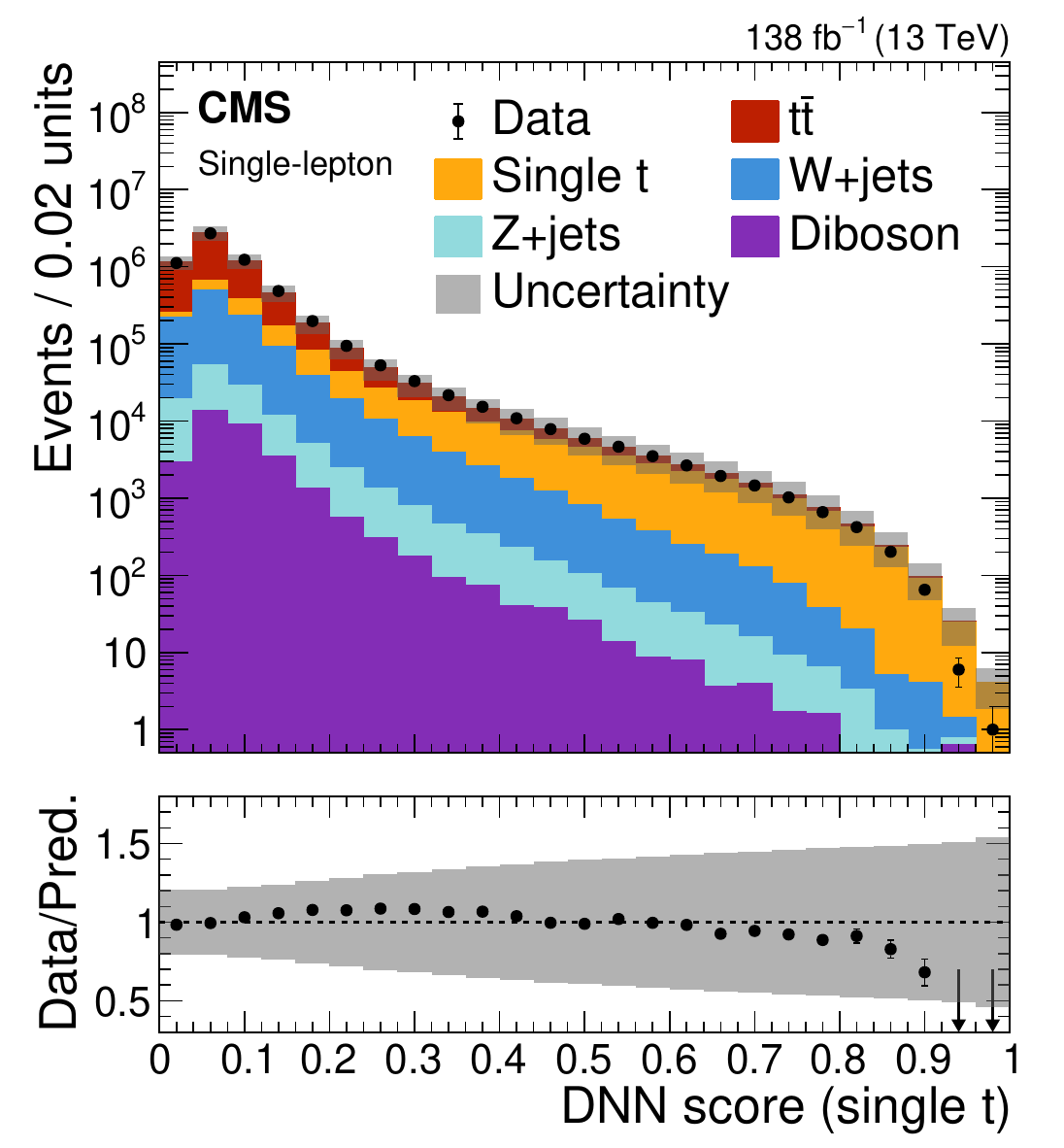} \\
\includegraphics[width=0.45\textwidth]{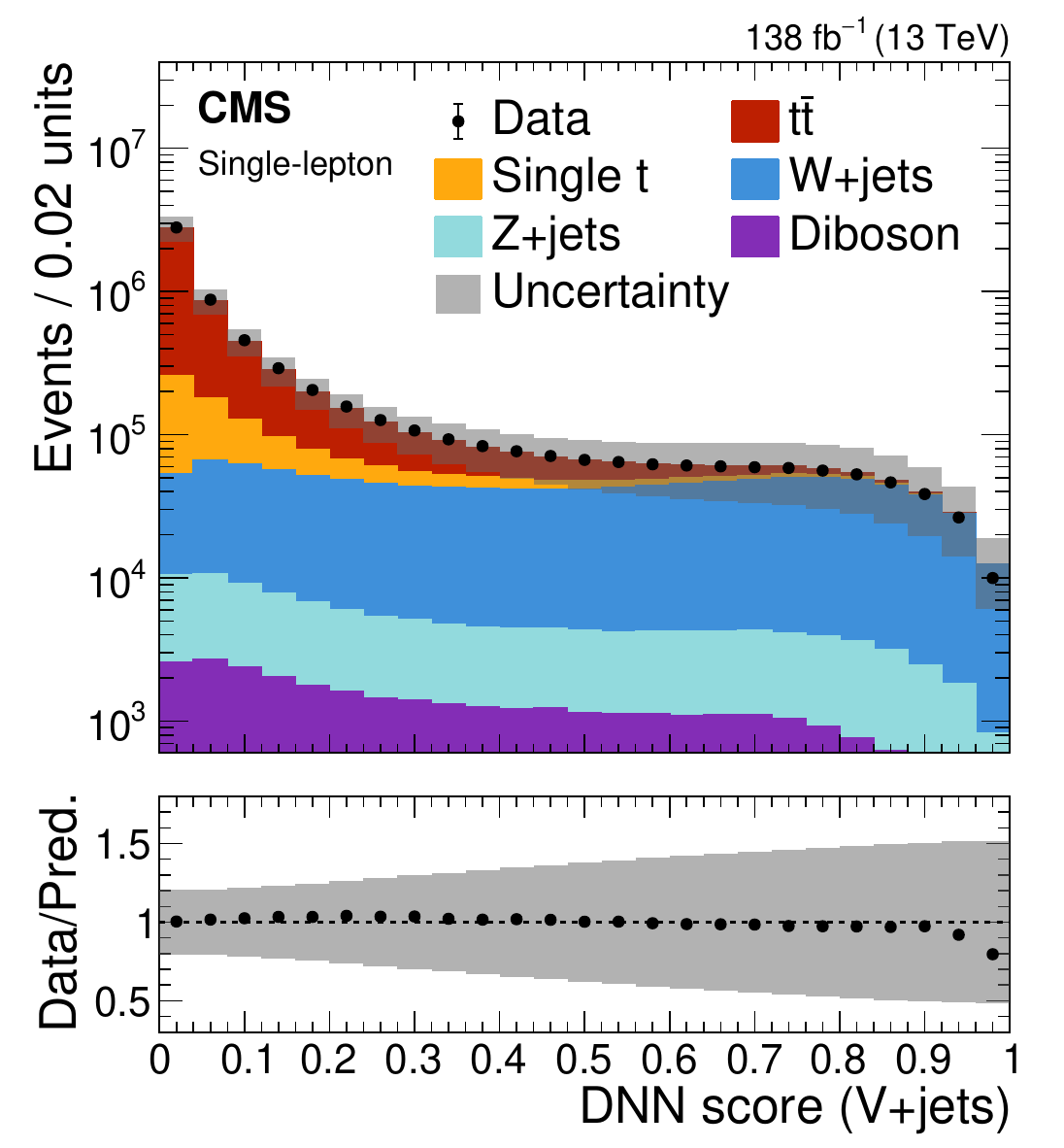}
\caption{The DNN score distributions for the combined muon and electron channels in the single-lepton channel:
\ttbar score (upper left), single \PQt score (upper right), and \Vjets score (lower).
The lower panels show the ratio of the data to the total SM background prediction.
The gray bands represent the uncertainty, computed by summing in quadrature the statistical uncertainty
and the systematic uncertainties affecting the normalization of each process. These observables are not fitted to extract the final results; the uncertainties are the prefit values.}
\label{fig:1L_dnn_outputnodes_stacked}
\end{figure}

The DNN score distributions for the combined electron and muon channels
for each class are shown in Fig.~\ref{fig:1L_dnn_outputnodes_stacked}.
The three DNN scores are used to classify events into a \ttbar-dominated SR,
and two CRs enriched in events originating from single \PQt and \Vjets processes.
Each event is assigned exclusively to a single category based on its highest output score.
In the SR, events are further categorized into the merged and resolved categories based on the presence of a \ttagged \akeight jet.
Events in the resolved SRs and in the CRs have a minimum value of \mtt of 350 \GeV, while events in the merged
SRs have a minimum \mtt value of 600\GeV.

\subsubsection{Event categorization based on spin correlation}

\begin{figure}[!ph]
\centering
\includegraphics[width=0.45\textwidth]{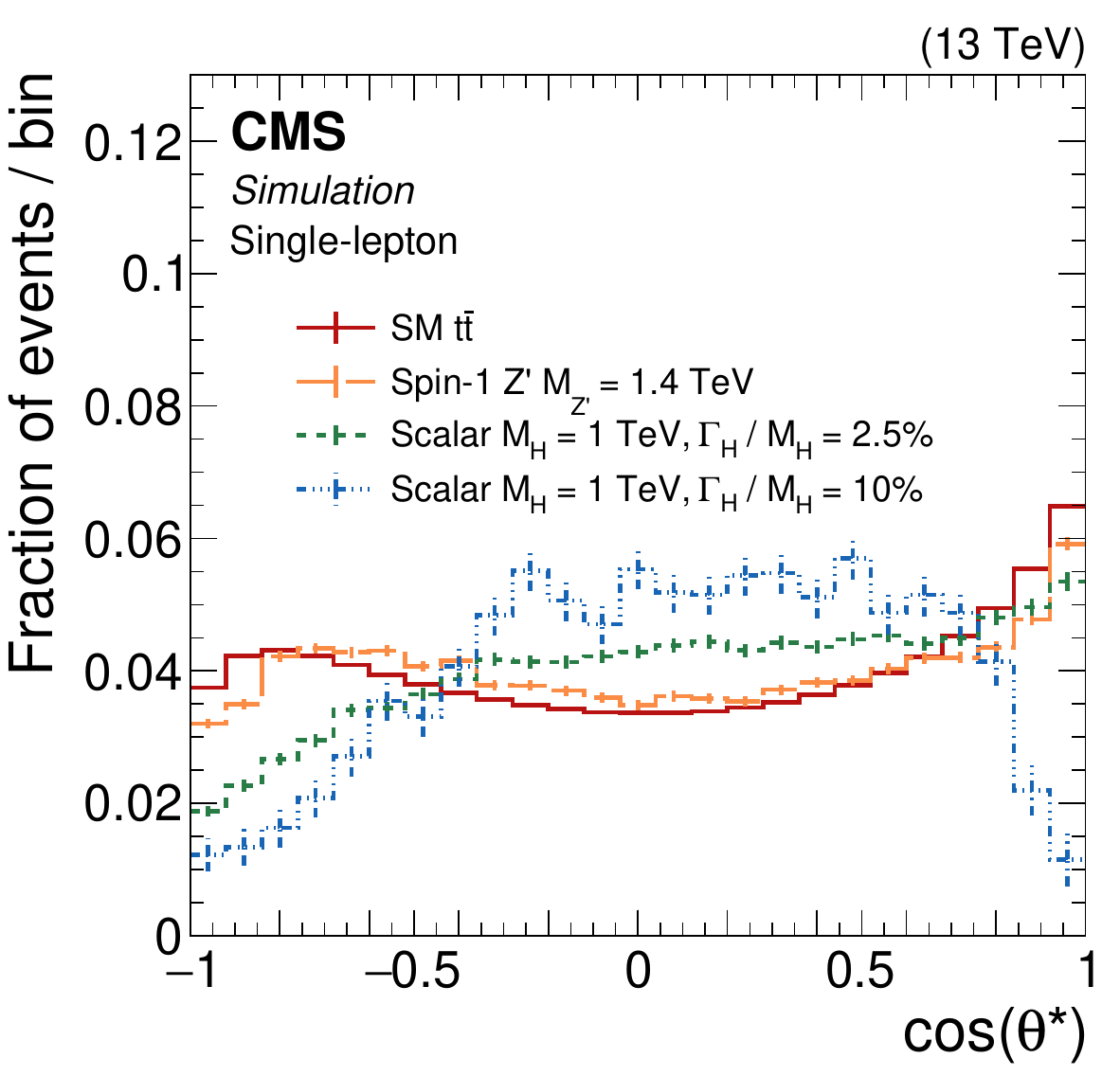}
\caption{Distribution of \costhetastar for different processes in simulation in the single-lepton channel:
SM \ttbar (solid red),
\Zprime with $\mtt=1.4\TeV$ (long-dashed orange),
scalar \PH with $\mH=1\TeV$ and 2.5\% relative width (short-dashed green), and
scalar \PH with $\mH=1\TeV$ and 10\% relative width (dash-dotted blue).
All distributions are normalized to unit area.}
\label{fig:1L_thetastar}
\end{figure}

To further improve the separation between the SM \ttbar background and the signal,
events in the SR are split according to their value of \costhetastar,
defined as the cosine of the angle between the momentum of the semileptonically decaying top quark
in the \ttbar rest frame and the momentum of the \ttbar system in the laboratory frame~\cite{CMS:HIG17027}.
This angular variable exploits spin correlation effects in top quark pair production resulting
in a shape difference between background and signal processes, as shown in Fig.~\ref{fig:1L_thetastar}.
While background events peak at values of $\costhetastar={\pm}1$, with a preference for positive values,
the distribution for spin-0 signal events is more isotropic.
For spin-1 resonances, the shape of the distribution is similar to the one for \ttbar,
and it varies depending on the mass and width of the resonance.
Six categories are defined based on the value of \costhetastar, with bin edges $[-1, -0.7, -0.5, 0.0, 0.5, 0.7, 1]$.
The binning is chosen to reflect shape differences between the signal and the SM \ttbar background,
with finer bins used in the regions where larger relative variations are observed.
The final event categorization includes ten SRs, given by the six bins in \costhetastar, four of which are further divided into
merged and resolved categories, and two CRs.
The last two bins, from 0.5 to 1 in \costhetastar, are not split into resolved and merged categories
because of a small number of expected events.

\begin{figure}[!ph]
\centering
\includegraphics[width=0.45\textwidth]{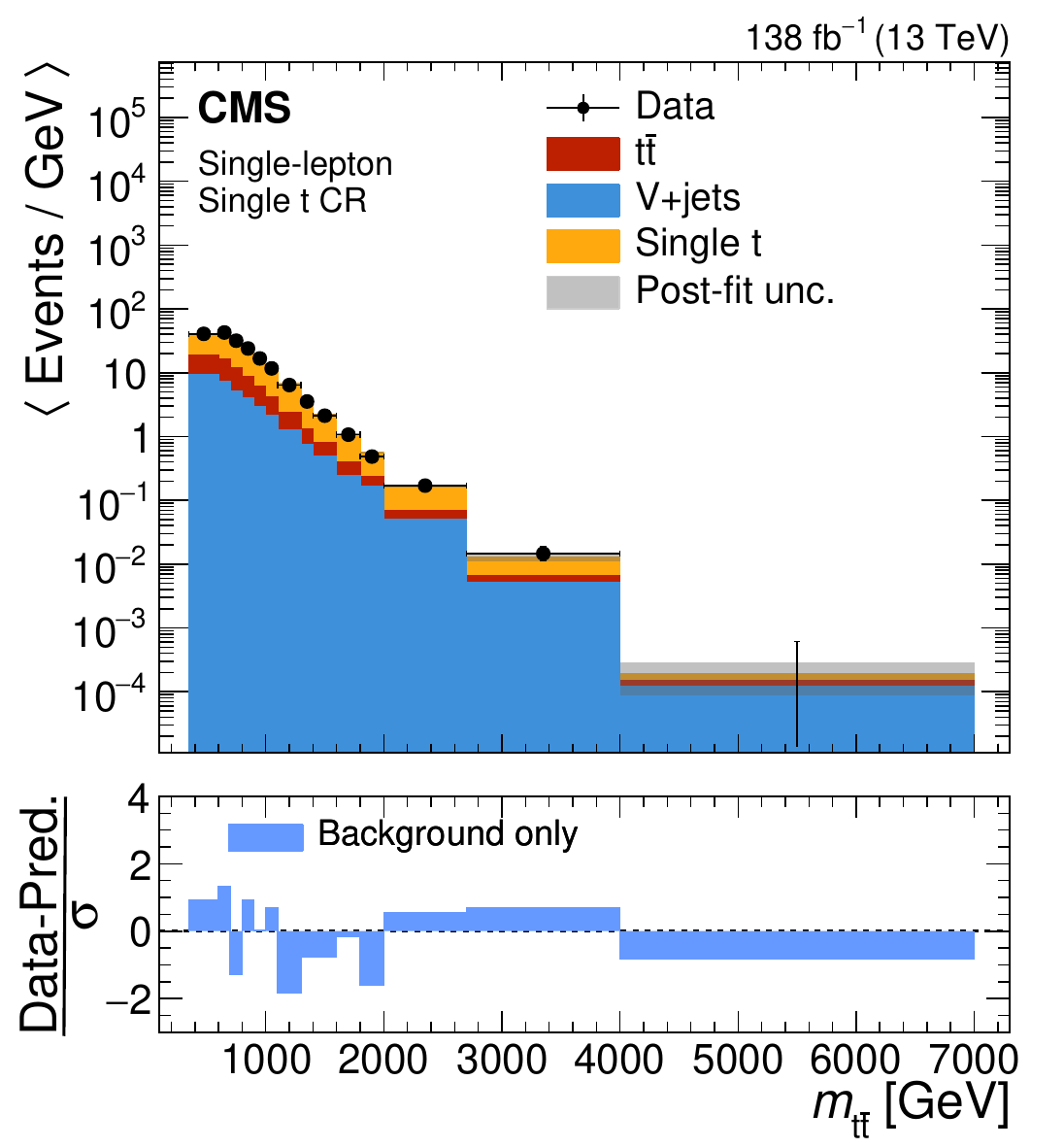}
\hfill
\includegraphics[width=0.45\textwidth]{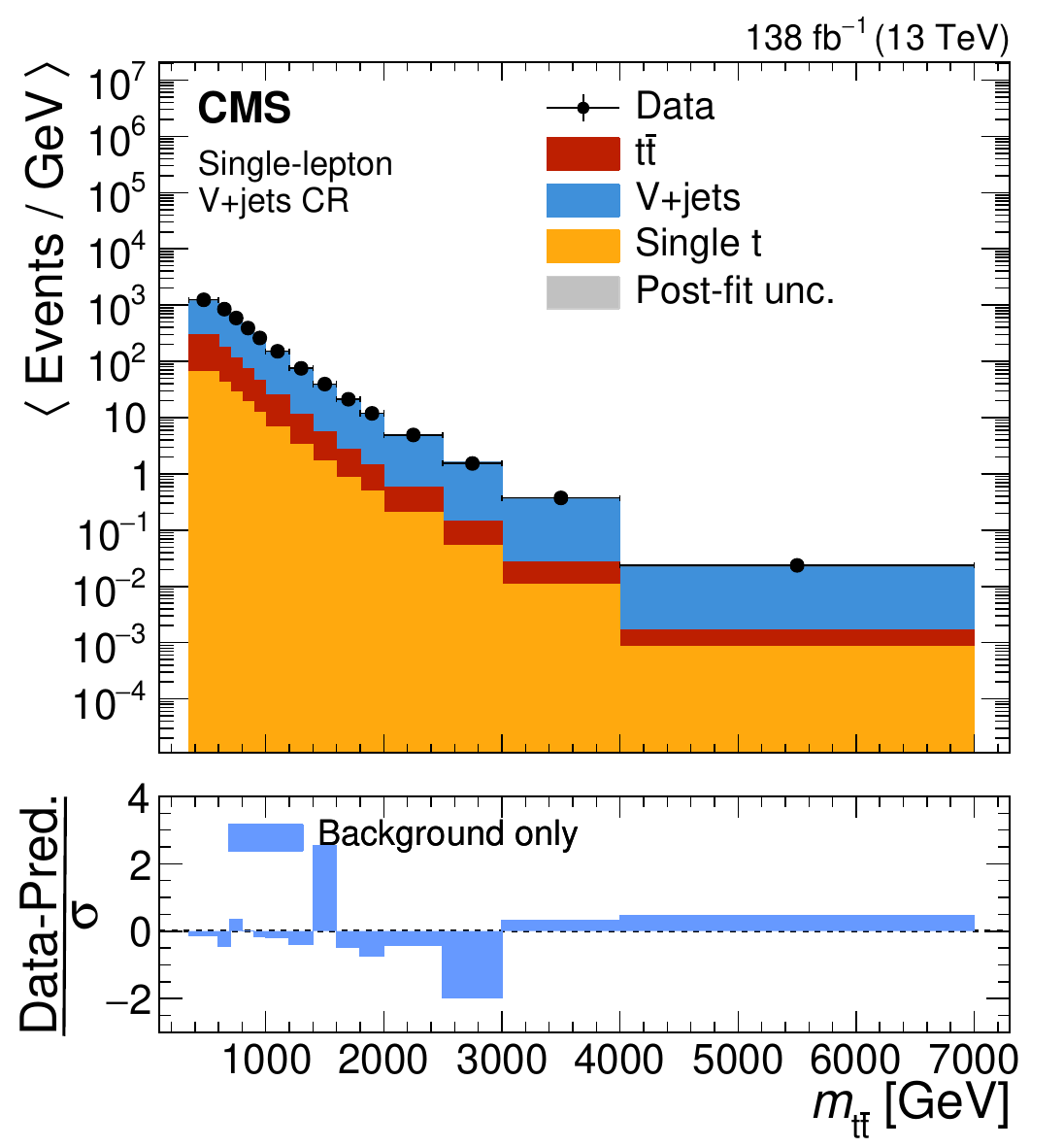}
\caption{Postfit distributions in \mtt in the single-lepton channel for data and simulation in the single \PQt (left) and \Vjets (right) CRs, under the background-only hypothesis. The horizontal bars on the data points indicate the bin width.
The lower panels show the pulls, defined as $(\text{Data}-\text{Prediction})/\sigma$, where $\sigma$ denotes the total postfit uncertainty in each bin, relative to the SM prediction.}
\label{fig:1L_mtt_CR}
\end{figure}

\begin{figure}[!ph]
\centering
\includegraphics[width=0.4\textwidth]{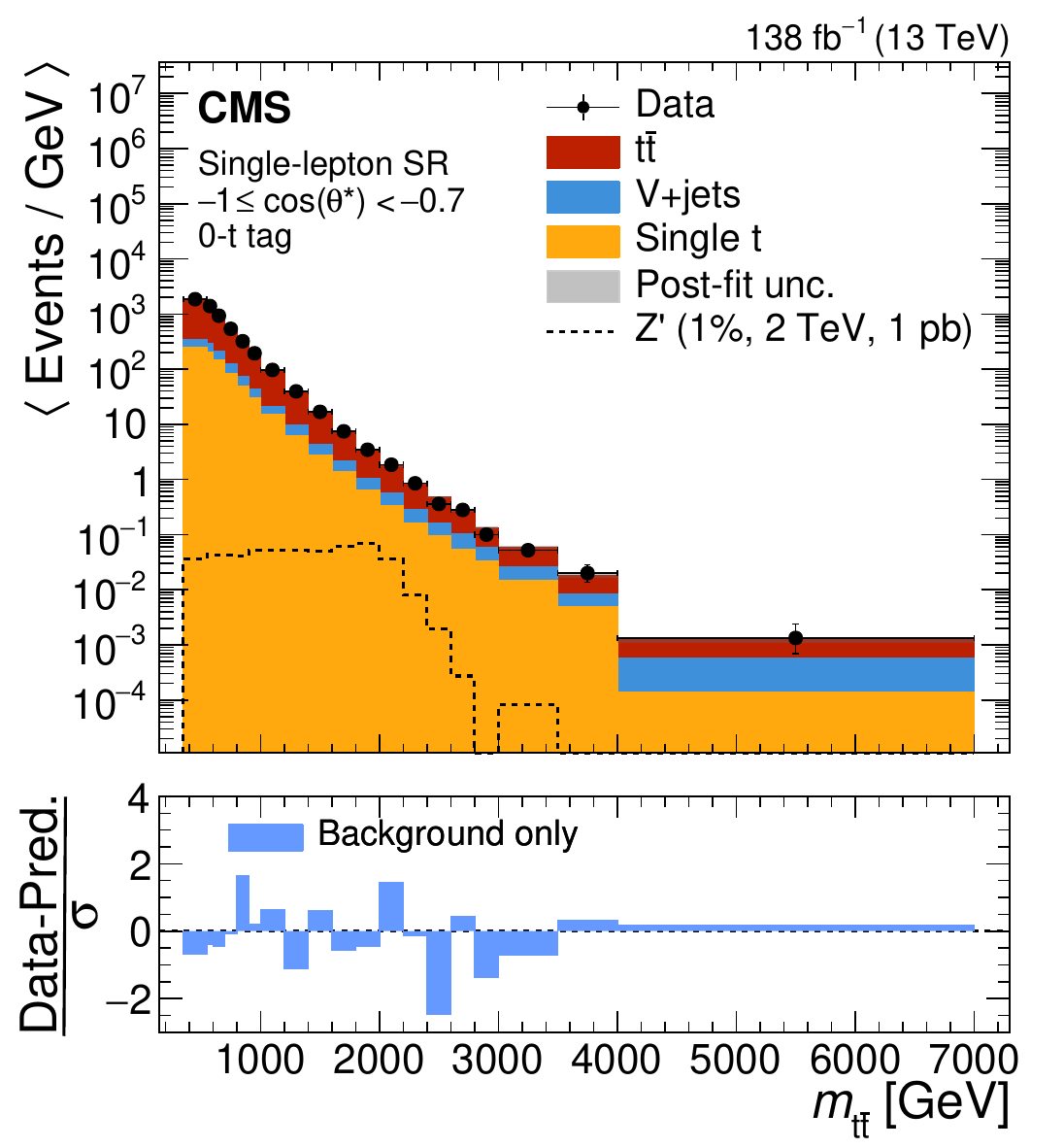}
\hspace{0.1\textwidth}
\includegraphics[width=0.4\textwidth]{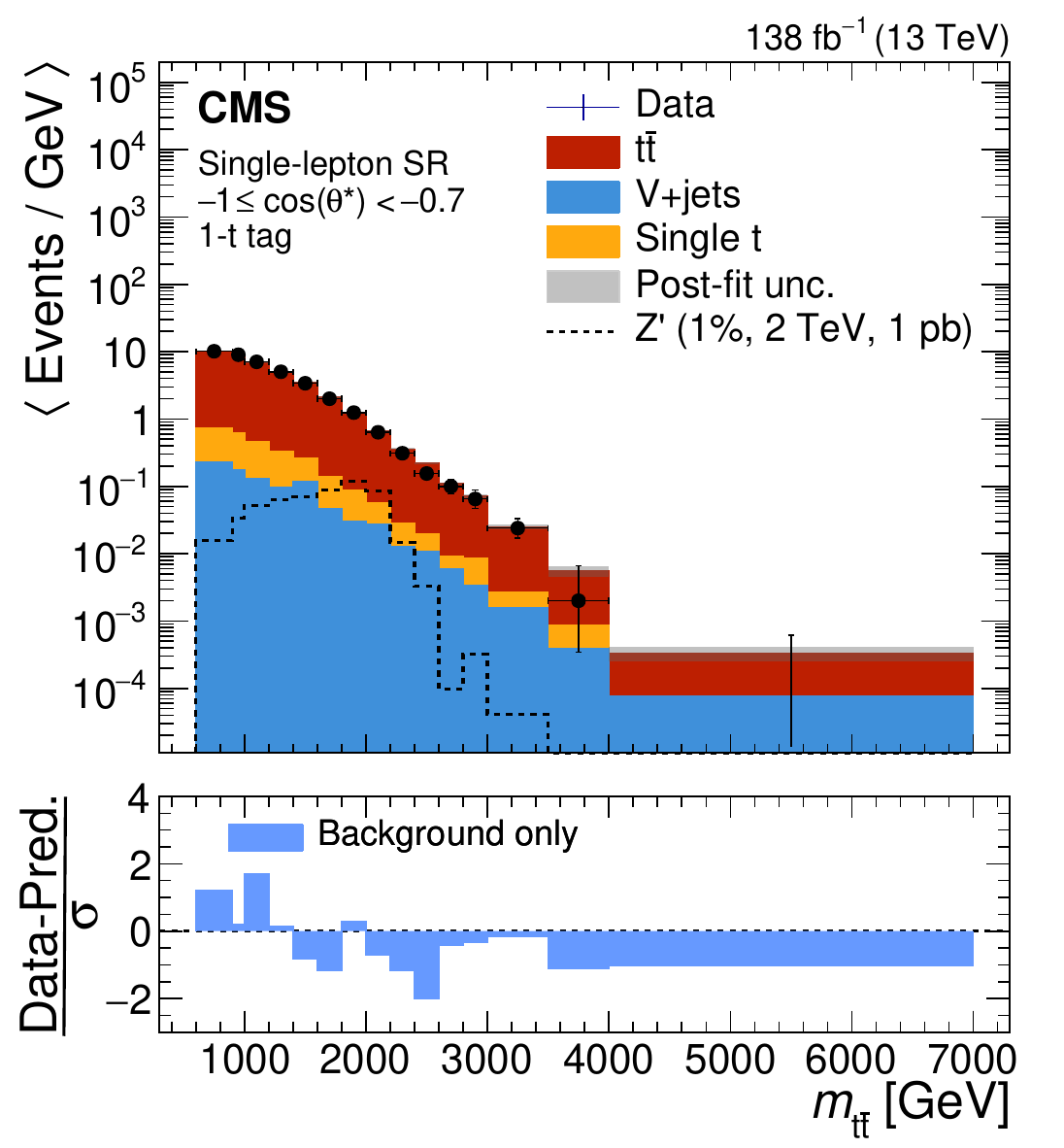} \\
\includegraphics[width=0.4\textwidth]{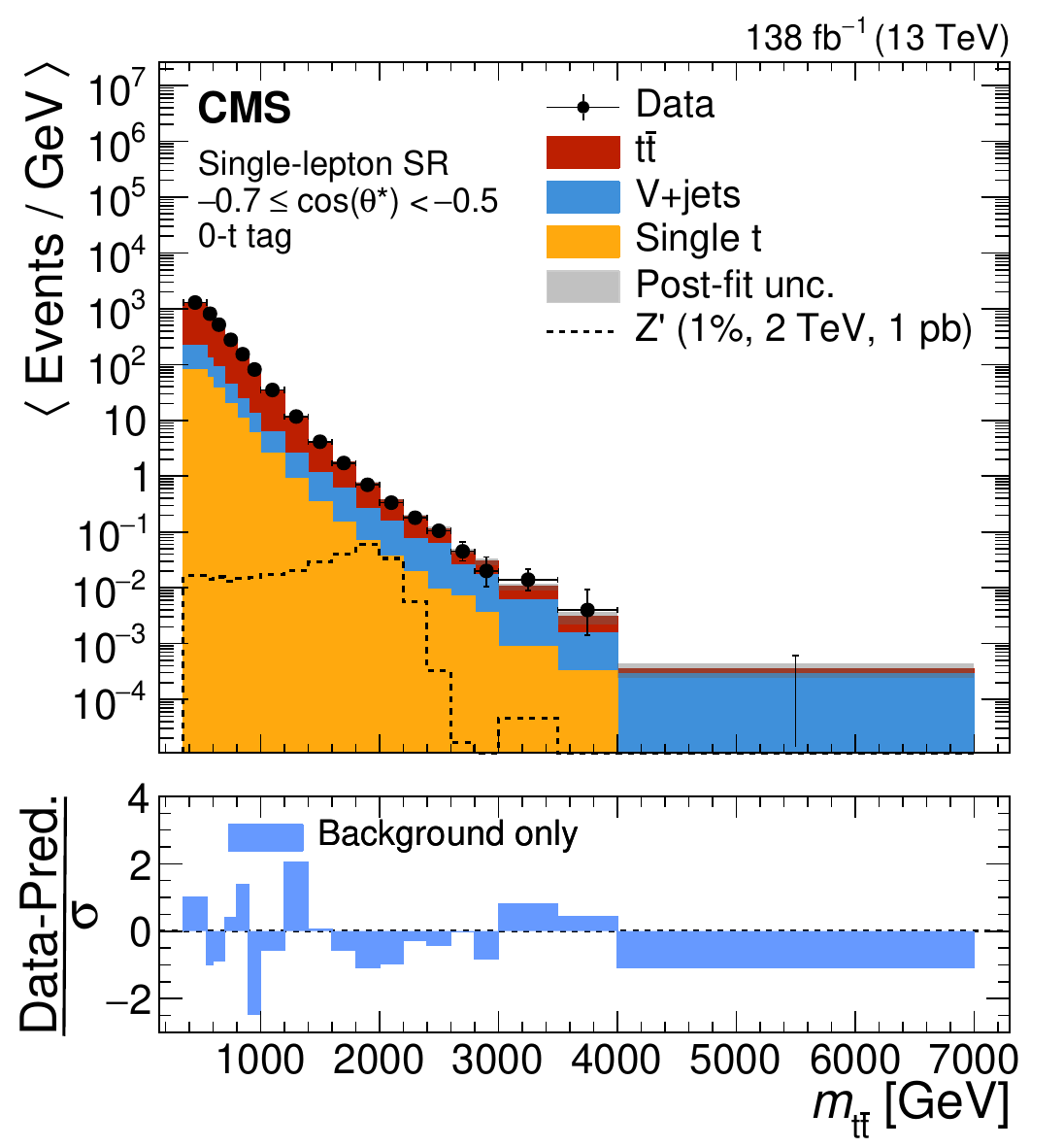}
\hspace{0.1\textwidth}
\includegraphics[width=0.4\textwidth]{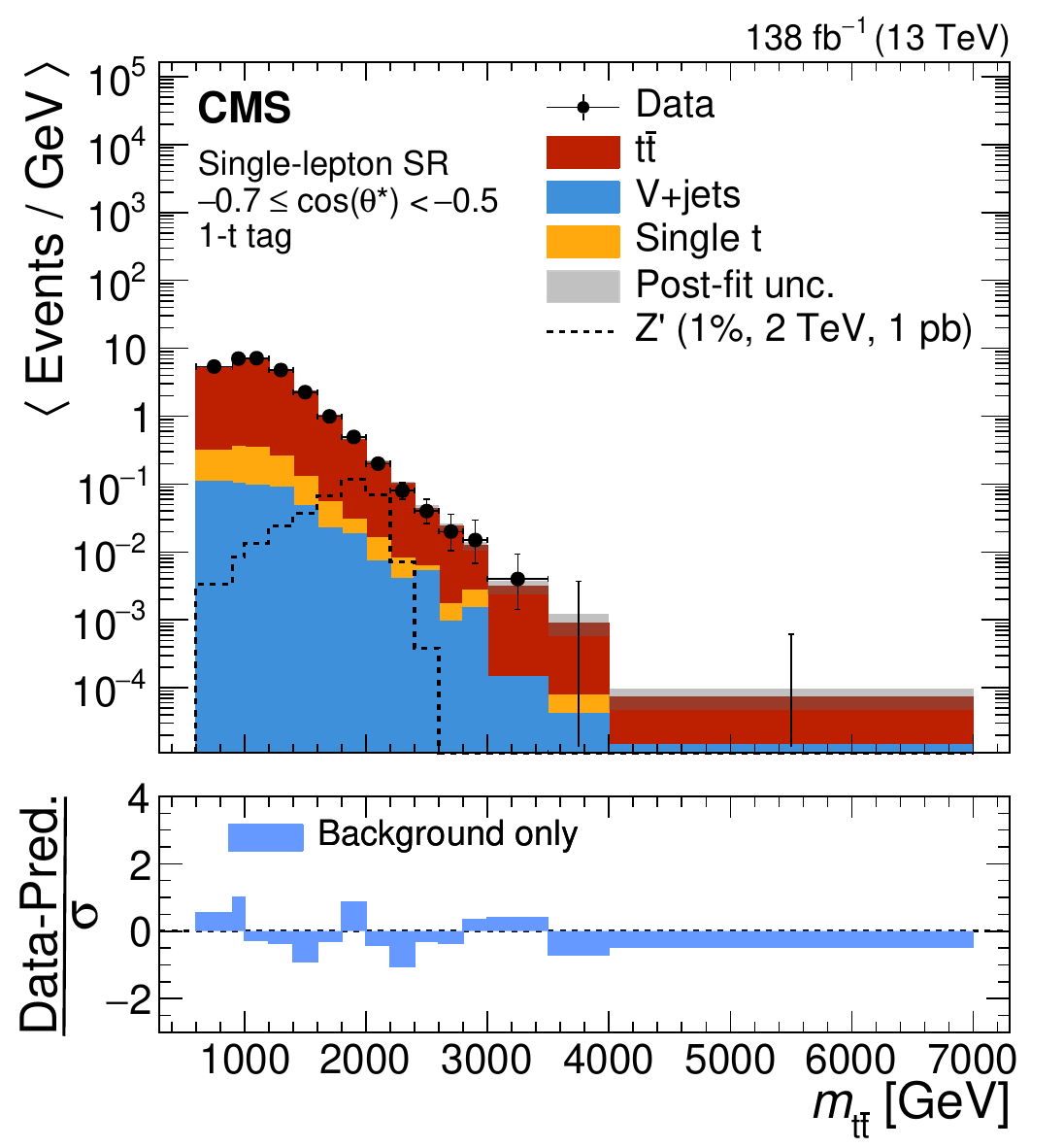} \\
\includegraphics[width=0.4\textwidth]{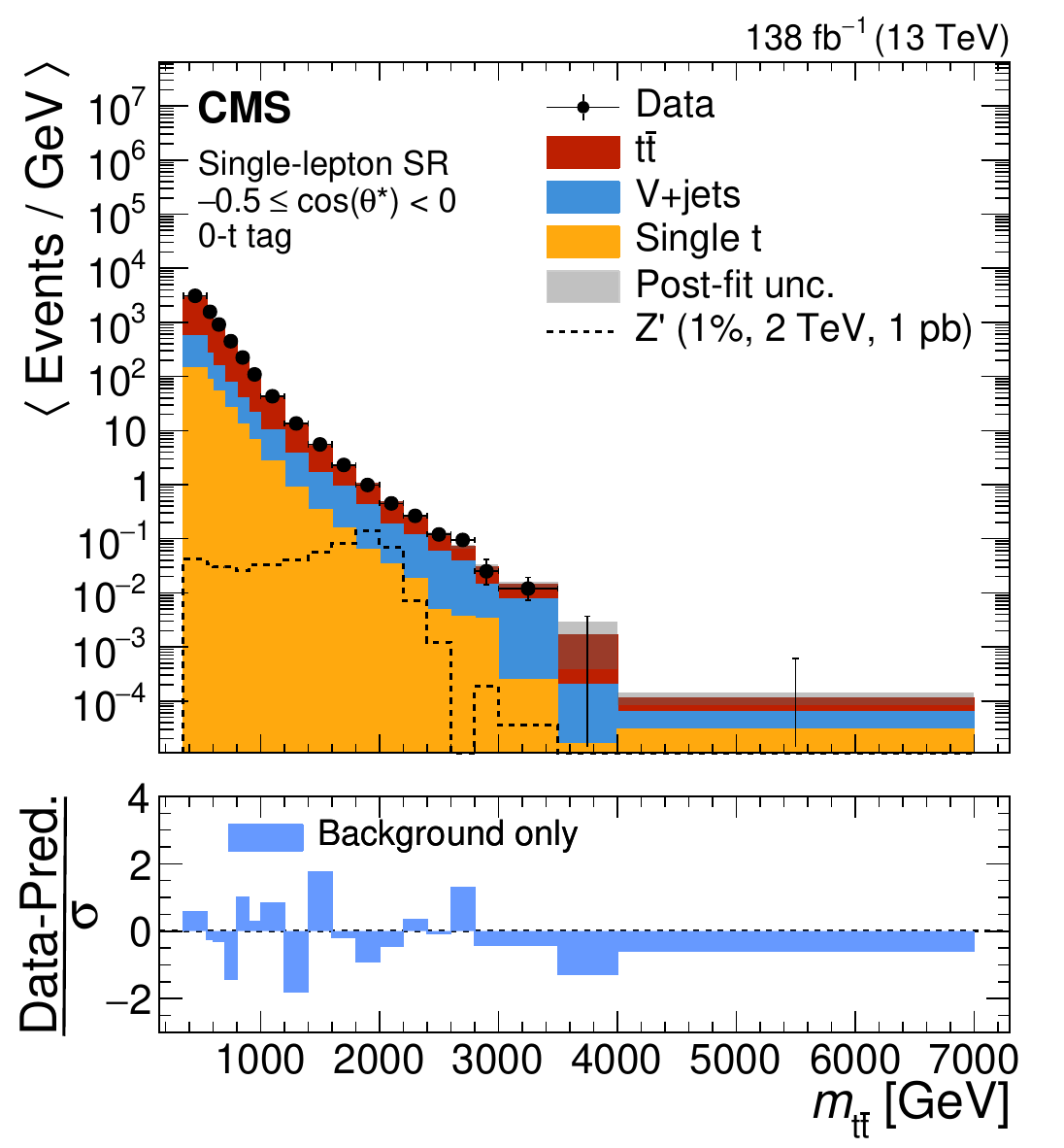}
\hspace{0.1\textwidth}
\includegraphics[width=0.4\textwidth]{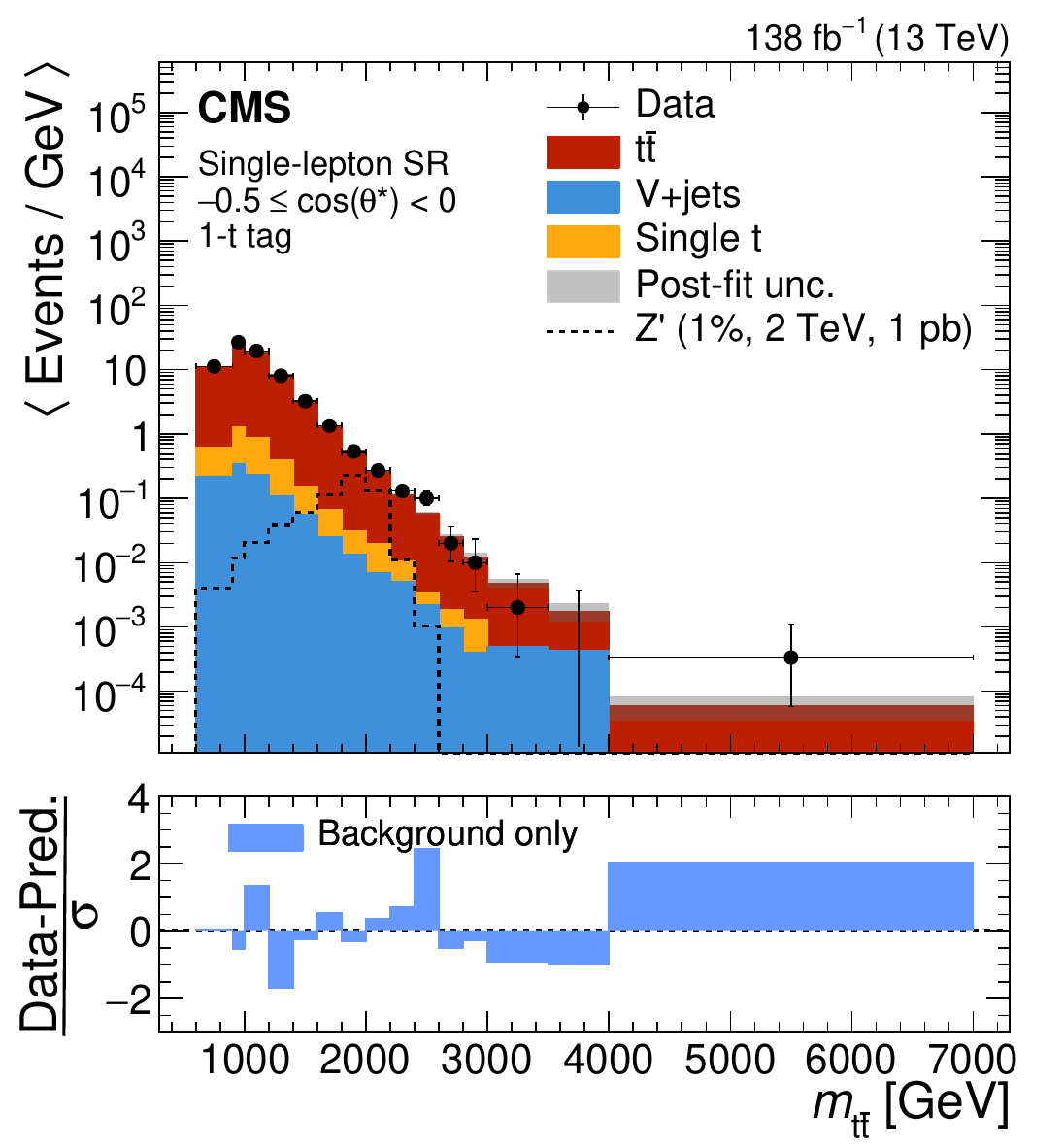}
\caption{Postfit distributions in \mtt in the single-lepton channel for data and simulation in the first three bins of \costhetastar in the \ttbar SR,
shown for the resolved (\zttag, left) and merged (\ottag, right) categories, under the background-only hypothesis. The horizontal bars on the data points indicate the bin width. For illustrative purposes, the \Zprime boson signal with a relative width of 1\% and a mass of 2\TeV is normalized to a cross section of 1\unit{pb} and overlaid to the backgrounds.
The lower panels show the pulls, defined as $(\text{Data}-\text{Prediction})/\sigma$, where $\sigma$ denotes the total postfit uncertainty in each bin, relative to the SM prediction.}
\label{fig:1L_mtt_SR1}
\end{figure}

\begin{figure}[!ht]
\centering
\includegraphics[width=0.45\textwidth]{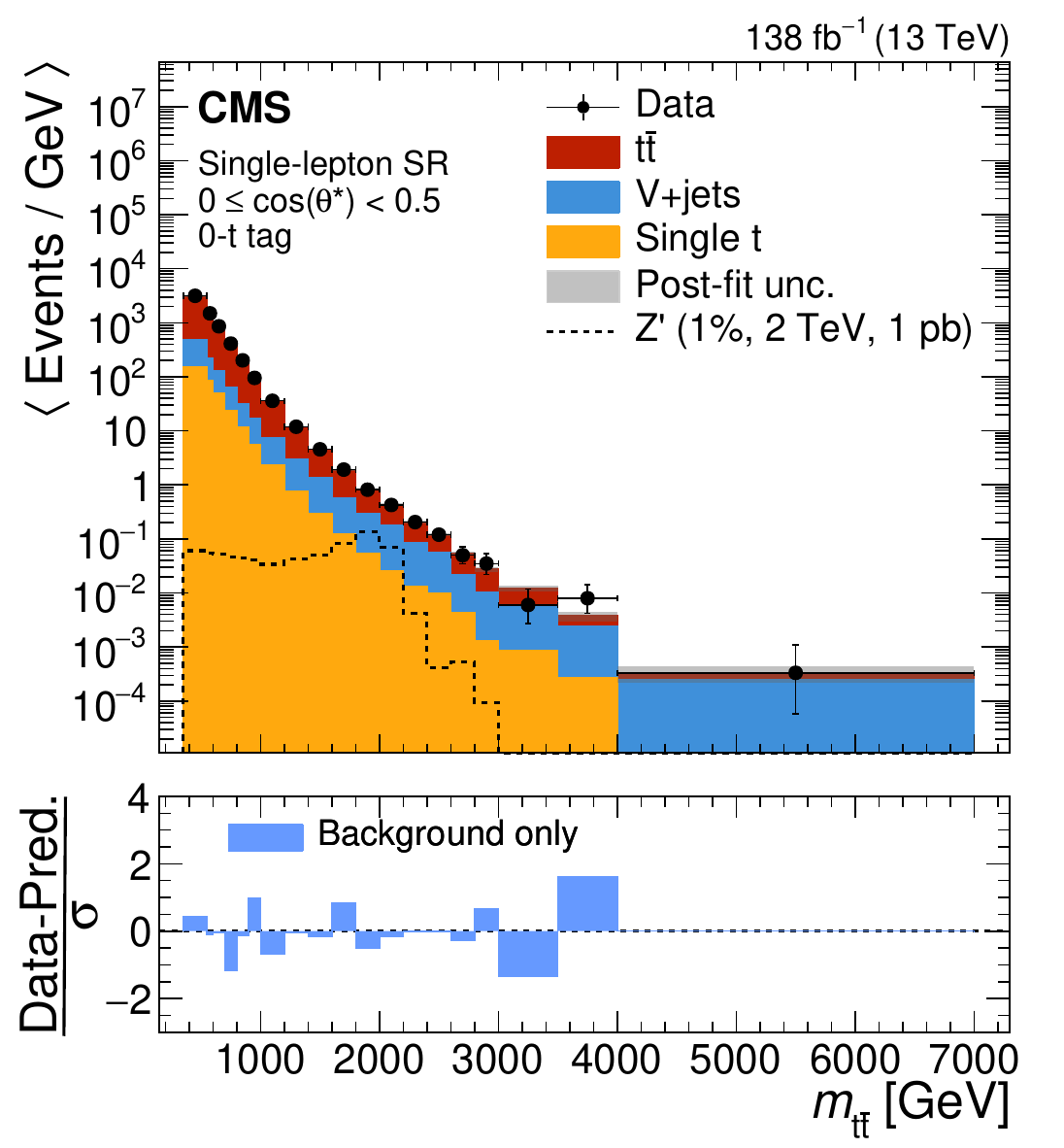}
\hfill
\includegraphics[width=0.45\textwidth]{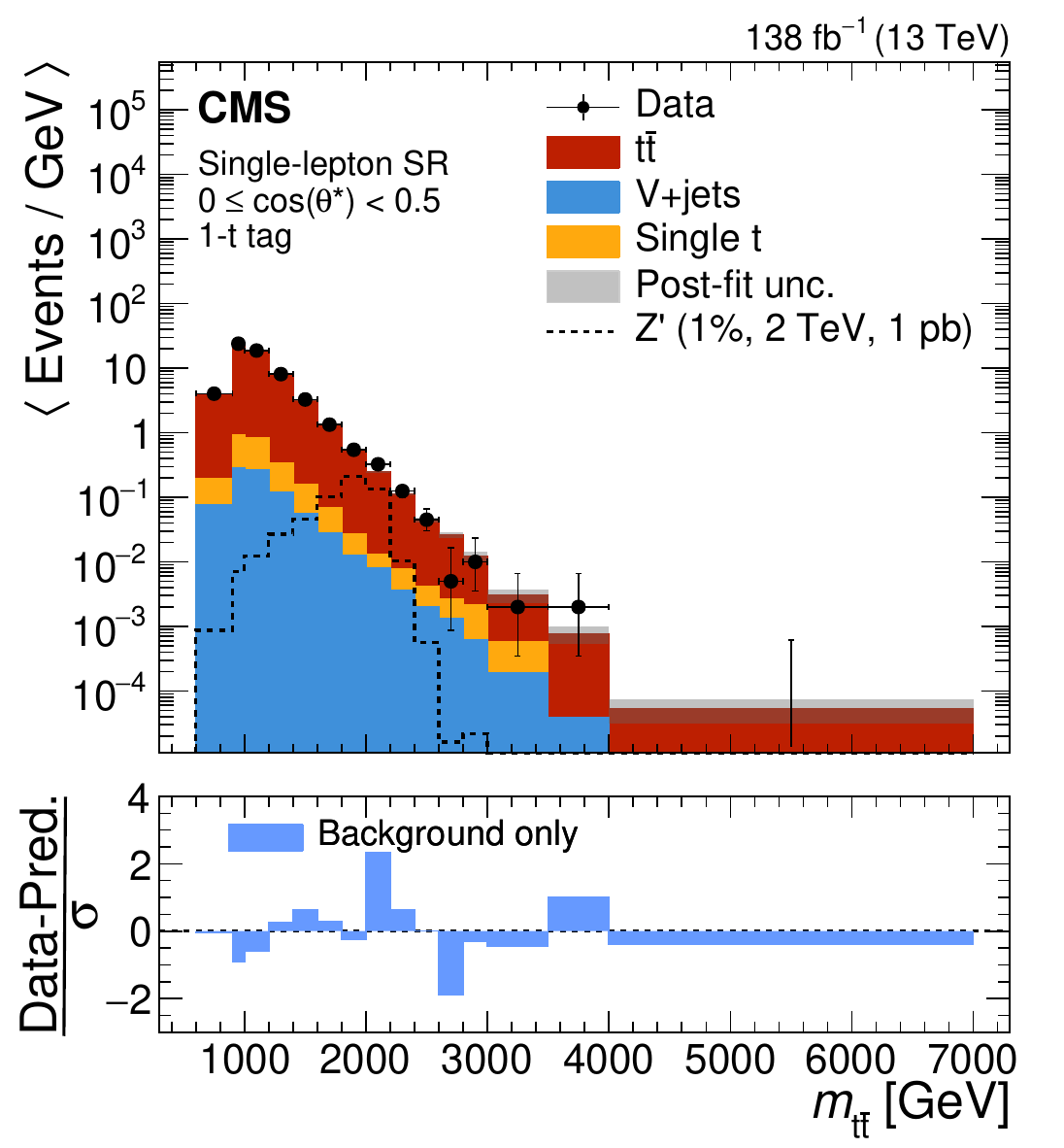} \\
\includegraphics[width=0.45\textwidth]{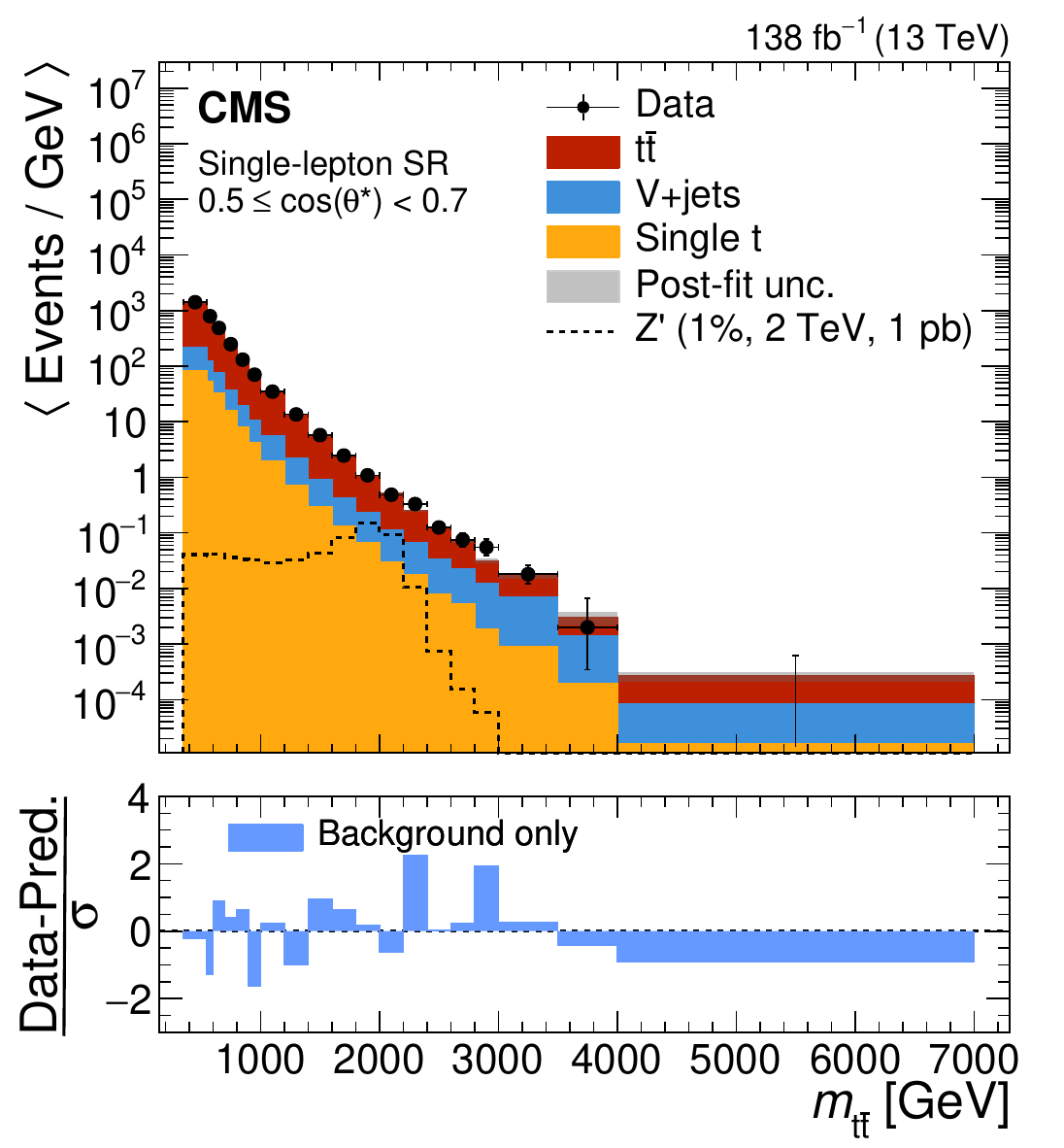}
\hfill
\includegraphics[width=0.45\textwidth]{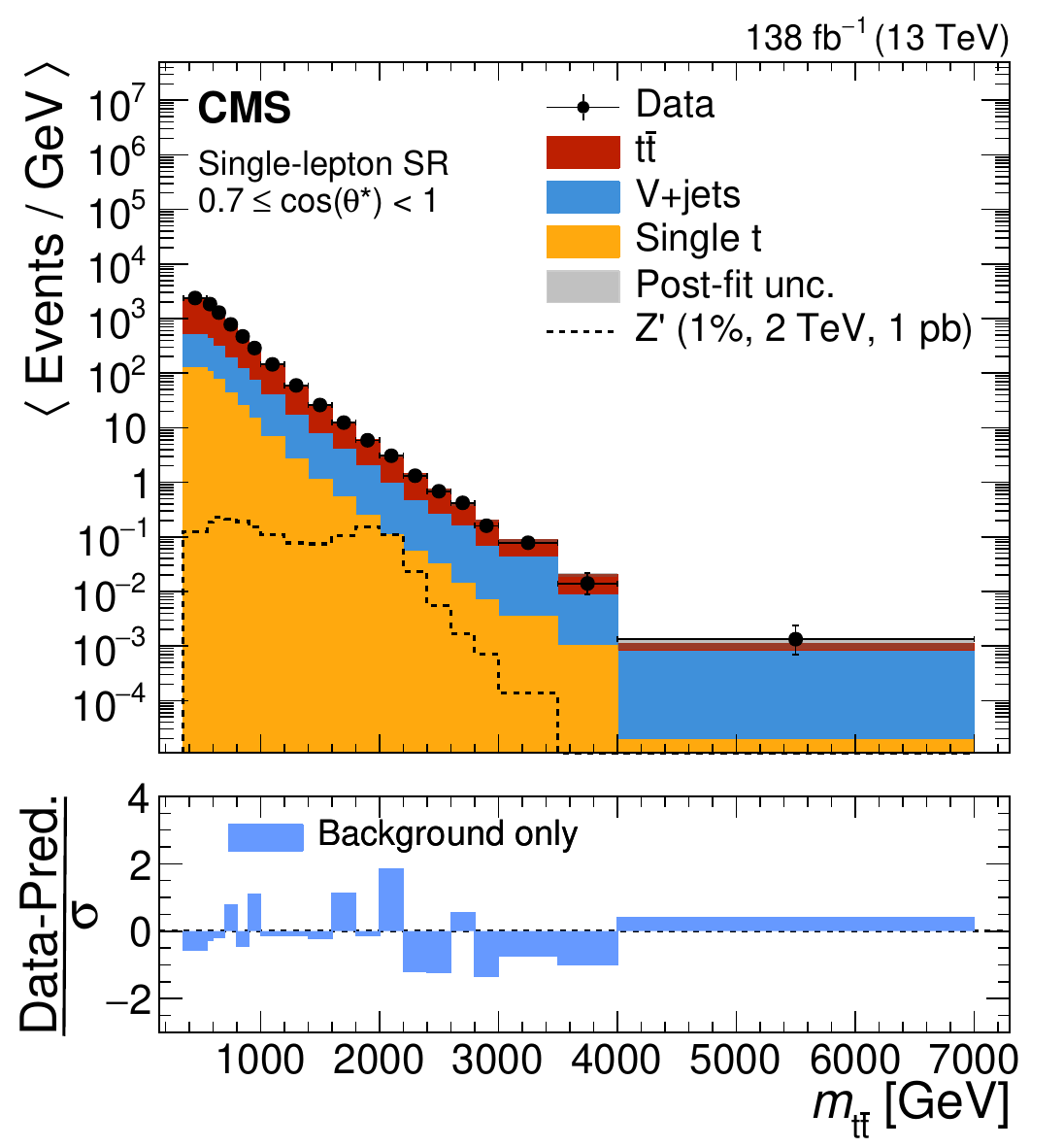}
\caption{Postfit distributions in \mtt in the single-lepton channel for data and simulation in the last three bins of \costhetastar in the \ttbar SR,
shown for the resolved (\zttag, left) and merged (\ottag, right) categories, under the background-only hypothesis. The horizontal bars on the data points indicate the bin width. In the last two \costhetastar bins (lower row) the resolved and merged categories are combined. For illustrative purposes, the \Zprime boson signal with a relative width of 1\% and a mass of 2\TeV is normalized to a cross section of 1\unit{pb} and overlaid to the backgrounds.
The lower panels show the pulls, defined as $(\text{Data}-\text{Prediction})/\sigma$, where $\sigma$ denotes the total postfit uncertainty
in each bin, relative to the SM prediction.}
\label{fig:1L_mtt_SR2}
\end{figure}

The postfit distributions in \mtt, obtained under the background-only hypothesis as described in Section~\ref{sec:results},
are shown in Fig.~\ref{fig:1L_mtt_CR} for the CRs, and in Figs.~\ref{fig:1L_mtt_SR1} and~\ref{fig:1L_mtt_SR2}
for the SRs. A good agreement between data and the total background prediction is observed.

\subsection{Dilepton channel}

This channel targets final states in which both the top quark and antiquark decay semileptonically ($\PQt\to\PQb\PW\to\PQb\Pell\PGn_{\Pell}$).
Selected events are required to contain two oppositely charged leptons (muons or electrons), at least two \akfour jets, with at least one identified as originating from a \PQb quark, and a significant amount of missing transverse momentum from undetected neutrinos.
The lepton selection criteria are not specifically optimized to identify muons or electrons originating from leptonic tau decays; however, such leptons are not explicitly vetoed.

Events are categorized into three channels based on the lepton flavour: \mumu, \emu, and \ee.
For the \mumu and \emu channels, the triggers require at least one muon with $\pt>50\GeV$ and $\abs{\eta}<2.4$, seeded by hits in either the muon chambers or the inner tracker.
The \ee channel uses a dielectron trigger requiring two electrons with $\pt>33\GeV$ and $\abs{\eta}<2.5$, without any isolation requirements at the trigger level.

Both resolved and merged topologies are considered. In the resolved regime, the decay products of the top quarks are well separated and reconstructed as individual objects.
In the merged regime, where the top quarks have high momenta, their decay products become collimated, and the lepton may be close in \DR to the jet originating from the \PQb quark. The sum of the angular distances between the leptons and the jets closest to them (\DRsum) is used to separate events into merged ($\DRsum<1.0$), resolved ($1.0<\DRsum<2.0$), and control regions ($\DRsum>2.0$), where \DRsum is defined as
\begin{equation}
    \DRsum = \DR_{\text{min}}(\Pell_1,\PQj_1)+\DR_{\text{min}}(\Pell_2,\PQj_2).
\end{equation}
In the \mumu and \emu channels, events must contain a muon with $\pt>53\GeV$.
For \mumu (\emu) events, the second muon (electron) must have $\pt>25\GeV$.
In the \ee channel, both electrons are required to have $\pt>36\GeV$, and electrons in the ECAL transition region between the barrel and endcap are excluded. These selections are driven by the trigger thresholds.

All leptons must satisfy the same two-dimensional isolation requirement as described in Eq.~\ref{eq:2dcut}.
This requirement helps suppress backgrounds from QCD multijet and \Wjets processes.

To reduce backgrounds from low-mass resonances and $\PZ/\PGg^\ast\to\Pell\Pell$ production in same-flavour dilepton events, the invariant mass of the dilepton pair is required to be above 20\GeV and outside the \PZ boson mass window: $76<m_{\Pell\Pell}<106\GeV$.

Events are further required to contain at least two \akfour jets with $\abs{\eta}<2.4$, and $\pt>100$ and 50\GeV for the leading and subleading jets, respectively.
At least one of the two leading jets must be \btagged.
In addition, \ptmiss is required to exceed 30\GeV to reject backgrounds from $\PZ/\PGg^\ast\to\Pell\Pell$ and QCD multijet processes.

The selected event sample is dominated by the irreducible \ttbar background.
Less important backgrounds arise from \Zjets, single top quark, and diboson production.
These backgrounds are modeled using simulated samples.
The normalization of the background contributions is based on theoretical cross sections, while their rates and shapes are allowed to vary within prior uncertainties during the fit.

In dilepton \ttbar events, the scalar sum of the transverse momenta of all reconstructed objects (\st) is chosen as the search variable, as it provides optimal sensitivity to high-mass resonances, which suffer from inefficiencies in reconstructing \mtt in the presence of two neutrinos in the highly Lorentz-boosted regime.

\begin{figure}[!ph]
\centering
\includegraphics[width=0.45\textwidth]{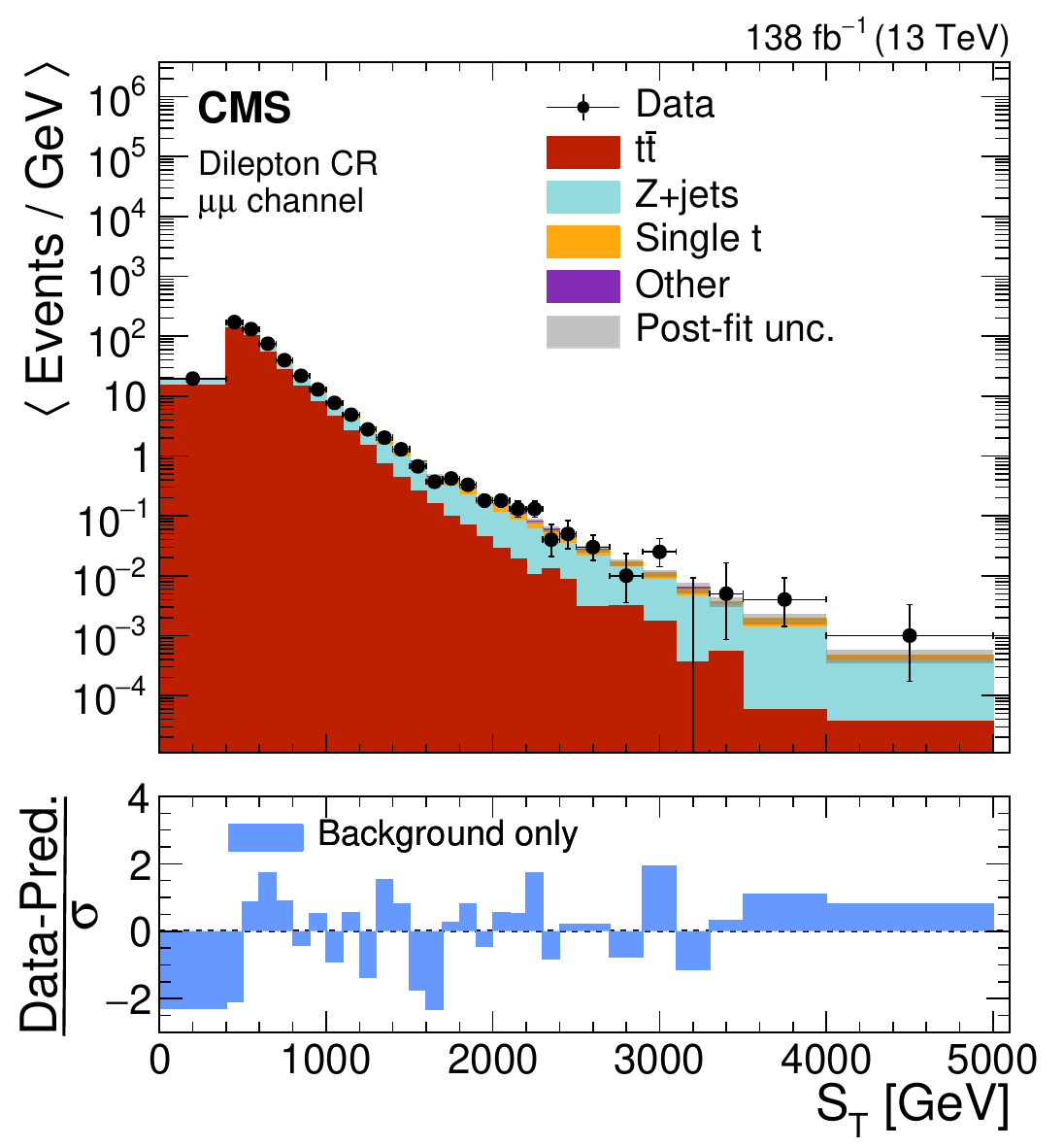}
\hfill
\includegraphics[width=0.45\textwidth]{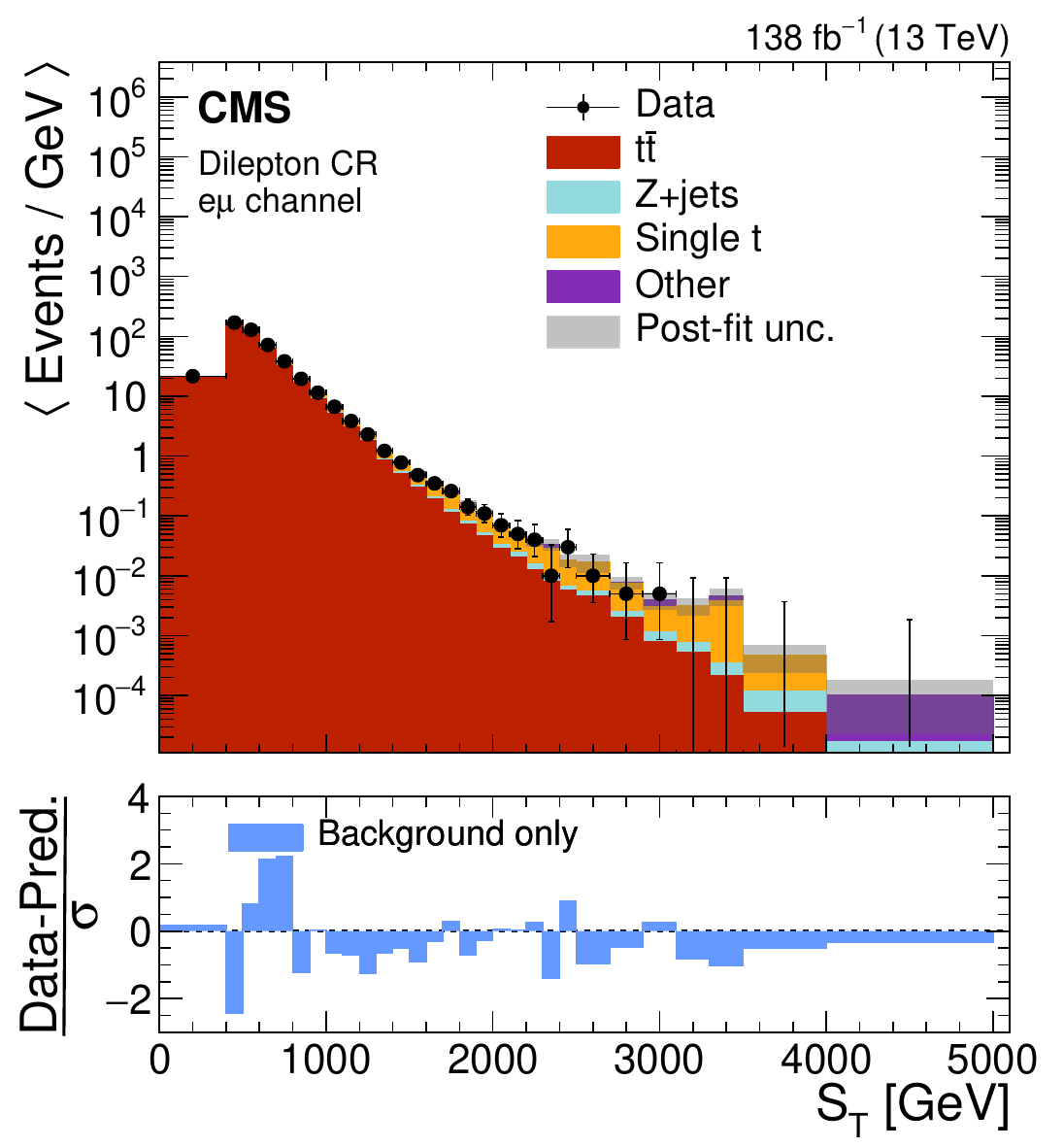} \\
\includegraphics[width=0.45\textwidth]{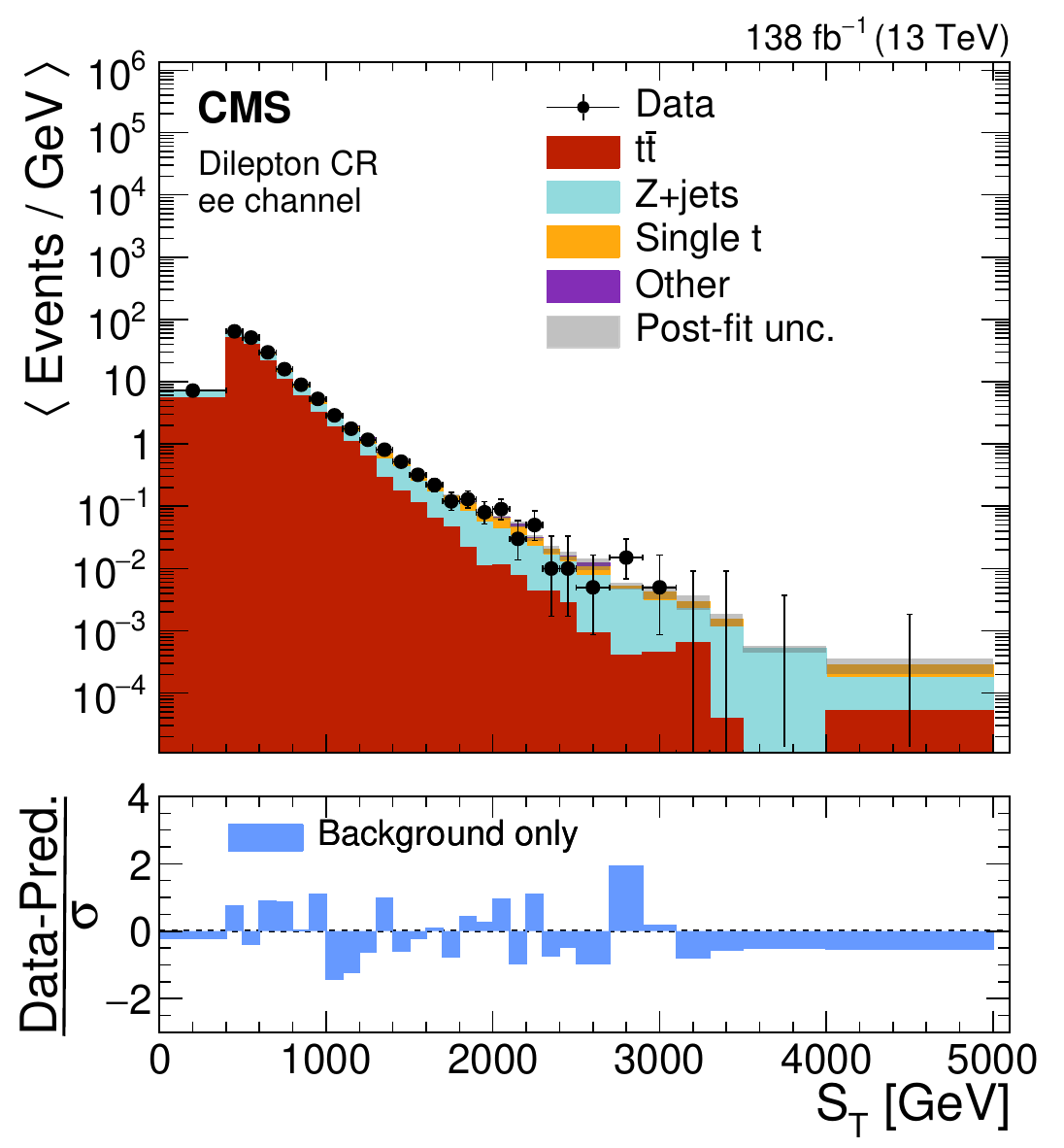}
\caption{Postfit distributions in \st for data and simulation in the CR for the dilepton channel.
Distributions are shown for the \mumu (upper left), \emu (upper right), and \ee (lower) channels, under the background-only hypothesis. The horizontal bars on the data points indicate the bin width.
The lower panels show the pulls, defined as $(\text{Data}-\text{Prediction})/\sigma$, where $\sigma$ denotes the total postfit uncertainty
in each bin, relative to the SM prediction.}
\label{fig:2L_postfit_CR}
\end{figure}

\begin{figure}[!ph]
\centering
\includegraphics[width=0.4\textwidth]{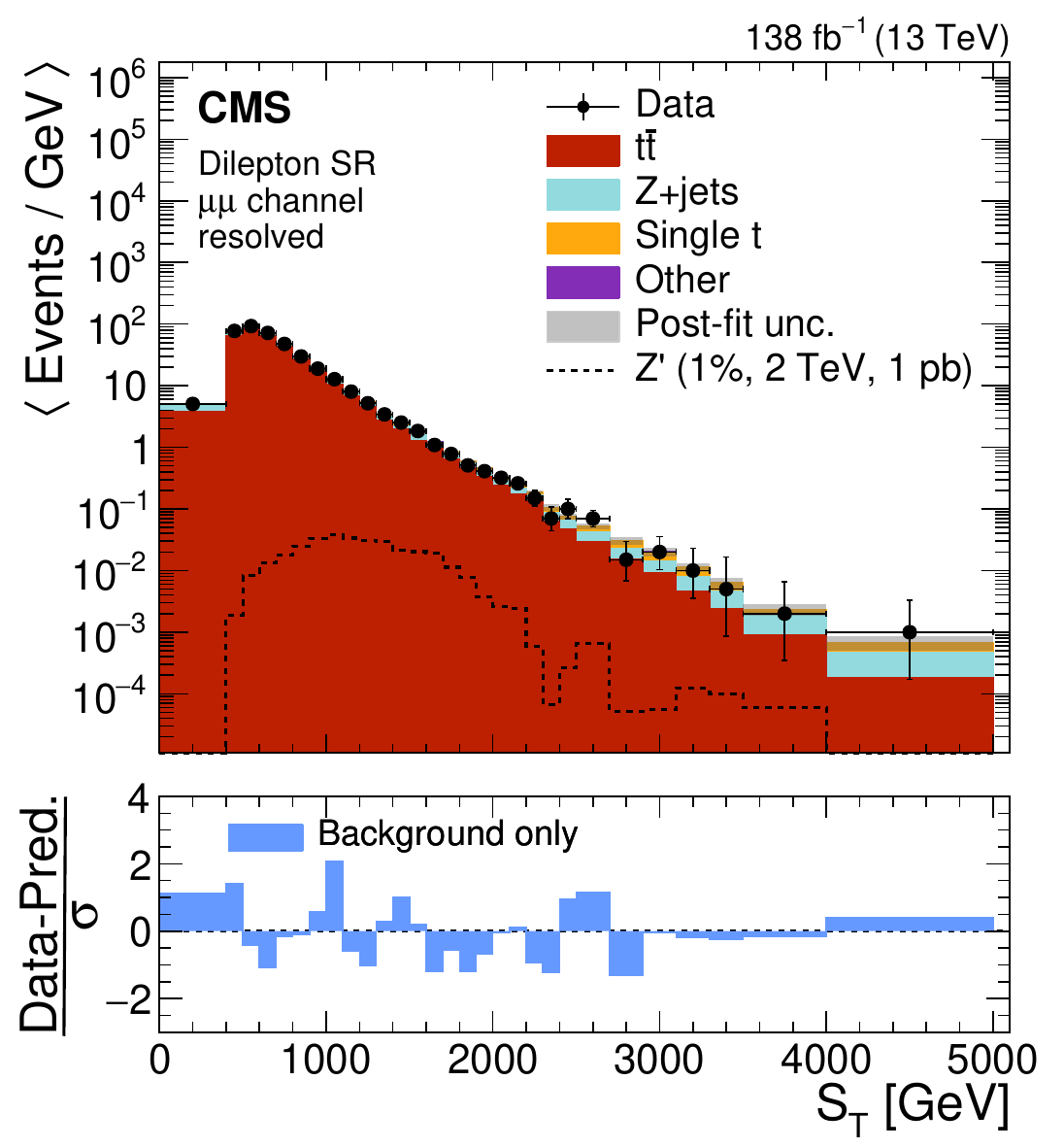}
\hspace{0.1\textwidth}
\includegraphics[width=0.4\textwidth]{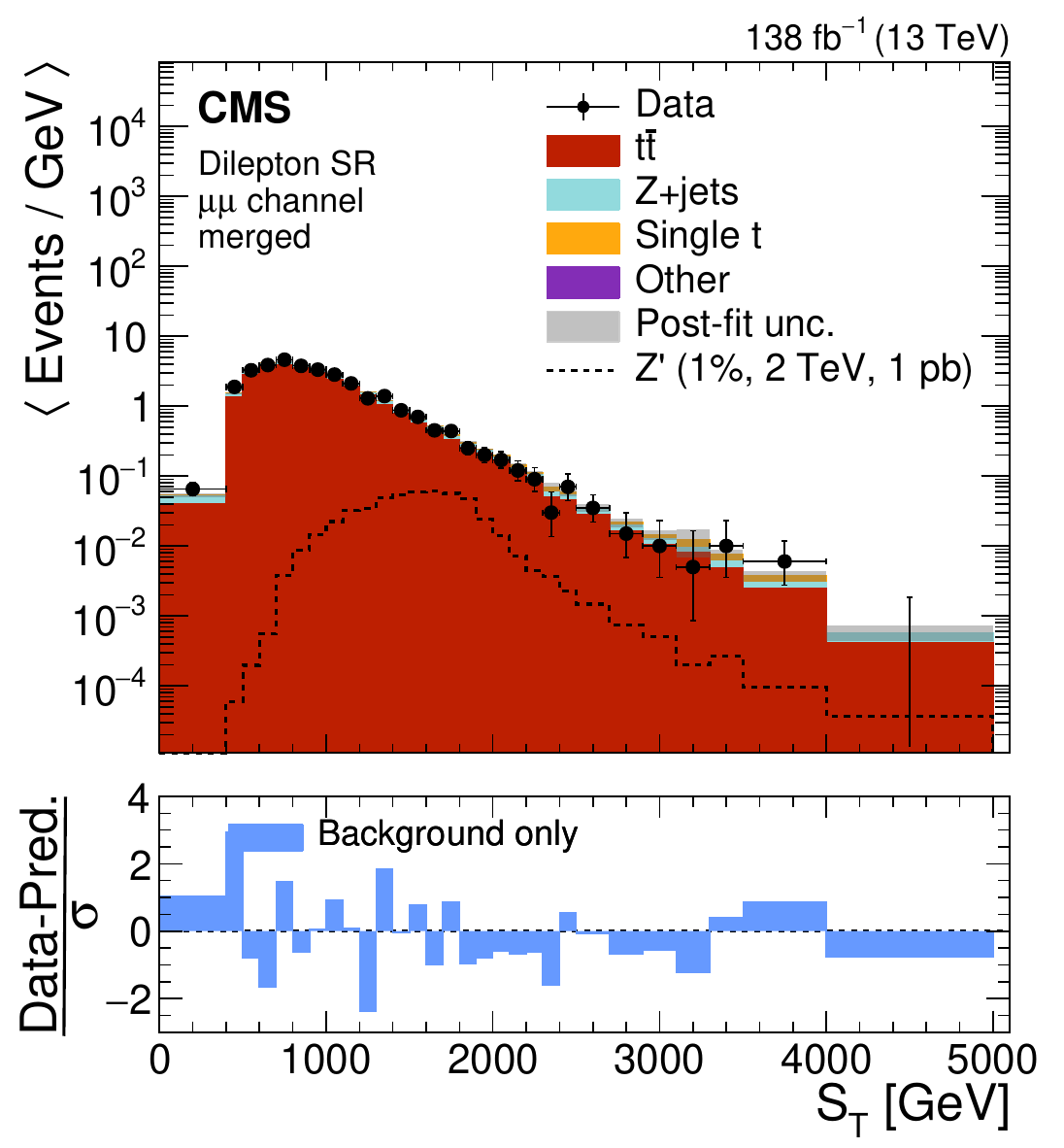} \\
\includegraphics[width=0.4\textwidth]{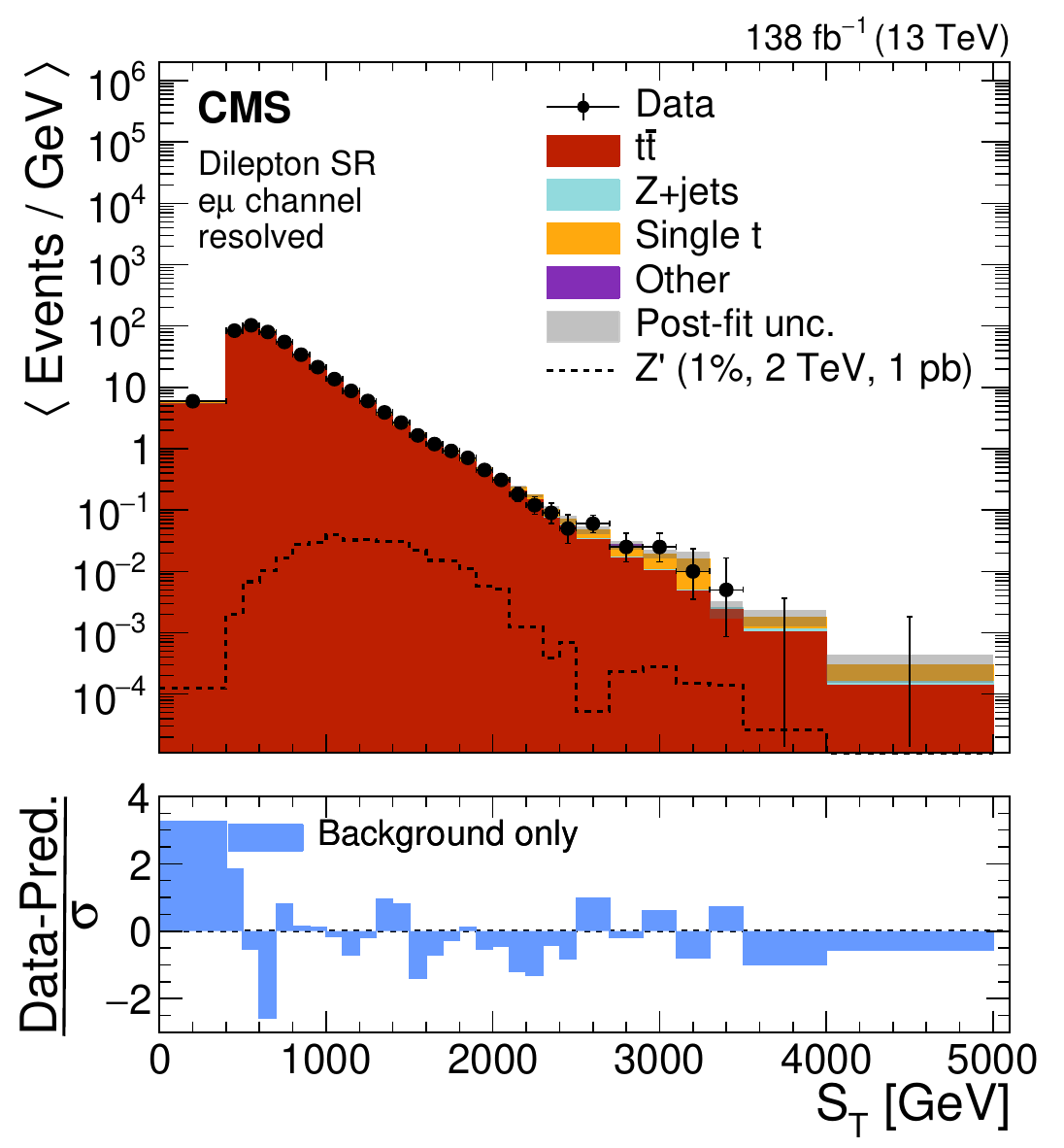}
\hspace{0.1\textwidth}
\includegraphics[width=0.4\textwidth]{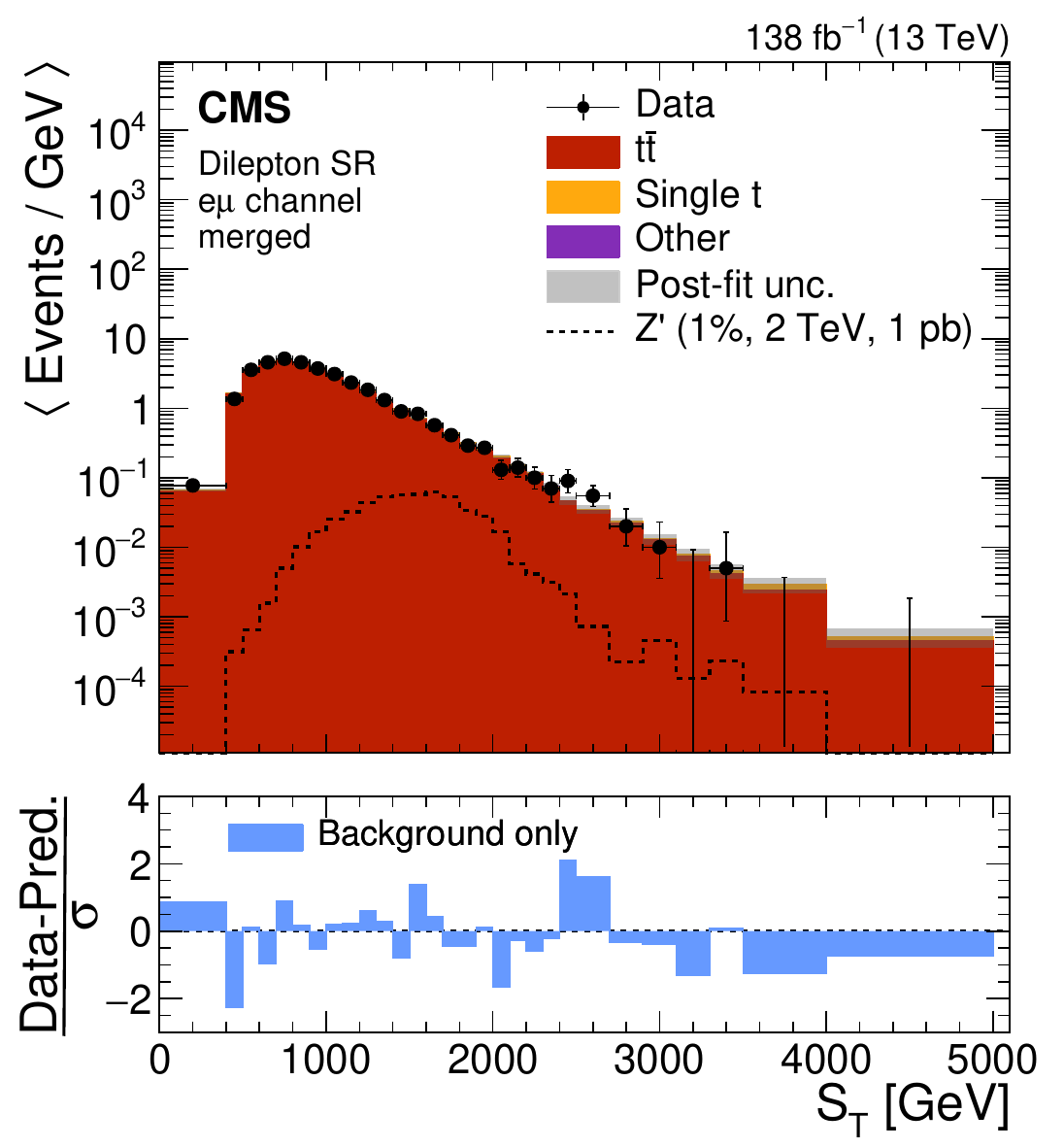} \\
\includegraphics[width=0.4\textwidth]{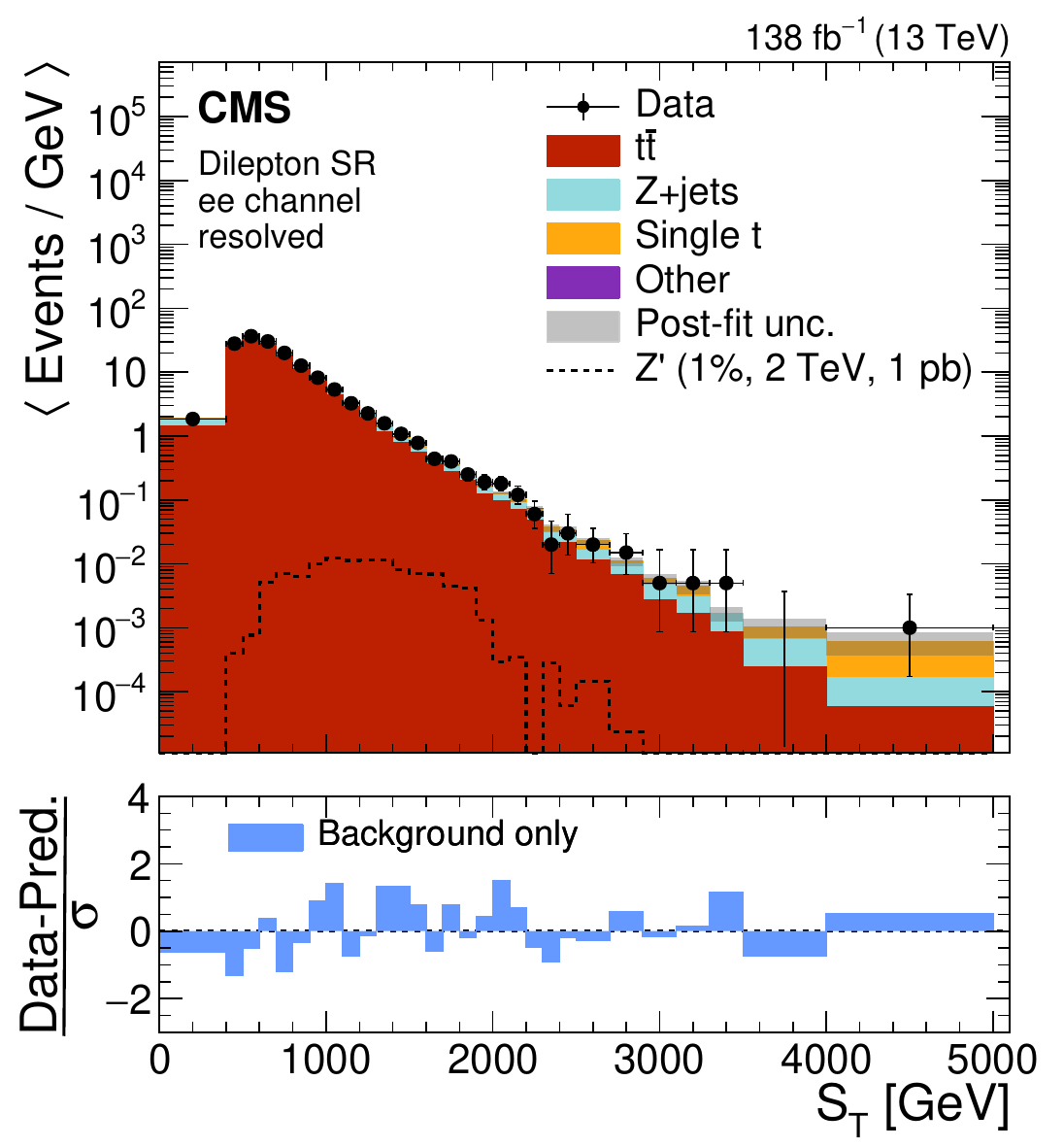}
\hspace{0.1\textwidth}
\includegraphics[width=0.4\textwidth]{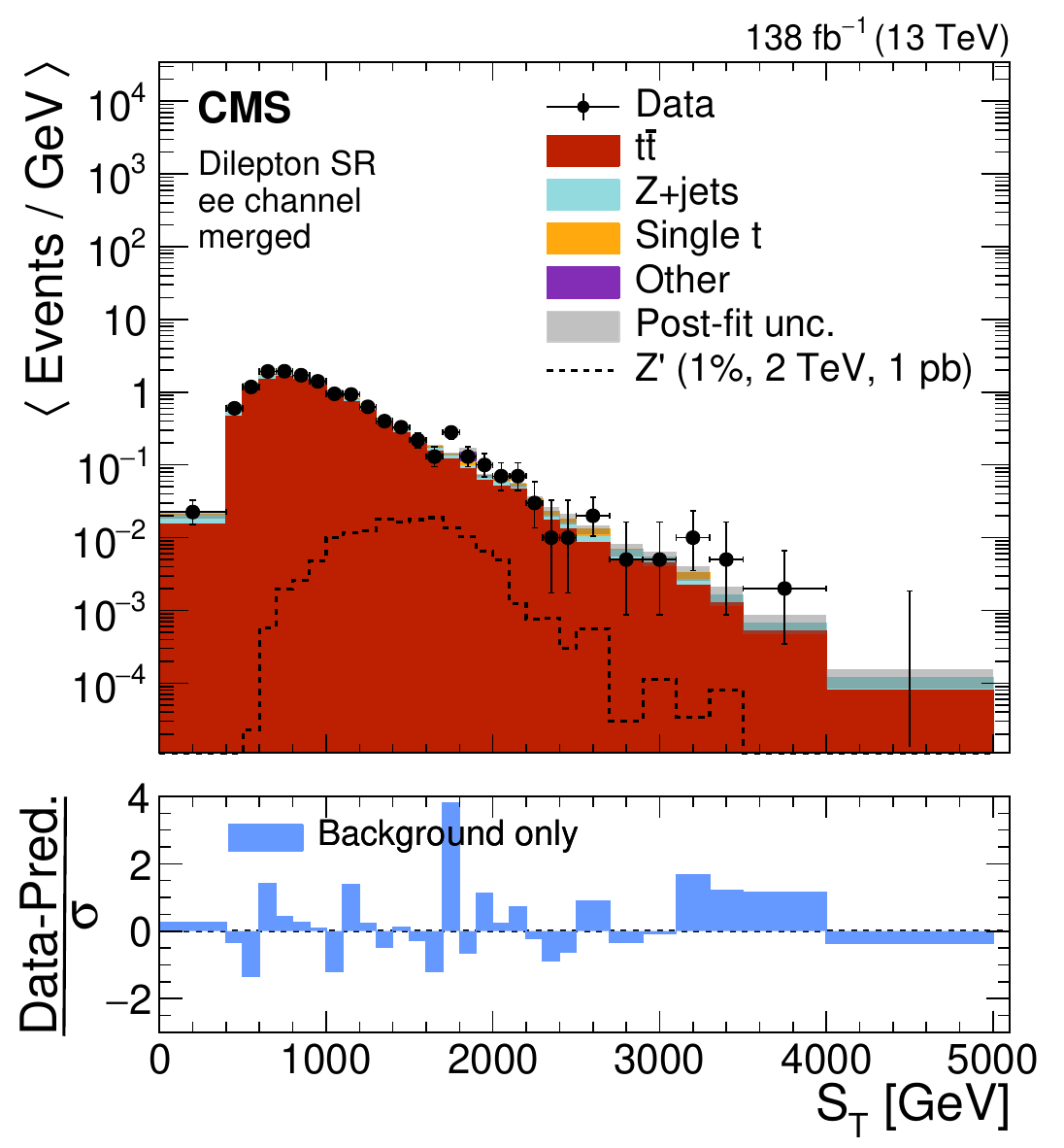}
\caption{Postfit distributions in \st for data and simulation for the resolved (left) and merged  (right) categories
for the dilepton channel.
Distributions are shown for the \mumu (upper), \emu (middle), and \ee (lower) channels, under the background-only hypothesis. The horizontal bars on the data points indicate the bin width. For illustrative purposes, the \Zprime boson signal with a relative width of 1\% and a mass of 2\TeV is normalized to a cross section of 1\unit{pb} and overlaid to the backgrounds.
The lower panels show the pulls, defined as $(\text{Data}-\text{Prediction})/\sigma$, where $\sigma$ denotes the total postfit uncertainty
in each bin, relative to the SM prediction.}
\label{fig:2L_postfit}
\end{figure}

The postfit distributions in \st, obtained under the background-only hypothesis as described in Section~\ref{sec:results},
are shown in Fig.~\ref{fig:2L_postfit_CR} for the CRs and in Fig.~\ref{fig:2L_postfit} for the SRs. A good agreement between data and the total background prediction is observed.

\section{Systematic uncertainties}
\label{sec:systematic_uncertainties}

Several sources of systematic uncertainty affect the normalization and shape of the observables considered in the search.
The sources of such uncertainties are summarized in Table~\ref{tab:syst} and described below.

\begin{table}[!bh]
\topcaption{Sources of systematic uncertainties and correlations between them. The correlations take various forms: among data-taking years, among different processes (such as \ttbar, single top quark production, etc), and/or channels (0\Pell, 1\Pell, and 2\Pell). The \Zprime signal with a relative width of 1\% and a mass of 2\TeV is used as a benchmark. The ``\ttbar rate'' row corresponds to the overall prior uncertainty in the \ttbar production cross section. The ``\ttbar shape'' row corresponds to differences in shapes between the NLO simulation and measured values of the \ttbar \pt spectrum at large momentum due to destructive interference from higher-order terms that are not present in the simulation.}
\centering
\renewcommand{\arraystretch}{1.1}
\cmsTable{\begin{tabular}{lcc}
Source & Type & Form of correlation \\
\hline
\ttbar rate  & norm.          & across years and all channels \\
\ttbar shape          & shape          & across channels (0\Pell and 1\Pell) \\
Drell--Yan rate & norm. & across years \\
\Wjets rate & norm. & across years \\
$\PV\PV$ rate & norm. & years \\
Single \PQt rate & norm. & across years \\
QCD estimate            & norm. \& shape & none (0\Pell), across years (1\Pell) \\
PDFs                    & shape & across years \\
\muR and \muF     & shape & across years \\
Jet energy scale        & norm. \& shape & across processes and all channels \\
Jet energy resolution   & norm. \& shape & across processes and all channels \\
\btagging               & norm. \& shape & across processes \\
\ttagging efficiency    & norm. \& shape & across processes (1\Pell and 2\Pell)\\
\ttagging mistag rate   & norm. \& shape & across years and processes (only in 1\Pell) \\
Lepton ID and iso       & norm. \& shape & across years, processes and channels (1\Pell and 2\Pell) \\
Lepton reconstruction   & norm. \& shape & across years, processes and channels (1\Pell and 2\Pell) \\
Trigger                 & norm. \& shape & across years and processes \\
L1 ECAL trigger inefficiency & norm. \& shape & across years, processes and channels \\
Pileup reweighting      & norm. \& shape & across years, processes and channels \\
Int. luminosity              & norm. & across processes \\
\end{tabular}}
\label{tab:syst}
\end{table}

Normalization uncertainties of 20 and 50\% are assigned to the production cross sections
of \ttbar and other processes, respectively.
These are set to conservative values as inputs to the maximum likelihood fit, which
extracts the normalizations of each process in situ, so the constraints
are set to very large values to avoid biasing the fit.
The specific values are adopted from the previously published analysis~\cite{Sirunyan2019},
and reflect the limited modeling accuracy of the backgrounds in the merged regime,
where jets have transverse momenta exceeding 400\GeV.
Separate uncertainties are also added for variations in
renormalization (\muR) and factorization (\muF) scales,
and uncertainties in the PDFs.
Their impact is estimated following the methodology described in Ref.~\cite{Butterworth_2016}.
These uncertainties are treated as uncorrelated across different background processes and signals.

The uncertainties in the simulation can include overall rates for the background processes (\ttbar, Drell--Yan, \Wjets and $\PV\PV$, and single top quark production). Drell--Yan, \Wjets, $\PV\PV$ and single top quark production are present in single-lepton and dilepton channels, but not correlated. The \ttbar transverse momentum spectrum is known to be mismodeled by the NLO generators due to destructive interference with higher-order terms in the perturbative expansion~\cite{Kidonakis:2012rm}. We correct for this effect in the dilepton channel explicitly, whereas in the all-hadronic and single-lepton channels, we account for this effect with the uncertainties in the estimation.

Differences in the selection efficiencies between data and simulation are corrected with
data-to-simulation scale factors (SFs). The corresponding uncertainties are obtained by varying
each SF independently within its uncertainty and propagating the resulting change through the analysis.
The jet energy scale and resolution uncertainties are evaluated as functions of the jet \pt and $\eta$,
and their variations are propagated to \ptvecmiss.
The uncertainties related to the \btagging efficiency are split based on the jet flavor, as well as jet \pt and $\eta$.
This analysis is sensitive to the \ttagging efficiency, which is extracted directly from the data.
The misidentification efficiency of the \ttagging selection in the single-lepton channel is measured directly in data, using the CR dominated by \Vjets events,
and applied to simulation. The systematic uncertainty due to this correction is determined by varying the SF value within its uncertainty.
For the all-hadronic channel, the two-dimensional sideband fit accounts
for misidentification, as described in Section~\ref{sec:background_estimation_0l}.

Uncertainties in the trigger efficiency, lepton identification, and reconstruction
are considered separately for muon and electron channels.
Additional systematic uncertainties originate from the L1 ECAL trigger inefficiency
due to detector timing issues~\cite{CMS:2020cmk}, and the pileup reweighting procedure, which is
evaluated by varying the total inelastic cross section by $\pm$4.6\%~\cite{Sirunyan:2018nqx}.
Lastly, the total integrated luminosity of 138\fbinv is assigned
a normalization uncertainty of 1.6\%~\cite{CMS:LUM-17-003,CMS:LUM-17-004,CMS:LUM-18-002}.

\section{Results}
\label{sec:results}

A binned maximum likelihood fit is performed simultaneously across the SRs and CRs using the \mtt, \mtop, and \st distributions.
The fit includes nuisance parameters that account for systematic uncertainties affecting both the normalization and shape
of the signal and background processes.
The CRs are used to constrain the nuisance parameters associated with the normalization and shape variations of the non-\ttbar backgrounds contributions.
The binning choice is optimized to maximize sensitivity, taking into account the available event counts across regions and bins.
It differs between the CRs and SRs due to the different minimum \mtt values used in the resolved and merged categories.
The observed data are found to be consistent with the postfit background predictions.

The modified frequentist approach~\cite{CMS-NOTE-2011-005,CLS1,CLS2},
known as the \CLs criterion with the profile likelihood ratio as the test statistic,
is used to set limits on the potential presence of a signal.
We use the asymptotic approximation to the profile likelihood test statistic~\cite{Cowan:2010js}.
The following results have been determined using the CMS statistical analysis tool \textsc{combine}~\cite{CAT-23-001},
which is based on the \textsc{RooFit}~\cite{Verkerke:2003ir} and \textsc{RooStats}~\cite{Moneta:2010pm} frameworks.

\subsection{Heavy resonance interpretations}

\begin{figure}[!ht]
\centering
\includegraphics[width=0.45\textwidth]{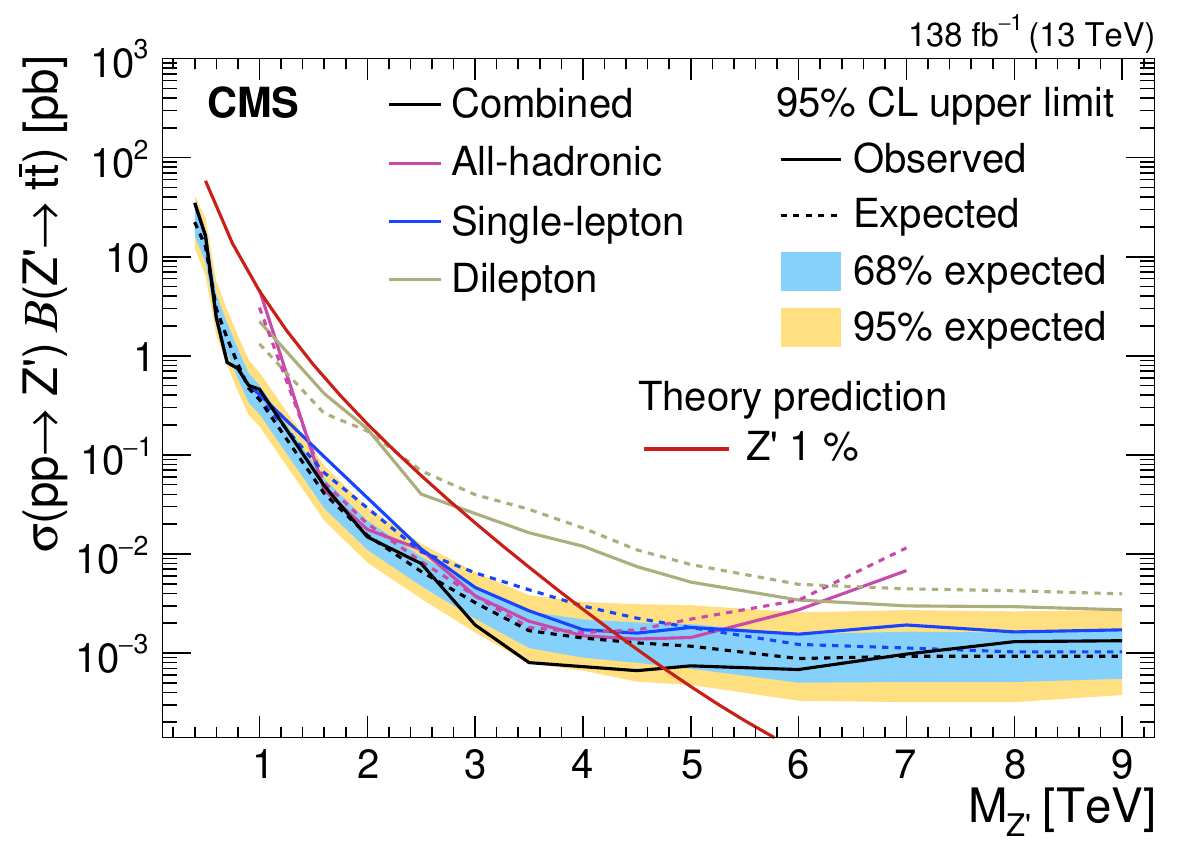}
\hfill
\includegraphics[width=0.45\textwidth]{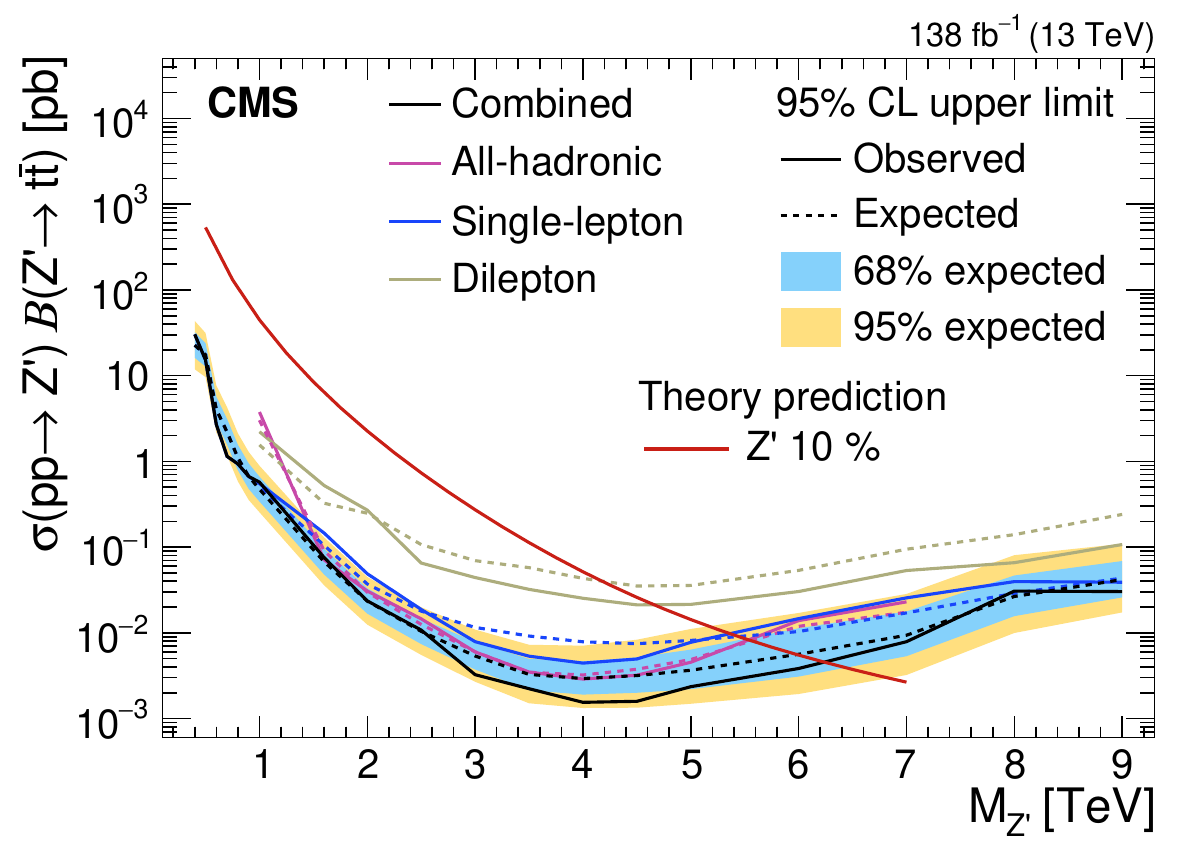} \\
\includegraphics[width=0.45\textwidth]{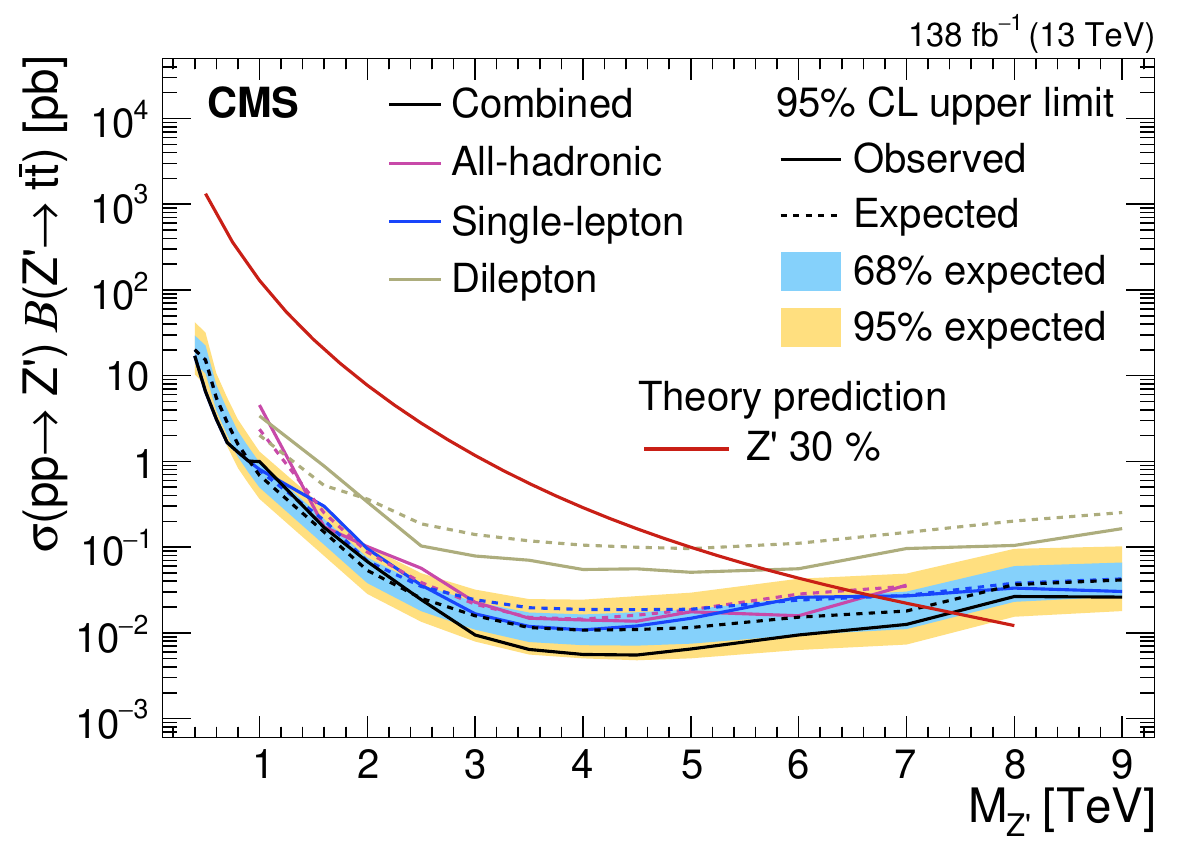}
\caption{Expected and observed upper limits at 95\% \CL on the product of
the production cross section and branching fraction as functions of the resonance mass.
The limits are shown for \Zprime bosons with 1 (upper left), 10 (upper right), and 30\% (lower) relative widths.
In each panel we plot the expected combined upper limit on the signal strength times branching ratio (black dashed line)
together with the 68 (light blue) and 95\% (yellow) uncertainty bands, and the corresponding observed upper limit (black solid line).
The expected (dashed lines) and observed (solid lines) limits from the single channels are overlaid: all-hadronic (purple), single-lepton (blue), and dilepton (light brown).
The limits are compared with the respective theory predictions shown by the solid red curves.
The rise in the limits seen at high mass for the \Zprime boson interpretation at 1\% relative width (upper left) for the all-hadronic case arises from the limited number of events available to estimate the background.}
\label{fig:limits_Zprime}
\end{figure}

\begin{figure}[!ht]
\centering
\includegraphics[width=0.45\textwidth]{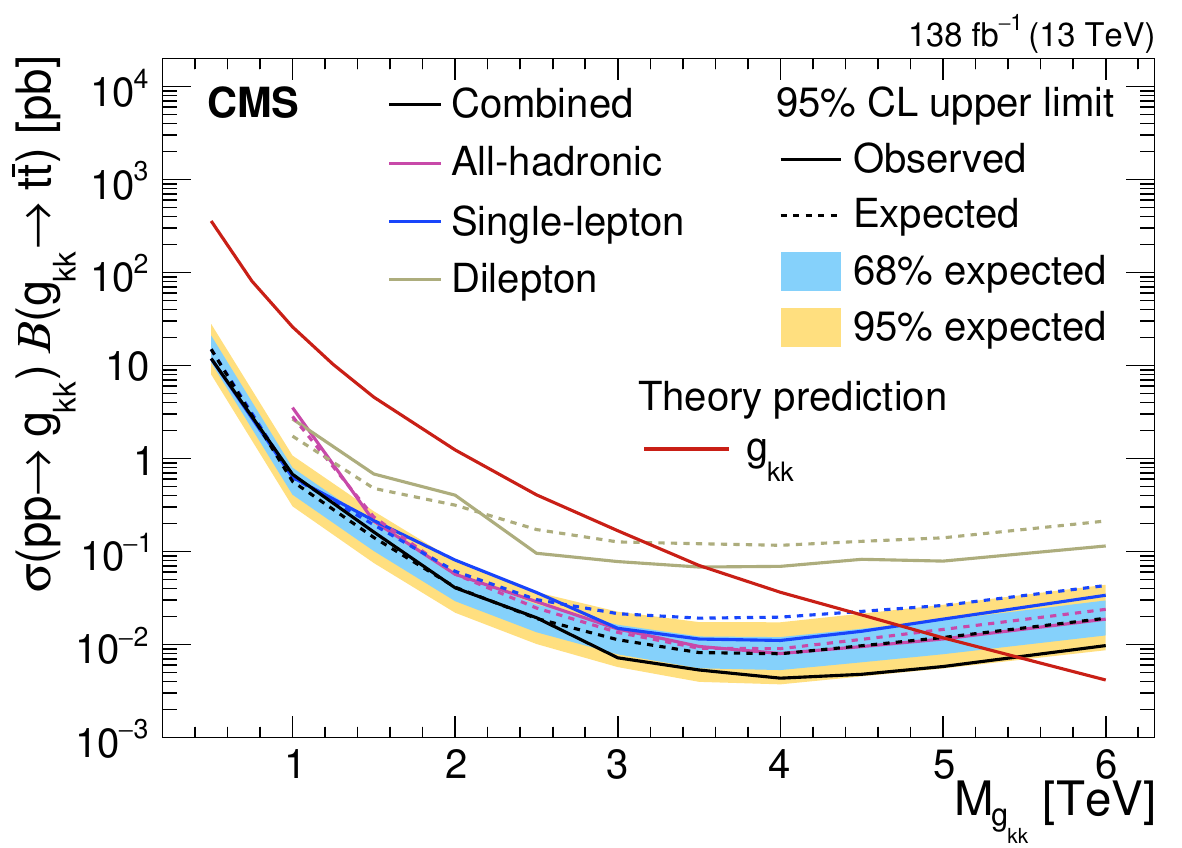}
\hfill
\includegraphics[width=0.45\textwidth]{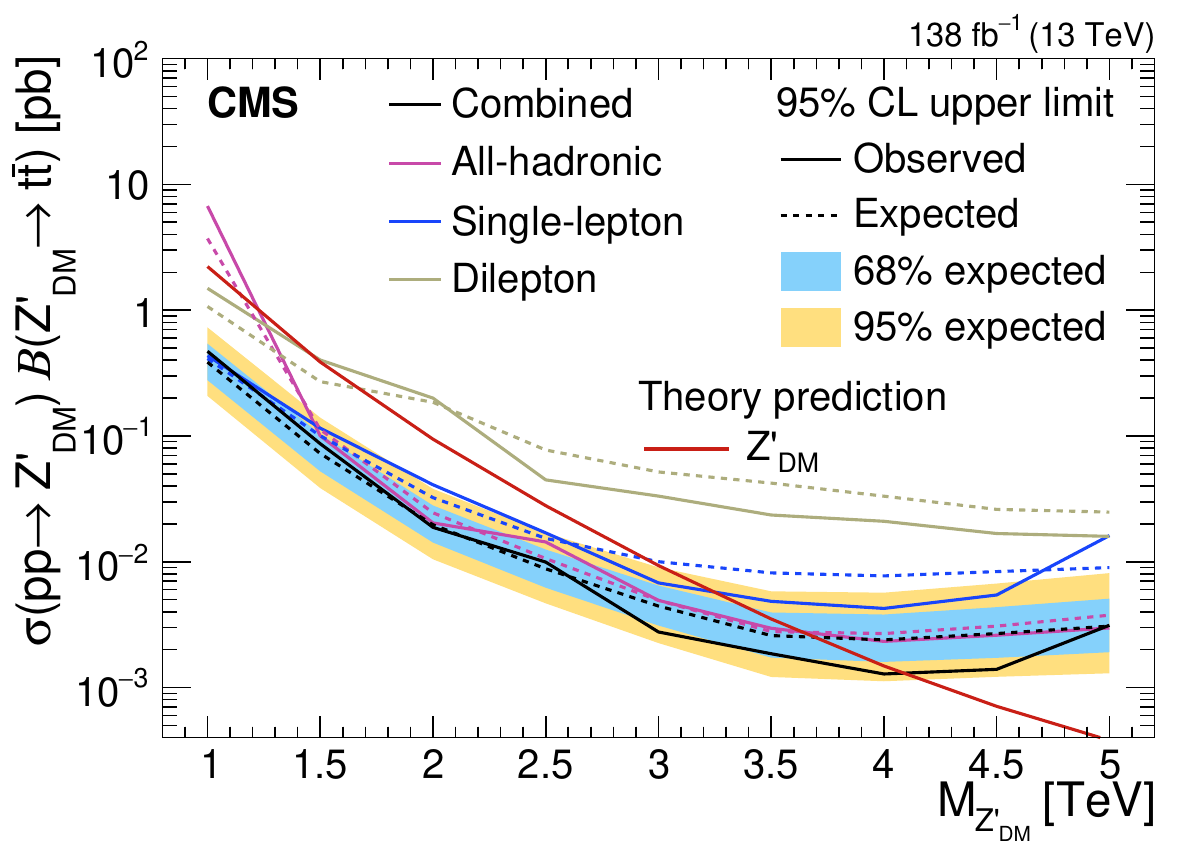}
\caption{Expected and observed upper limits at 95\% \CL on the product of
the production cross section and branching fraction as functions of the resonance mass.
The limits are shown for the Kaluza--Klein gluon (left) and dark matter (right) scenarios.
In each panel we plot the expected combined upper limit on the signal strength times branching fraction (black dashed line)
together with the 68 (light blue) and 95\% (yellow) uncertainty bands, and the corresponding observed upper limit (black solid line).
The expected (dashed lines) and observed (solid lines) limits from the individual channels are overlaid: all-hadronic (purple), single-lepton (blue), and dilepton (light brown).
The limits are compared with the respective theory predictions shown by the solid red curves.}
\label{fig:limits_gkk_dm}
\end{figure}

We set upper limits at 95\% confidence level (\CL) on
the product  $\sigma(\pp\to\PX)\BR(\PX\to\ttbar)$ as functions of the resonance mass,
where \PX denotes a \Zprime, \ZprimeDM, or \gKK signal.
The expected and observed exclusion limits at 95\% \CL are shown in Figs.~\ref{fig:limits_Zprime} and~\ref{fig:limits_gkk_dm}. Overall, the single-lepton channel is the leading contributor to the combination for masses below 1.5\TeV. Above this value, the all-hadronic and single-lepton channels exhibit a complementary sensitivity, contributing  with a comparable weight to the combination.
For \Zprime bosons, the excluded mass ranges derived from this analysis are at 0.4--4.8, 0.4--6.2, and 0.4--7.4\TeV,
assuming relative widths of 1, 10, and 30\%, respectively.
The presence of the \gKK resonances and DM mediators is excluded for masses between 0.5--5.5\TeV and 1--4.2\TeV, respectively.

To quantify the impact of individual sources of uncertainty on the upper limits, nuisance parameters are grouped into disjoint categories. The parameters in each group are then individually fixed to their postfit values, and the resulting reduction in the signal strength uncertainty is used to determine the contribution of that group to the total variance. Table~\ref{tab:breakdown} summarizes the relative contributions of the dominant sources of uncertainty to the total variance of the upper limits, evaluated using a \Zprime signal with a 1\% relative width and a mass of 2\TeV as a benchmark. The uncertainty is found to be dominated by the statistical component. The largest contribution from systematic uncertainties arises from nuisance parameters affecting both the shape and normalization of the \ttbar background.

\begin{table}[!ht]
\topcaption{Relative contribution of the dominant sources of uncertainty to the total variance of the upper limits. The benchmark scenario corresponds to a \Zprime signal with a 1\% relative width and a mass of 2\TeV. The top-quark modeling category includes nuisance parameters associated with the \ttbar production rate, \ttagging efficiency, \ttagging mistag rate, and modeling of the \ttbar transverse momentum spectrum.}
\centering
\renewcommand{\arraystretch}{1.1}
\begin{tabular}{lc}
    Source & Fraction of total variance (\%) \\
    \hline
    Top quark modeling & 22 \\
    \muR and \muF & 5 \\
    PDFs & 4 \\
    Jet energy scale and resolution & 3 \\
    Others & 6 \\[\cmsTabSkip]
    Systematic uncertainty & 40 \\
    Statistical uncertainty & 60
\end{tabular}
\label{tab:breakdown}
\end{table}

\subsection{Scalar and pseudoscalar interpretations}

The single-lepton channel has sensitivity at sufficiently low masses to allow investigation of scalar and pseudoscalar interpretations,
whereas the all-hadronic and dilepton channels do not have sensitivity there. We therefore only use the single-lepton channel for this interpretation.

\begin{figure}[!bh]
\centering
\includegraphics[width=0.44\textwidth]{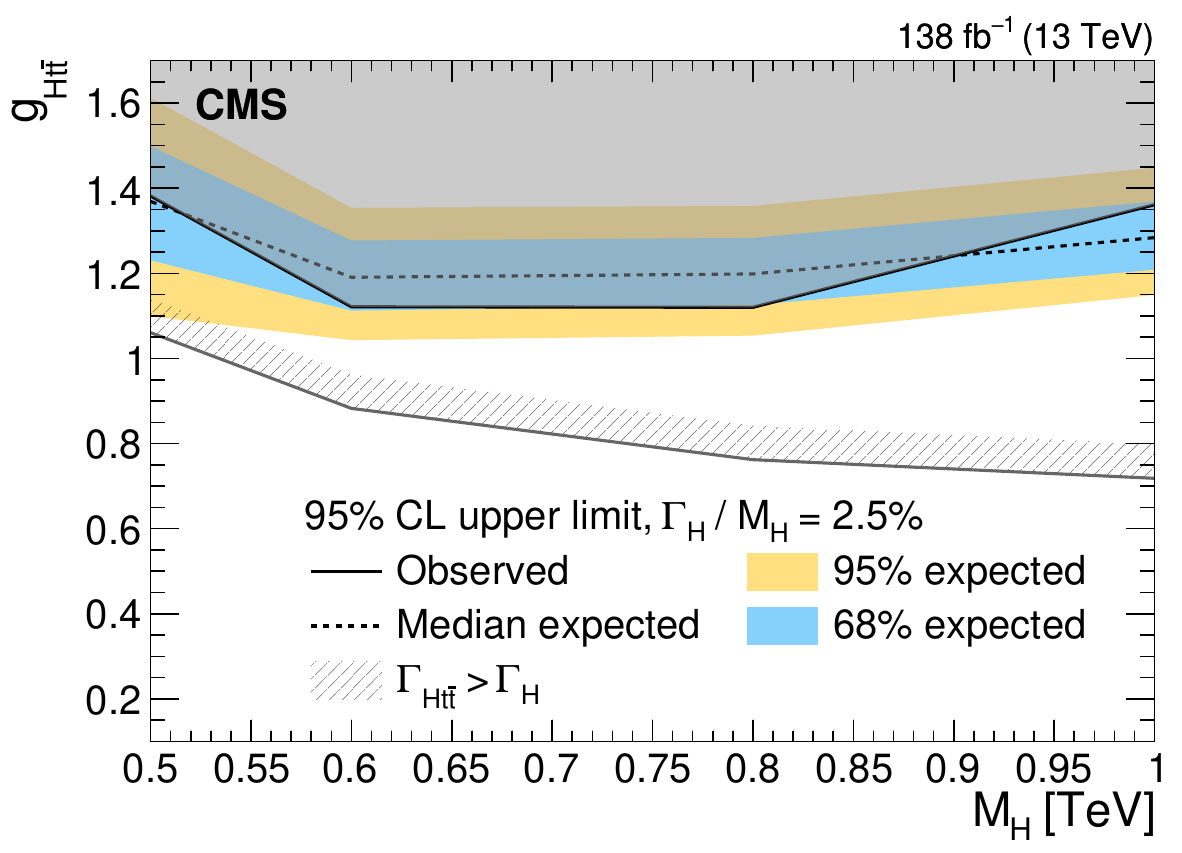}
\hspace{0.1\textwidth}
\includegraphics[width=0.44\textwidth]{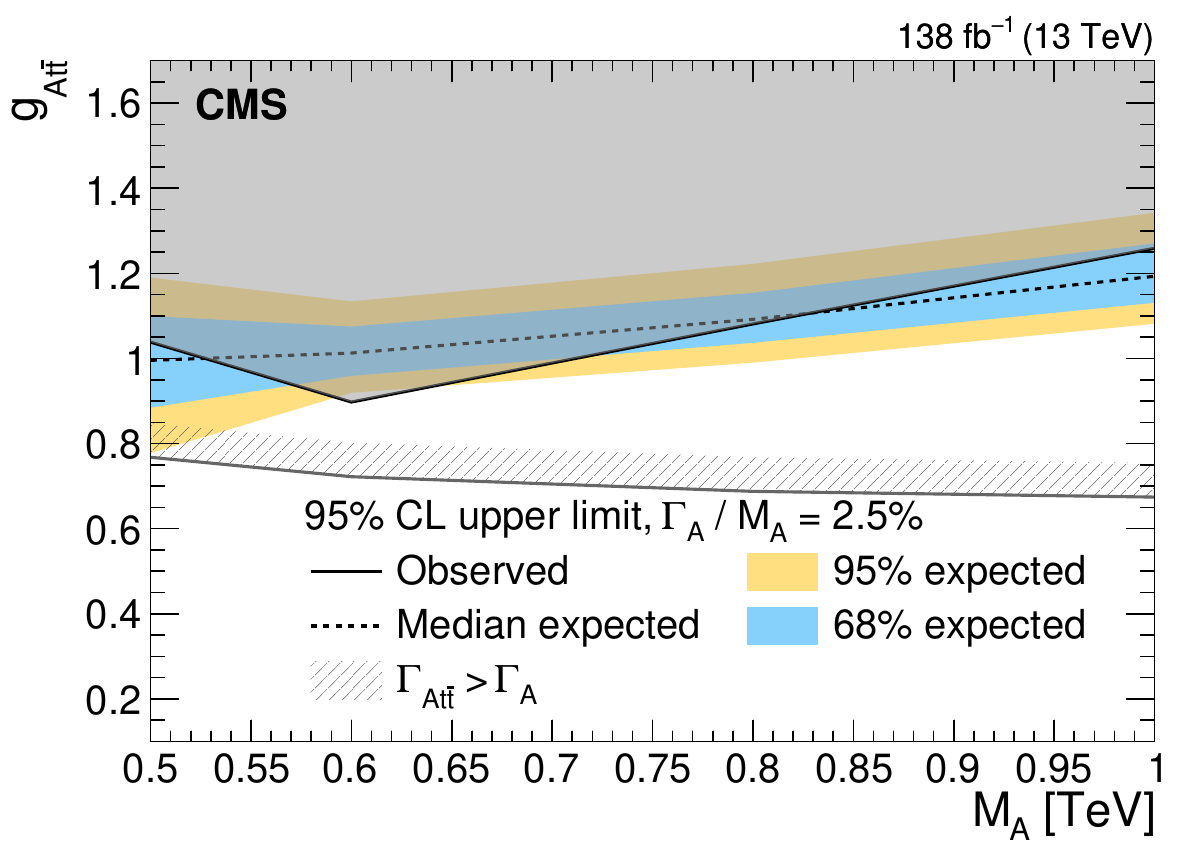} \\
\includegraphics[width=0.44\textwidth]{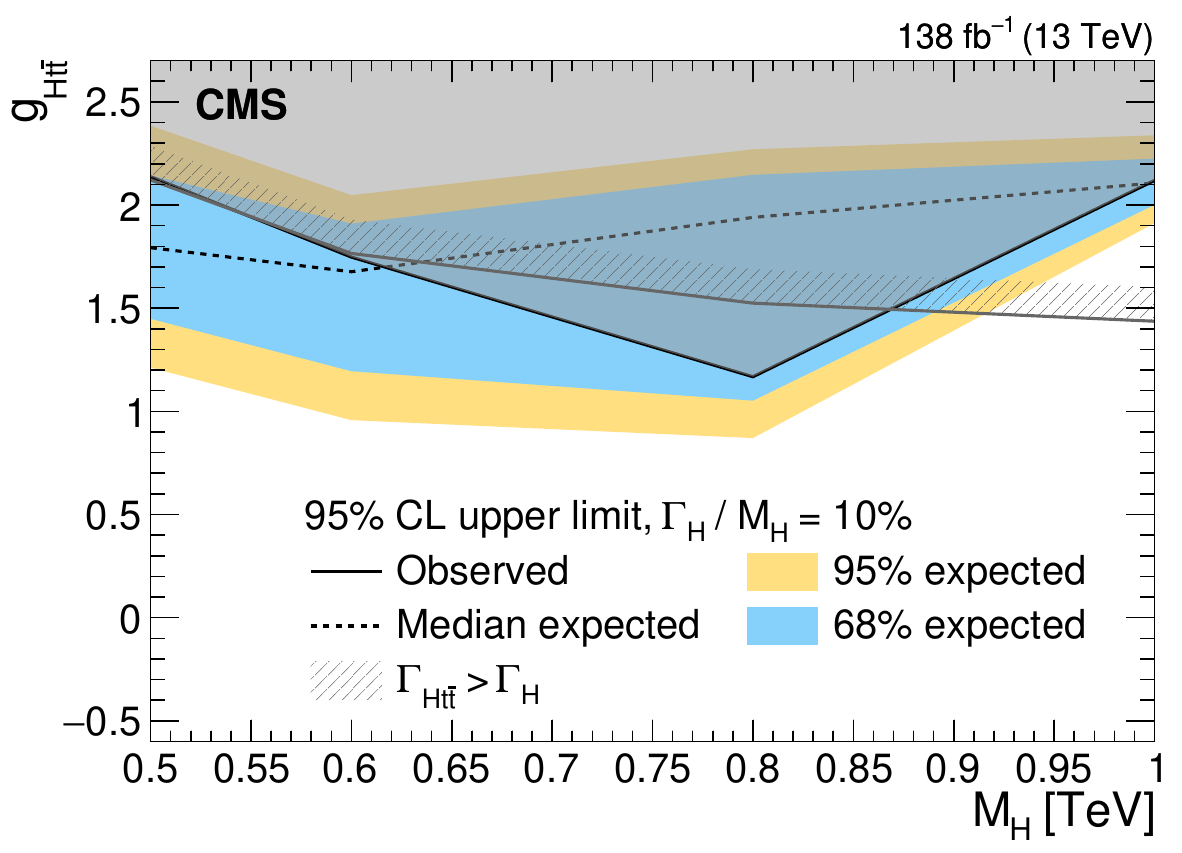}
\hspace{0.1\textwidth}
\includegraphics[width=0.44\textwidth]{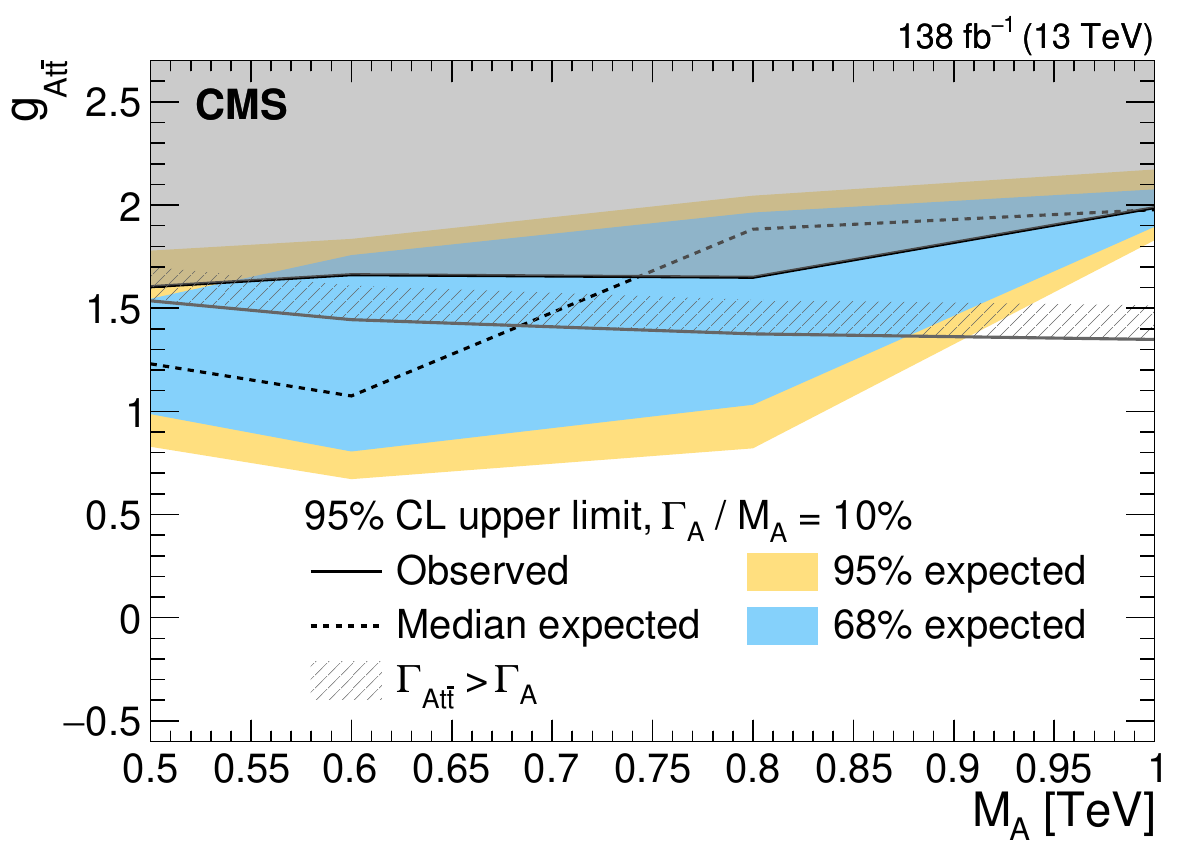} \\
\includegraphics[width=0.44\textwidth]{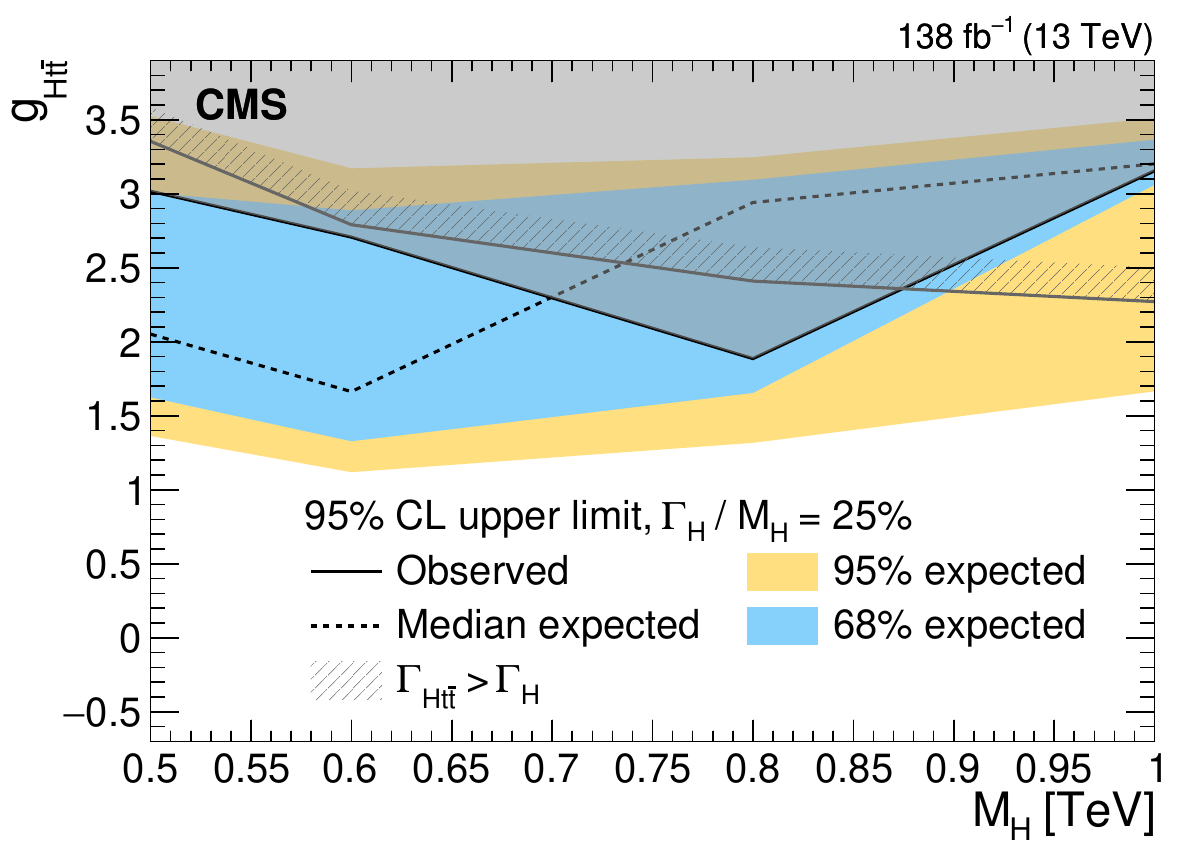}
\hspace{0.1\textwidth}
\includegraphics[width=0.44\textwidth]{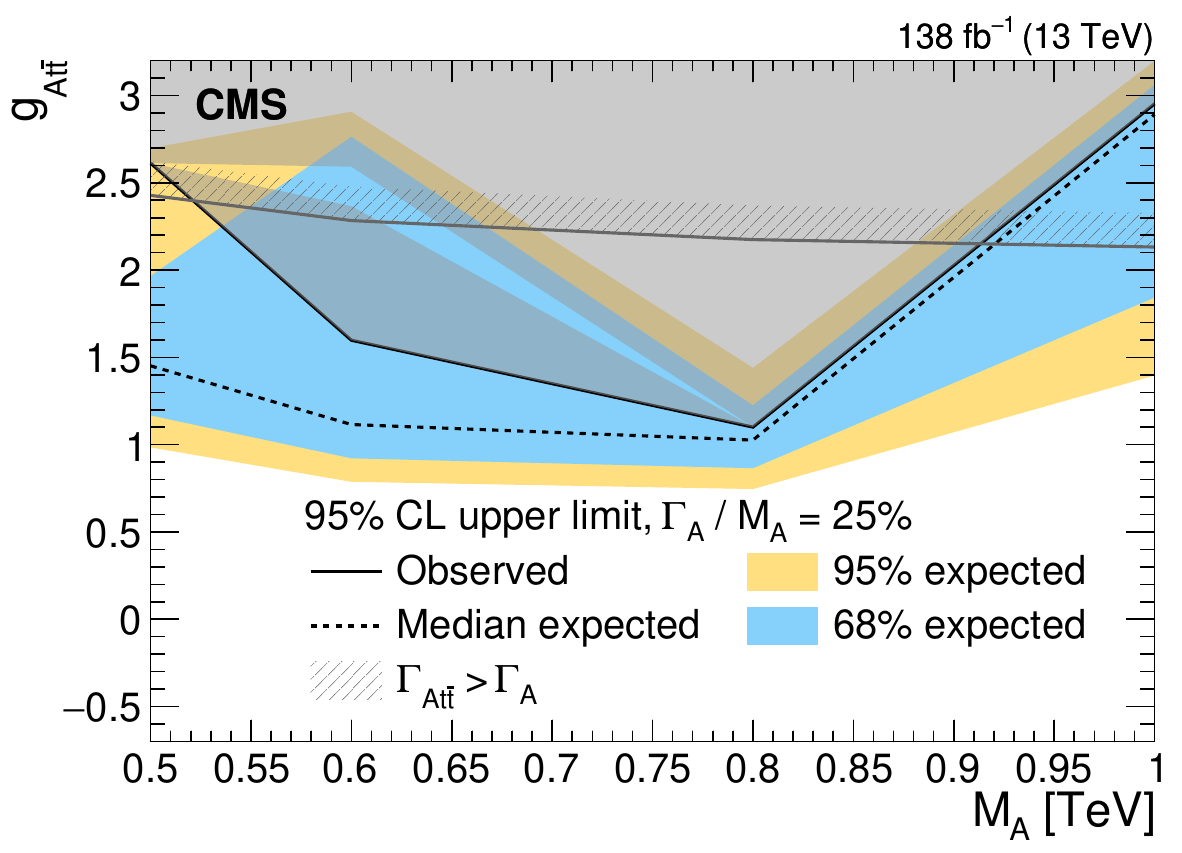}
\caption{Expected and observed upper limits at 95\% \CL on the coupling strength modifier
for scalar (\PH, left) and pseudoscalar (\PA, right) heavy Higgs bosons with relative widths of
2.5 (upper), 10 (middle), and 25\% (lower), respectively.
The solid gray shaded area denotes the parameter space excluded by this search.
The discontinuity in the shape of the excluded region,
observed for the 25\% width pseudoscalar signals with masses below 0.8\TeV,
arises from the behavior of the \CLs scan.
The gray hatched area indicates the unphysical parameter space where the partial width \Gammaptt
exceeds the total width \Gammap.}
\label{fig:limits_HA}
\end{figure}

In the two-Higgs doublet model (2HDM), we set upper limits at the 95\% \CL
on the coupling strength modifiers \gptt,
where \PGF represents either the scalar (\PH) or pseudoscalar (\PA) boson.
In the statistical analysis, both the resonant signal and its interference term are combined
yielding the following parametrization for the combined signal contribution \Stot
in each bin $i$ of the \mtt distribution
\begin{equation}
    \Stot = \mu\Sres+\sqrt{\mu}\Sint,
\end{equation}
where $\mu$ is the signal strength modifier, and \Sres and \Sint represent
the resonant and interference contributions to the signal, respectively.
In the 2HDM model, the signal strength modifier is directly related to
the coupling strength modifier as $\mu=\gptt^4$.
Exclusion limits at 95\% \CL are set on the coupling strength modifier,
covering resonance masses in the range 0.5--1\TeV and relative widths of 2.5, 10, and 25\%.
The expected and observed exclusion limits for both scalar and pseudoscalar scenarios,
across the different width hypotheses, are shown in Fig.~\ref{fig:limits_HA}.
By design, upper limits are set only for resonance masses above 500\GeV.
While the sensitivity of this search is inferior than that of the previous CMS analysis in Ref.~\cite{CMS:2025dzq}
in the mass region below 800\GeV, it becomes comparable to or better than
the previous CMS result for resonance masses above 800\GeV.
This improved sensitivity is achieved by selecting regions of phase space that were only partially covered by the previous CMS result;
at least 50\% of the selected phase space in this search lies outside the acceptance of the earlier single-lepton search.
This extended coverage is primarily driven by the different \btagging multiplicity and lepton isolation requirements
used in the event selection.

The previous CMS analyses in Refs.~\cite{CMS:2025dzq,CMS:2025kzt} reported a deviation from the background prediction,
near the \ttbar production threshold, consistent with the production of a \ttbar bound state
with a cross section of 8.8\unit{pb}.
To ensure that such a process near threshold does not bias the background estimation in this analysis,
a dedicated test was performed using simulated events with a cross section equal to the upper limit set by the previous result.
The resulting contribution was found to be less than 2\% of the total background in the affected mass range,
confirming its negligible impact on the background estimate.

\section{Summary}
\label{sec:summary}

A search for new particles decaying to a top quark-antiquark pair has been presented. The analysis uses
138\fbinv of data collected during 2016--2018 by the CMS experiment at a centre-of-mass energy of 13\TeV.
The analysis performs a model-independent search and is sensitive both to the resolved and the merged regimes of the top quark hadronic decay.
Upper limits at 95\% confidence level are placed for different benchmark models. Heavy \Zprime bosons in the leptophobic topcolor
model with relative widths of 1, 10, and 30\% are excluded for mass ranges 0.4--4.8, 0.4--6.2, and 0.4--7.4\TeV, respectively.
Additionally, Kaluza--Klein gluons in the Randall--Sundrum model
and dark-matter mediators are excluded for masses between 0.5--5.5 and 1.0--4.2\TeV, respectively. These results set the most stringent limits to date for the considered models.
Limits on the coupling strength modifier are set for scalar and pseudoscalar heavy Higgs bosons in two-Higgs-doublet models for 2.5, 10,
and 25\% relative widths in the mass range 0.5--1\TeV.

\begin{acknowledgments}
We congratulate our colleagues in the CERN accelerator departments for the excellent performance of the LHC and thank the technical and administrative staffs at CERN and at other CMS institutes for their contributions to the success of the CMS effort. In addition, we gratefully acknowledge the computing centres and personnel of the Worldwide LHC Computing Grid and other centres for delivering so effectively the computing infrastructure essential to our analyses. Finally, we acknowledge the enduring support for the construction and operation of the LHC, the CMS detector, and the supporting computing infrastructure provided by the following funding agencies: SC (Armenia), BMBWF and FWF (Austria); FNRS and FWO (Belgium); CNPq, CAPES, FAPERJ, FAPERGS, and FAPESP (Brazil); MES and BNSF (Bulgaria); CERN; CAS, MoST, and NSFC (China); MINCIENCIAS (Colombia); MSES and CSF (Croatia); RIF (Cyprus); SENESCYT (Ecuador); ERC PRG and PSG, TARISTU24-TK10 and MoER TK202 (Estonia); Academy of Finland, MEC, and HIP (Finland); CEA and CNRS/IN2P3 (France); SRNSF (Georgia); BMFTR, DFG, and HGF (Germany); GSRI (Greece); NKFIH (Hungary); DAE and DST (India); IPM (Iran); SFI (Ireland); INFN (Italy); MSIT and NRF (Republic of Korea); MES (Latvia); LMTLT (Lithuania); MOE and UM (Malaysia); BUAP, CINVESTAV, CONACYT, LNS, SEP, and UASLP-FAI (Mexico); MOS (Montenegro); MBIE (New Zealand); PAEC (Pakistan); MES, NSC, and NAWA (Poland); FCT (Portugal);  MESTD (Serbia); MICIU/AEI and PCTI (Spain); MOSTR (Sri Lanka); Swiss Funding Agencies (Switzerland); MST (Taipei); MHESI (Thailand); TUBITAK and TENMAK (T\"{u}rkiye); NASU (Ukraine); STFC (United Kingdom); DOE and NSF (USA).

\hyphenation{Rachada-pisek} Individuals have received support from the Marie-Curie programme and the European Research Council and Horizon 2020 Grant, contract Nos.\ 675440, 724704, 752730, 758316, 765710, 824093, 101115353, 101002207, 101001205, and COST Action CA16108 (European Union); the Leventis Foundation; the Alfred P.\ Sloan Foundation; the Alexander von Humboldt Foundation; the Science Committee, project no. 22rl-037 (Armenia); the Fonds pour la Formation \`a la Recherche dans l'Industrie et dans l'Agriculture (FRIA) and Fonds voor Wetenschappelijk Onderzoek contract No. 1228724N (Belgium); the Beijing Municipal Science \& Technology Commission, No. Z191100007219010, the Fundamental Research Funds for the Central Universities, the Ministry of Science and Technology of China under Grant No. 2023YFA1605804, the Natural Science Foundation of China under Grant No. 12535004, and USTC Research Funds of the Double First-Class Initiative No.\ YD2030002017 (China); the Ministry of Education, Youth and Sports (MEYS) of the Czech Republic; the Shota Rustaveli National Science Foundation, grant FR-22-985 (Georgia); the Deutsche Forschungsgemeinschaft (DFG), among others, under Germany's Excellence Strategy -- EXC 2121 ``Quantum Universe" -- 390833306, and under project number 400140256 - GRK2497; the Hellenic Foundation for Research and Innovation (HFRI), Project Number 2288 (Greece); the Hungarian Academy of Sciences, the New National Excellence Program - \'UNKP, the NKFIH research grants K 131991, K 133046, K 138136, K 143460, K 143477, K 146913, K 146914, K 147048, 2020-2.2.1-ED-2021-00181, TKP2021-NKTA-64, and 2025-1.1.5-NEMZ\_KI-2025-00004 (Hungary); the Council of Science and Industrial Research, India; ICSC -- National Research Centre for High Performance Computing, Big Data and Quantum Computing, FAIR -- Future Artificial Intelligence Research, and CUP I53D23001070006 (Mission 4 Component 1), funded by the NextGenerationEU program, the Italian Ministry of University and Research (MUR) under Bando PRIN 2022 -- CUP I53C24002390006, PRIN PRIMULA 2022RBYK7T (Italy); the Latvian Council of Science; the Ministry of Education and Science, project no. 2022/WK/14, and the National Science Center, contracts Opus 2021/41/B/ST2/01369, 2021/43/B/ST2/01552, 2023/49/B/ST2/03273, and the NAWA contract BPN/PPO/2021/1/00011 (Poland); the Funda\c{c}\~ao para a Ci\^encia e a Tecnologia (Portugal); the National Priorities Research Program by Qatar National Research Fund; MICIU/AEI/10.13039/501100011033, ERDF/EU, ``European Union NextGenerationEU/PRTR", projects PID2022-142604OB-C21, PID2022-139519OB-C21, PID2023-147706NB-I00, PID2023-148896NB-I00, PID2023-146983NB-I00, PID2023-147115NB-I00, PID2023-148418NB-C41, PID2023-148418NB-C42, PID2023-148418NB-C43, PID2023-148418NB-C44, PID2024-158190NB-C22, RYC2021-033305-I, RYC2024-048719-I, CNS2023-144781, CNS2024-154769 and Plan de Ciencia, Tecnolog{\'i}a e Innovaci{\'o}n de Asturias, Spain; the Chulalongkorn Academic into Its 2nd Century Project Advancement Project, the National Science, Research and Innovation Fund program IND\_FF\_68\_369\_2300\_097, and the Program Management Unit for Human Resources \& Institutional Development, Research and Innovation, grant B39G680009 (Thailand); the Eric \& Wendy Schmidt Fund for Strategic Innovation through the CERN Next Generation Triggers project under grant agreement number SIF-2023-004; the Kavli Foundation; the Nvidia Corporation; the SuperMicro Corporation; the Welch Foundation, contract C-1845; and the Weston Havens Foundation (USA).
\end{acknowledgments}\section*{Data availability} Release and preservation of data used by the CMS Collaboration as the basis for publications is guided by the  \href{https://doi.org/10.7483/OPENDATA.CMS.1BNU.8V1W}{CMS data preservation, re-use and open access policy}.

\bibliography{auto_generated}
\cleardoublepage \appendix\section{The CMS Collaboration \label{app:collab}}\begin{sloppypar}\hyphenpenalty=5000\widowpenalty=500\clubpenalty=5000\input{B2G-25-009-public-authorlist.tex}\end{sloppypar}
\end{document}

%% file: B2G-25-009-public-authorlist.tex
\cmsinstitute{Yerevan Physics Institute, Yerevan, Armenia}
{\tolerance=6000
A.~Gevorgyan\cmsorcid{0000-0003-2751-9489}, A.~Hayrapetyan, V.~Makarenko\cmsorcid{0000-0002-8406-8605}, A.~Tumasyan\cmsAuthorMark{1}\cmsorcid{0009-0000-0684-6742}
\par}
\cmsinstitute{Marietta Blau Institute for Particle Physics, Vienna, Austria}
{\tolerance=6000
W.~Adam\cmsorcid{0000-0001-9099-4341}, L.~Benato\cmsorcid{0000-0001-5135-7489}, T.~Bergauer\cmsorcid{0000-0002-5786-0293}, M.~Dragicevic\cmsorcid{0000-0003-1967-6783}, P.S.~Hussain\cmsorcid{0000-0002-4825-5278}, M.~Jeitler\cmsAuthorMark{2}\cmsorcid{0000-0002-5141-9560}, N.~Krammer\cmsorcid{0000-0002-0548-0985}, A.~Li\cmsorcid{0000-0002-4547-116X}, D.~Liko\cmsorcid{0000-0002-3380-473X}, M.~Matthewman, J.~Schieck\cmsAuthorMark{2}\cmsorcid{0000-0002-1058-8093}, R.~Sch\"{o}fbeck\cmsAuthorMark{2}\cmsorcid{0000-0002-2332-8784}, M.~Shooshtari\cmsorcid{0009-0004-8882-4887}, M.~Sonawane\cmsorcid{0000-0003-0510-7010}, N.~Van~Den~Bossche\cmsorcid{0000-0003-2973-4991}, W.~Waltenberger\cmsorcid{0000-0002-6215-7228}, C.-E.~Wulz\cmsAuthorMark{2}\cmsorcid{0000-0001-9226-5812}
% ...
\par}
\vskip\cmsinstskip
\dag:~Deceased\\
$^{1}$Also at Yerevan State University, Yerevan, Armenia\\
$^{2}$Also at TU Wien, Vienna, Austria\\
% ...


%% file: auto_generated.bib
@ARTICLE{zprime_Rosner,
	AUTHOR=	"Rosner, J. L.",
	TITLE=	"Prominent decay modes of a leptophobic {\PZpr}",
	EPRINT=	"hep-ph/9607207",
	ARCHIVEPREFIX=	"arXiv",
	DOI=	"10.1016/0370-2693(96)01022-2",
	JOURNAL=	"Phys. Lett. B",
	VOLUME=	"387",
	PAGES=	"113",
	YEAR=	"1996",
}

@ARTICLE{zprime_Lynch,
	AUTHOR=	"Lynch, K. R. and Mrenna, S. and Narain, M. and Simmons, E. H.",
	TITLE=	"Finding {\PZpr} bosons coupled preferentially to the third family at {CERN} {LEP} and the {Fermilab} {Tevatron}",
	EPRINT=	"hep-ph/0007286",
	ARCHIVEPREFIX=	"arXiv",
	DOI=	"10.1103/PhysRevD.63.035006",
	JOURNAL=	"Phys. Rev. D",
	VOLUME=	"63",
	PAGES=	"035006",
	YEAR=	"2001",
}

@ARTICLE{zprime_Carena,
	AUTHOR=	"Carena, M. and Daleo, A. and Dobrescu, B. A. and Tait, T. M. P.",
	TITLE=	"{\PZpr} gauge bosons at the {Fermilab} {Tevatron}",
	EPRINT=	"hep-ph/0408098",
	ARCHIVEPREFIX=	"arXiv",
	DOI=	"10.1103/PhysRevD.70.093009",
	JOURNAL=	"Phys. Rev. D",
	VOLUME=	"70",
	PAGES=	"093009",
	YEAR=	"2004",
}

@ARTICLE{Hill1991419,
	AUTHOR=	"Hill, C. T.",
	TITLE=	"Topcolor: top quark condensation in a gauge extension of the standard model",
	DOI=	"10.1016/0370-2693(91)91061-Y",
	JOURNAL=	"Phys. Lett. B",
	VOLUME=	"266",
	PAGES=	"419",
	YEAR=	"1991",
}

@ARTICLE{Hill:1993hs,
	AUTHOR=	"Hill, C. T. and Parke, S. J.",
	TITLE=	"Top quark production: Sensitivity to new physics",
	EPRINT=	"hep-ph/9312324",
	ARCHIVEPREFIX=	"arXiv",
	DOI=	"10.1103/PhysRevD.49.4454",
	JOURNAL=	"Phys. Rev. D",
	VOLUME=	"49",
	PAGES=	"4454",
	YEAR=	"1994",
}

@ARTICLE{Hill:1994hp,
	AUTHOR=	"Hill, C. T.",
	TITLE=	"Topcolor assisted technicolor",
	EPRINT=	"hep-ph/9411426",
	ARCHIVEPREFIX=	"arXiv",
	DOI=	"10.1016/0370-2693(94)01660-5",
	JOURNAL=	"Phys. Lett. B",
	VOLUME=	"345",
	PAGES=	"483",
	YEAR=	"1995",
}

@ARTICLE{Jain11124928,
	AUTHOR=	"Harris, R. M. and Jain, S.",
	TITLE=	"Cross sections for leptophobic topcolor {\PZpr} decaying to top-antitop",
	EPRINT=	"1112.4928",
	ARCHIVEPREFIX=	"arXiv",
	PRIMARYCLASS=	"hep-ph",
	DOI=	"10.1140/epjc/s10052-012-2072-4",
	JOURNAL=	"Eur. Phys. J. C",
	VOLUME=	"72",
	PAGES=	"2072",
	YEAR=	"2012",
}

@ARTICLE{axigluon,
	AUTHOR=	"Frampton, P. H. and Glashow, S. L.",
	TITLE=	"Chiral color: An alternative to the standard model",
	DOI=	"10.1016/0370-2693(87)90859-8",
	JOURNAL=	"Phys. Lett. B",
	VOLUME=	"190",
	PAGES=	"157",
	YEAR=	"1987",
}

@ARTICLE{Choudhury:2007ux,
	AUTHOR=	"Choudhury, D. and Godbole, R. M. and Singh, R. K. and Wagh, K.",
	TITLE=	"Top production at the {Tevatron}/{LHC} and nonstandard, strongly interacting spin one particles",
	EPRINT=	"0705.1499",
	ARCHIVEPREFIX=	"arXiv",
	PRIMARYCLASS=	"hep-ph",
	DOI=	"10.1016/j.physletb.2007.09.057",
	JOURNAL=	"Phys. Lett. B",
	VOLUME=	"657",
	PAGES=	"69",
	YEAR=	"2007",
}

@INPROCEEDINGS{Godbole:2008qw,
	AUTHOR=	"Godbole, R. M. and Choudhury, D.",
	TITLE=	"Nonstandard, strongly interacting spin one \ttbar resonances",
	BOOKTITLE=	"{Proc. 34th International Conference on High Energy Physics (ICHEP 2008): Philadelphia PA, USA, July 30--August 05, 2008}",
	EPRINT=	"0810.3635",
	ARCHIVEPREFIX=	"arXiv",
	PRIMARYCLASS=	"hep-ph",
	YEAR=	"2008",
}

@ARTICLE{Agashe:2006hk,
	AUTHOR=	"Agashe, K. and Belyaev, A. and Krupovnickas, T. and Perez, G. and Virzi, J.",
	TITLE=	"{CERN} {LHC} signals from warped extra dimensions",
	EPRINT=	"hep-ph/0612015",
	ARCHIVEPREFIX=	"arXiv",
	DOI=	"10.1103/PhysRevD.77.015003",
	JOURNAL=	"Phys. Rev. D",
	VOLUME=	"77",
	PAGES=	"015003",
	YEAR=	"2008",
}

@ARTICLE{Davoudiasl:1999jd,
	AUTHOR=	"Davoudiasl, H. and Hewett, J. L. and Rizzo, T. G.",
	TITLE=	"Phenomenology of the {Randall}--{Sundrum} gauge hierarchy model",
	EPRINT=	"hep-ph/9909255",
	ARCHIVEPREFIX=	"arXiv",
	DOI=	"10.1103/PhysRevLett.84.2080",
	JOURNAL=	"Phys. Rev. Lett.",
	VOLUME=	"84",
	PAGES=	"2080",
	YEAR=	"2000",
}

@ARTICLE{Randall:1999ee,
	AUTHOR=	"Randall, L. and Sundrum, R.",
	TITLE=	"A large mass hierarchy from a small extra dimension",
	EPRINT=	"hep-ph/9905221",
	ARCHIVEPREFIX=	"arXiv",
	DOI=	"10.1103/PhysRevLett.83.3370",
	JOURNAL=	"Phys. Rev. Lett.",
	VOLUME=	"83",
	PAGES=	"3370",
	YEAR=	"1999",
}

@ARTICLE{PhysRevLett.83.4690,
	AUTHOR=	"Randall, L. and Sundrum, R.",
	TITLE=	"An alternative to compactification",
	EPRINT=	"hep-th/9906064",
	ARCHIVEPREFIX=	"arXiv",
	DOI=	"10.1103/PhysRevLett.83.4690",
	JOURNAL=	"Phys. Rev. Lett.",
	VOLUME=	"83",
	PAGES=	"4690",
	YEAR=	"1999",
}

@UNPUBLISHED{ATLAS_run2,
	AUTHOR=	"{ATLAS Collaboration}",
	TITLE=	"Search for \ttbar resonances in final states with exactly one or two leptons using 140\fbinv of ${\Pp\Pp}$ collision data at $\sqrt{s}={13\TeV}$ with the {ATLAS} experiment",
	EPRINT=	"2512.17856",
	ARCHIVEPREFIX=	"arXiv",
	PRIMARYCLASS=	"hep-ex",
	YEAR=	"2025",
	NOTE=	"Submitted to \textit{JHEP}",
}

@ARTICLE{Sirunyan2019,
	AUTHOR=	"{CMS Collaboration}",
	TITLE=	"Search for resonant \ttbar production in proton-proton collisions at $\sqrt{s}={13\TeV}$",
	EPRINT=	"1810.05905",
	ARCHIVEPREFIX=	"arXiv",
	PRIMARYCLASS=	"hep-ex",
	DOI=	"10.1007/JHEP04(2019)031",
	JOURNAL=	"JHEP",
	VOLUME=	"04",
	PAGES=	"031",
	YEAR=	"2019",
}

@ARTICLE{CMS:2025kzt,
	AUTHOR=	"{CMS Collaboration}",
	TITLE=	"Observation of a pseudoscalar excess at the top quark pair production threshold",
	EPRINT=	"2503.22382",
	ARCHIVEPREFIX=	"arXiv",
	PRIMARYCLASS=	"hep-ex",
	DOI=	"10.1088/1361-6633/adf7d3",
	JOURNAL=	"Rep. Prog. Phys.",
	VOLUME=	"88",
	PAGES=	"087801",
	YEAR=	"2025",
}

@ARTICLE{CMS:2025dzq,
	AUTHOR=	"{CMS Collaboration}",
	TITLE=	"Search for heavy pseudoscalar and scalar bosons decaying to a top quark pair in proton-proton collisions at $\sqrt{s}={13\TeV}$",
	EPRINT=	"2507.05119",
	ARCHIVEPREFIX=	"arXiv",
	PRIMARYCLASS=	"hep-ex",
	DOI=	"10.1088/1361-6633/ae2207",
	JOURNAL=	"Rep. Prog. Phys.",
	VOLUME=	"88",
	PAGES=	"127801",
	YEAR=	"2025",
}

@ARTICLE{DeepAK8,
	AUTHOR=	"{CMS Collaboration}",
	TITLE=	"Identification of heavy, energetic, hadronically decaying particles using machine-learning techniques",
	EPRINT=	"2004.08262",
	ARCHIVEPREFIX=	"arXiv",
	PRIMARYCLASS=	"hep-ex",
	DOI=	"10.1088/1748-0221/15/06/P06005",
	JOURNAL=	"JINST",
	VOLUME=	"15",
	PAGES=	"P06005",
	YEAR=	"2020",
}

@MISC{hepdata,
	HOWPUBLISHED=	"{HEPData} record for this analysis",
	DOI=	"10.17182/hepdata.168403",
	YEAR=	"2026",
}

@ARTICLE{Barger:1980ix,
	AUTHOR=	"Barger, V. D. and Keung, W.-Y. and Ma, E.",
	TITLE=	"A gauge model with light {\PW} and {\PZ} bosons",
	DOI=	"10.1103/PhysRevD.22.727",
	JOURNAL=	"Phys. Rev. D",
	VOLUME=	"22",
	PAGES=	"727",
	YEAR=	"1980",
}

@ARTICLE{Contino:2011np,
	AUTHOR=	"Contino, R. and Pappadopulo, D. and Marzocca, D. and Rattazzi, R.",
	TITLE=	"On the effect of resonances in composite {Higgs} phenomenology",
	EPRINT=	"1109.1570",
	ARCHIVEPREFIX=	"arXiv",
	PRIMARYCLASS=	"hep-ph",
	DOI=	"10.1007/JHEP10(2011)081",
	JOURNAL=	"JHEP",
	VOLUME=	"10",
	PAGES=	"081",
	YEAR=	"2011",
}

@ARTICLE{Bellazzini:2014yua,
	AUTHOR=	"Bellazzini, B. and Cs{\'a}ki, C. and Serra, J.",
	TITLE=	"Composite {Higgses}",
	EPRINT=	"1401.2457",
	ARCHIVEPREFIX=	"arXiv",
	PRIMARYCLASS=	"hep-ph",
	DOI=	"10.1140/epjc/s10052-014-2766-x",
	JOURNAL=	"Eur. Phys. J. C",
	VOLUME=	"74",
	PAGES=	"2766",
	YEAR=	"2014",
}

@ARTICLE{Albert:2017,
	AUTHOR=	"Albert, A. and others",
	TITLE=	"Recommendations of the {LHC Dark Matter Working Group}: Comparing {LHC} searches for dark matter mediators in visible and invisible decay channels and calculations of the thermal relic density",
	EPRINT=	"1703.05703",
	ARCHIVEPREFIX=	"arXiv",
	PRIMARYCLASS=	"hep-ex",
	DOI=	"10.1016/j.dark.2019.100377",
	JOURNAL=	"Phys. Dark Univ.",
	VOLUME=	"26",
	PAGES=	"100377",
	YEAR=	"2019",
}

@ARTICLE{Bonciani:2015hgv,
	AUTHOR=	"Bonciani, R. and Je{\v{z}}o, T. and Klasen, M. and Lyonnet, F. and Schienbein, I.",
	TITLE=	"Electroweak top-quark pair production at the {LHC} with {\PZpr} bosons to {NLO} {QCD} in {\POWHEG}",
	EPRINT=	"1511.08185",
	ARCHIVEPREFIX=	"arXiv",
	PRIMARYCLASS=	"hep-ph",
	DOI=	"10.1007/JHEP02(2016)141",
	JOURNAL=	"JHEP",
	VOLUME=	"02",
	PAGES=	"141",
	YEAR=	"2016",
}

@ARTICLE{Lee:1973iz,
	AUTHOR=	"Lee, T. D.",
	TITLE=	"A theory of spontaneous ${T}$ violation",
	DOI=	"10.1103/PhysRevD.8.1226",
	JOURNAL=	"Phys. Rev. D",
	VOLUME=	"8",
	PAGES=	"1226",
	YEAR=	"1973",
}

@ARTICLE{Branco:2011iw,
	AUTHOR=	"Branco, G. C. and Ferreira, P. M. and Lavoura, L. and Rebelo, M. N. and Sher, M. and Silva, J. P.",
	TITLE=	"Theory and phenomenology of two-{Higgs}-doublet models",
	EPRINT=	"1106.0034",
	ARCHIVEPREFIX=	"arXiv",
	PRIMARYCLASS=	"hep-ph",
	DOI=	"10.1016/j.physrep.2012.02.002",
	JOURNAL=	"Phys. Rept.",
	VOLUME=	"516",
	PAGES=	"1",
	YEAR=	"2012",
}

@ARTICLE{Haber:2015pua,
	AUTHOR=	"Haber, H. E. and St{\r{a}}l, O.",
	TITLE=	"New {LHC} benchmarks for the ${CP}$-conserving two-{Higgs}-doublet model",
	EPRINT=	"1507.04281",
	ARCHIVEPREFIX=	"arXiv",
	PRIMARYCLASS=	"hep-ph",
	DOI=	"10.1140/epjc/s10052-015-3697-x",
	JOURNAL=	"Eur. Phys. J. C",
	VOLUME=	"75",
	PAGES=	"491",
	YEAR=	"2015",
	NOTE=	"[Erratum: \DOI{10.1140/epjc/s10052-016-4151-4}]",
}

@ARTICLE{Kling:2016opi,
	AUTHOR=	"Kling, F. and No, J. M. and Su, S.",
	TITLE=	"Anatomy of exotic {Higgs} decays in {2HDM}",
	EPRINT=	"1604.01406",
	ARCHIVEPREFIX=	"arXiv",
	PRIMARYCLASS=	"hep-ph",
	DOI=	"10.1007/JHEP09(2016)093",
	JOURNAL=	"JHEP",
	VOLUME=	"09",
	PAGES=	"093",
	YEAR=	"2016",
}

@ARTICLE{pseudohiggs,
	AUTHOR=	"Dicus, D. and Stange, A. and Willenbrock, S.",
	TITLE=	"Higgs decay to top quarks at hadron colliders",
	EPRINT=	"hep-ph/9404359",
	ARCHIVEPREFIX=	"arXiv",
	DOI=	"10.1016/0370-2693(94)91017-0",
	JOURNAL=	"Phys. Lett. B",
	VOLUME=	"333",
	PAGES=	"126",
	YEAR=	"1994",
}

@ARTICLE{Chatrchyan:2008aa,
	AUTHOR=	"{CMS Collaboration}",
	TITLE=	"The {CMS} experiment at the {CERN} {LHC}",
	DOI=	"10.1088/1748-0221/3/08/S08004",
	JOURNAL=	"JINST",
	VOLUME=	"3",
	PAGES=	"S08004",
	YEAR=	"2008",
}

@ARTICLE{Hayrapetyan_2024,
	AUTHOR=	"{CMS Collaboration}",
	TITLE=	"Development of the {CMS} detector for the {CERN} {LHC} \mbox{Run 3}",
	EPRINT=	"2309.05466",
	ARCHIVEPREFIX=	"arXiv",
	PRIMARYCLASS=	"physics.ins-det",
	DOI=	"10.1088/1748-0221/19/05/P05064",
	JOURNAL=	"JINST",
	VOLUME=	"19",
	PAGES=	"P05064",
	YEAR=	"2024",
}

@ARTICLE{CMS:2020uim,
	AUTHOR=	"{CMS Collaboration}",
	TITLE=	"Electron and photon reconstruction and identification with the {CMS} experiment at the {CERN} {LHC}",
	EPRINT=	"2012.06888",
	ARCHIVEPREFIX=	"arXiv",
	PRIMARYCLASS=	"hep-ex",
	DOI=	"10.1088/1748-0221/16/05/P05014",
	JOURNAL=	"JINST",
	VOLUME=	"16",
	PAGES=	"P05014",
	YEAR=	"2021",
}

@ARTICLE{CMS:2018rym,
	AUTHOR=	"{CMS Collaboration}",
	TITLE=	"Performance of the {CMS} muon detector and muon reconstruction with proton-proton collisions at $\sqrt{s}={13\TeV}$",
	EPRINT=	"1804.04528",
	ARCHIVEPREFIX=	"arXiv",
	PRIMARYCLASS=	"physics.ins-det",
	DOI=	"10.1088/1748-0221/13/06/P06015",
	JOURNAL=	"JINST",
	VOLUME=	"13",
	PAGES=	"P06015",
	YEAR=	"2018",
}

@ARTICLE{Chatrchyan:2014fea,
	AUTHOR=	"{CMS Collaboration}",
	TITLE=	"Description and performance of track and primary-vertex reconstruction with the {CMS} tracker",
	EPRINT=	"1405.6569",
	ARCHIVEPREFIX=	"arXiv",
	PRIMARYCLASS=	"physics.ins-det",
	DOI=	"10.1088/1748-0221/9/10/P10009",
	JOURNAL=	"JINST",
	VOLUME=	"9",
	PAGES=	"P10009",
	YEAR=	"2014",
}

@ARTICLE{Sirunyan:2017ulk,
	AUTHOR=	"{CMS Collaboration}",
	TITLE=	"Particle-flow reconstruction and global event description with the {CMS} detector",
	EPRINT=	"1706.04965",
	ARCHIVEPREFIX=	"arXiv",
	PRIMARYCLASS=	"physics.ins-det",
	DOI=	"10.1088/1748-0221/12/10/P10003",
	JOURNAL=	"JINST",
	VOLUME=	"12",
	PAGES=	"P10003",
	YEAR=	"2017",
}

@ARTICLE{Khachatryan:2016kdb,
	AUTHOR=	"{CMS Collaboration}",
	TITLE=	"Jet energy scale and resolution in the {CMS} experiment in ${\Pp\Pp}$ collisions at {8\TeV}",
	EPRINT=	"1607.03663",
	ARCHIVEPREFIX=	"arXiv",
	PRIMARYCLASS=	"hep-ex",
	DOI=	"10.1088/1748-0221/12/02/P02014",
	JOURNAL=	"JINST",
	VOLUME=	"12",
	PAGES=	"P02014",
	YEAR=	"2017",
}

@ARTICLE{CMS:2018jrd,
	AUTHOR=	"{CMS Collaboration}",
	TITLE=	"Performance of reconstruction and identification of {\PGt} leptons decaying to hadrons and {\PGnGt} in ${\Pp\Pp}$ collisions at $\sqrt{s}={13\TeV}$",
	EPRINT=	"1809.02816",
	ARCHIVEPREFIX=	"arXiv",
	PRIMARYCLASS=	"hep-ex",
	DOI=	"10.1088/1748-0221/13/10/P10005",
	JOURNAL=	"JINST",
	VOLUME=	"13",
	PAGES=	"P10005",
	YEAR=	"2018",
}

@ARTICLE{CMS:2019ctu,
	AUTHOR=	"{CMS Collaboration}",
	TITLE=	"Performance of missing transverse momentum reconstruction in proton-proton collisions at $\sqrt{s}={13\TeV}$ using the {CMS} detector",
	EPRINT=	"1903.06078",
	ARCHIVEPREFIX=	"arXiv",
	PRIMARYCLASS=	"hep-ex",
	DOI=	"10.1088/1748-0221/14/07/P07004",
	JOURNAL=	"JINST",
	VOLUME=	"14",
	PAGES=	"P07004",
	YEAR=	"2019",
}

@ARTICLE{CMS:2020cmk,
	AUTHOR=	"{CMS Collaboration}",
	TITLE=	"Performance of the {CMS} {\Lone} trigger in proton-proton collisions at $\sqrt{s}={13\TeV}$",
	EPRINT=	"2006.10165",
	ARCHIVEPREFIX=	"arXiv",
	PRIMARYCLASS=	"hep-ex",
	DOI=	"10.1088/1748-0221/15/10/P10017",
	JOURNAL=	"JINST",
	VOLUME=	"15",
	PAGES=	"P10017",
	YEAR=	"2020",
}

@ARTICLE{Khachatryan:2016bia,
	AUTHOR=	"{CMS Collaboration}",
	TITLE=	"The {CMS} trigger system",
	EPRINT=	"1609.02366",
	ARCHIVEPREFIX=	"arXiv",
	PRIMARYCLASS=	"physics.ins-det",
	DOI=	"10.1088/1748-0221/12/01/P01020",
	JOURNAL=	"JINST",
	VOLUME=	"12",
	PAGES=	"P01020",
	YEAR=	"2017",
}

@ARTICLE{CMS:2024aqx,
	AUTHOR=	"{CMS Collaboration}",
	TITLE=	"Performance of the {CMS} high-level trigger during {LHC} \mbox{Run 2}",
	EPRINT=	"2410.17038",
	ARCHIVEPREFIX=	"arXiv",
	PRIMARYCLASS=	"physics.ins-det",
	DOI=	"10.1088/1748-0221/19/11/P11021",
	JOURNAL=	"JINST",
	VOLUME=	"19",
	PAGES=	"P11021",
	YEAR=	"2024",
}

@TECHREPORT{CMS-TDR-15-02,
	AUTHOR=	"{CMS Collaboration}",
	TITLE=	"Technical proposal for the {Phase-II} upgrade of the {Compact Muon Solenoid}",
	TYPE=	"CMS Technical Proposal",
	NUMBER=	"CERN-LHCC-2015-010, CMS-TDR-15-02",
	YEAR=	"2015",
	DOI=	"10.17181/CERN.VU8I.D59J",
}

@ARTICLE{Bertolini:2014bba,
	AUTHOR=	"Bertolini, D. and Harris, P. and Low, M. and Tran, N.",
	TITLE=	"Pileup per particle identification",
	EPRINT=	"1407.6013",
	ARCHIVEPREFIX=	"arXiv",
	PRIMARYCLASS=	"hep-ph",
	DOI=	"10.1007/JHEP10(2014)059",
	JOURNAL=	"JHEP",
	VOLUME=	"10",
	PAGES=	"059",
	YEAR=	"2014",
}

@ARTICLE{Sirunyan:2020foa,
	AUTHOR=	"{CMS Collaboration}",
	TITLE=	"Pileup mitigation at {CMS} in {13\TeV} data",
	EPRINT=	"2003.00503",
	ARCHIVEPREFIX=	"arXiv",
	PRIMARYCLASS=	"hep-ex",
	DOI=	"10.1088/1748-0221/15/09/P09018",
	JOURNAL=	"JINST",
	VOLUME=	"15",
	PAGES=	"P09018",
	YEAR=	"2020",
}

@ARTICLE{Cacciari:2011ma,
	AUTHOR=	"Cacciari, M. and Salam, G. P. and Soyez, G.",
	TITLE=	"{\FASTJET} user manual",
	EPRINT=	"1111.6097",
	ARCHIVEPREFIX=	"arXiv",
	PRIMARYCLASS=	"hep-ph",
	DOI=	"10.1140/epjc/s10052-012-1896-2",
	JOURNAL=	"Eur. Phys. J. C",
	VOLUME=	"72",
	PAGES=	"1896",
	YEAR=	"2012",
}

@ARTICLE{Cacciari:2008gp,
	AUTHOR=	"Cacciari, M. and Salam, G. P. and Soyez, G.",
	TITLE=	"The anti-\kt jet clustering algorithm",
	EPRINT=	"0802.1189",
	ARCHIVEPREFIX=	"arXiv",
	PRIMARYCLASS=	"hep-ph",
	DOI=	"10.1088/1126-6708/2008/04/063",
	JOURNAL=	"JHEP",
	VOLUME=	"04",
	PAGES=	"063",
	YEAR=	"2008",
}

@ARTICLE{Sirunyan:2017ezt,
	AUTHOR=	"{CMS Collaboration}",
	TITLE=	"Identification of heavy-flavour jets with the {CMS} detector in ${\Pp\Pp}$ collisions at {13\TeV}",
	EPRINT=	"1712.07158",
	ARCHIVEPREFIX=	"arXiv",
	PRIMARYCLASS=	"physics.ins-det",
	DOI=	"10.1088/1748-0221/13/05/P05011",
	JOURNAL=	"JINST",
	VOLUME=	"13",
	PAGES=	"P05011",
	YEAR=	"2018",
}

@ARTICLE{Bols:2020bkb,
	AUTHOR=	"Bols, E. and Kieseler, J. and Verzetti, M. and Stoye, M. and Stakia, A.",
	TITLE=	"Jet flavour classification using {DeepJet}",
	EPRINT=	"2008.10519",
	ARCHIVEPREFIX=	"arXiv",
	PRIMARYCLASS=	"hep-ex",
	DOI=	"10.1088/1748-0221/15/12/P12012",
	JOURNAL=	"JINST",
	VOLUME=	"15",
	PAGES=	"P12012",
	YEAR=	"2020",
}

@ARTICLE{Larkoski:2014wba,
	AUTHOR=	"Larkoski, A. J. and Marzani, S. and Soyez, G. and Thaler, J.",
	TITLE=	"Soft drop",
	EPRINT=	"1402.2657",
	ARCHIVEPREFIX=	"arXiv",
	PRIMARYCLASS=	"hep-ph",
	DOI=	"10.1007/JHEP05(2014)146",
	JOURNAL=	"JHEP",
	VOLUME=	"05",
	PAGES=	"146",
	YEAR=	"2014",
}

@ARTICLE{Dasgupta:2013ihk,
	AUTHOR=	"Dasgupta, M. and Fregoso, A. and Marzani, S. and Salam, G. P.",
	TITLE=	"Towards an understanding of jet substructure",
	EPRINT=	"1307.0007",
	ARCHIVEPREFIX=	"arXiv",
	PRIMARYCLASS=	"hep-ph",
	DOI=	"10.1007/JHEP09(2013)029",
	JOURNAL=	"JHEP",
	VOLUME=	"09",
	PAGES=	"029",
	YEAR=	"2013",
}

@ARTICLE{Dokshitzer:1997in,
	AUTHOR=	"Dokshitzer, Y. L. and Leder, G. D. and Moretti, S. and Webber, B. R.",
	TITLE=	"Better jet clustering algorithms",
	EPRINT=	"hep-ph/9707323",
	ARCHIVEPREFIX=	"arXiv",
	DOI=	"10.1088/1126-6708/1997/08/001",
	JOURNAL=	"JHEP",
	VOLUME=	"08",
	PAGES=	"001",
	YEAR=	"1997",
}

@INPROCEEDINGS{Wobisch:1998wt,
	AUTHOR=	"Wobisch, M. and Wengler, T.",
	TITLE=	"Hadronization corrections to jet cross-sections in deep inelastic scattering",
	BOOKTITLE=	"{Proc. Workshop on Monte Carlo Generators for HERA Physics: Hamburg, Germany, April 27--30, 1998}",
	EPRINT=	"hep-ph/9907280",
	ARCHIVEPREFIX=	"arXiv",
	PAGES=	"270",
	YEAR=	"1998",
}

@ARTICLE{Nason:2004rx,
	AUTHOR=	"Nason, P.",
	TITLE=	"A new method for combining {NLO} {QCD} with shower {Monte Carlo} algorithms",
	EPRINT=	"hep-ph/0409146",
	ARCHIVEPREFIX=	"arXiv",
	DOI=	"10.1088/1126-6708/2004/11/040",
	JOURNAL=	"JHEP",
	VOLUME=	"11",
	PAGES=	"040",
	YEAR=	"2004",
}

@ARTICLE{Frixione:2007nw,
	AUTHOR=	"Frixione, S. and Ridolfi, G. and Nason, P.",
	TITLE=	"A positive-weight next-to-leading-order {Monte Carlo} for heavy flavour hadroproduction",
	EPRINT=	"0707.3088",
	ARCHIVEPREFIX=	"arXiv",
	PRIMARYCLASS=	"hep-ph",
	DOI=	"10.1088/1126-6708/2007/09/126",
	JOURNAL=	"JHEP",
	VOLUME=	"09",
	PAGES=	"126",
	YEAR=	"2007",
}

@ARTICLE{Frixione:2007vw,
	AUTHOR=	"Frixione, S. and Nason, P. and Oleari, C.",
	TITLE=	"Matching {NLO} {QCD} computations with parton shower simulations: the {\POWHEG} method",
	EPRINT=	"0709.2092",
	ARCHIVEPREFIX=	"arXiv",
	PRIMARYCLASS=	"hep-ph",
	DOI=	"10.1088/1126-6708/2007/11/070",
	JOURNAL=	"JHEP",
	VOLUME=	"11",
	PAGES=	"070",
	YEAR=	"2007",
}

@ARTICLE{Alioli:2010xd,
	AUTHOR=	"Alioli, S. and Nason, P. and Oleari, C. and Re, E.",
	TITLE=	"A general framework for implementing {NLO} calculations in shower {Monte Carlo} programs: the {\POWHEG} \textsc{box}",
	EPRINT=	"1002.2581",
	ARCHIVEPREFIX=	"arXiv",
	PRIMARYCLASS=	"hep-ph",
	DOI=	"10.1007/JHEP06(2010)043",
	JOURNAL=	"JHEP",
	VOLUME=	"06",
	PAGES=	"043",
	YEAR=	"2010",
}

@ARTICLE{Re:2010bp,
	AUTHOR=	"Re, E.",
	TITLE=	"Single-top ${\PW\PQt}$-channel production matched with parton showers using the {\POWHEG} method",
	EPRINT=	"1009.2450",
	ARCHIVEPREFIX=	"arXiv",
	PRIMARYCLASS=	"hep-ph",
	DOI=	"10.1140/epjc/s10052-011-1547-z",
	JOURNAL=	"Eur. Phys. J. C",
	VOLUME=	"71",
	PAGES=	"1547",
	YEAR=	"2011",
}

@ARTICLE{Czakon:2011xx,
	AUTHOR=	"Czakon, M. and Mitov, A.",
	TITLE=	"\textsc{top++}: a program for the calculation of the top-pair cross-section at hadron colliders",
	EPRINT=	"1112.5675",
	ARCHIVEPREFIX=	"arXiv",
	PRIMARYCLASS=	"hep-ph",
	DOI=	"10.1016/j.cpc.2014.06.021",
	JOURNAL=	"Comput. Phys. Commun.",
	VOLUME=	"185",
	PAGES=	"2930",
	YEAR=	"2014",
}

@ARTICLE{Alwall:2007fs,
	AUTHOR=	"Alwall, J. and H{\"o}che, S. and Krauss, F. and Lavesson, N. and L{\"o}nnblad, L. and Maltoni, F. and Mangano, M. L. and Moretti, M. and Papadopoulos, C. G. and Piccinini, F. and Schumann, S. and Treccani, M. and Winter, J. and Worek, M.",
	TITLE=	"Comparative study of various algorithms for the merging of parton showers and matrix elements in hadronic collisions",
	EPRINT=	"0706.2569",
	ARCHIVEPREFIX=	"arXiv",
	PRIMARYCLASS=	"hep-ph",
	DOI=	"10.1140/epjc/s10052-007-0490-5",
	JOURNAL=	"Eur. Phys. J. C",
	VOLUME=	"53",
	PAGES=	"473",
	YEAR=	"2008",
}

@ARTICLE{Alwall:2014hca,
	AUTHOR=	"Alwall, J. and Frederix, R. and Frixione, S. and Hirschi, V. and Maltoni, F. and Mattelaer, O. and Shao, H.-S. and Stelzer, T. and Torrielli, P. and Zaro, M.",
	TITLE=	"The automated computation of tree-level and next-to-leading order differential cross sections, and their matching to parton shower simulations",
	EPRINT=	"1405.0301",
	ARCHIVEPREFIX=	"arXiv",
	PRIMARYCLASS=	"hep-ph",
	DOI=	"10.1007/JHEP07(2014)079",
	JOURNAL=	"JHEP",
	VOLUME=	"07",
	PAGES=	"079",
	YEAR=	"2014",
}

@ARTICLE{Lindert:2017olm,
	AUTHOR=	"Lindert, J. M. and others",
	TITLE=	"Precise predictions for {\PV}+jets dark matter backgrounds",
	EPRINT=	"1705.04664",
	ARCHIVEPREFIX=	"arXiv",
	PRIMARYCLASS=	"hep-ph",
	DOI=	"10.1140/epjc/s10052-017-5389-1",
	JOURNAL=	"Eur. Phys. J. C",
	VOLUME=	"77",
	PAGES=	"829",
	YEAR=	"2017",
}

@ARTICLE{Sjostrand:2014zea,
	AUTHOR=	"Sj{\"o}strand, T. and Ask, S. and Christiansen, J. R. and Corke, R. and Desai, N. and Ilten, P. and Mrenna, S. and Prestel, S. and Rasmussen, C. O. and Skands, P. Z.",
	TITLE=	"An introduction to {\PYTHIA8.2}",
	EPRINT=	"1410.3012",
	ARCHIVEPREFIX=	"arXiv",
	PRIMARYCLASS=	"hep-ph",
	DOI=	"10.1016/j.cpc.2015.01.024",
	JOURNAL=	"Comput. Phys. Commun.",
	VOLUME=	"191",
	PAGES=	"159",
	YEAR=	"2015",
}

@ARTICLE{Kidonakis:2012rm,
	AUTHOR=	"Kidonakis, N.",
	TITLE=	"{NNLL} threshold resummation for top-pair and single-top production",
	EPRINT=	"1210.7813",
	ARCHIVEPREFIX=	"arXiv",
	PRIMARYCLASS=	"hep-ph",
	DOI=	"10.1134/S1063779614040091",
	JOURNAL=	"Phys. Part. Nucl.",
	VOLUME=	"45",
	PAGES=	"714",
	YEAR=	"2014",
}

@ARTICLE{Gao:2010bb,
	AUTHOR=	"Gao, J. and Li, C. S. and Li, B. H. and Zhu, H. X. and Yuan, C.-P.",
	TITLE=	"Next-to-leading order {QCD} corrections to the heavy resonance production and decay into top quark pair at the {LHC}",
	EPRINT=	"1004.0876",
	ARCHIVEPREFIX=	"arXiv",
	PRIMARYCLASS=	"hep-ph",
	DOI=	"10.1103/PhysRevD.82.014020",
	JOURNAL=	"Phys. Rev. D",
	VOLUME=	"82",
	PAGES=	"014020",
	YEAR=	"2010",
}

@ARTICLE{Hespel:2016qaf,
	AUTHOR=	"Hespel, B. and Maltoni, F. and Vryonidou, E.",
	TITLE=	"Signal background interference effects in heavy scalar production and decay to a top-anti-top pair",
	EPRINT=	"1606.04149",
	ARCHIVEPREFIX=	"arXiv",
	PRIMARYCLASS=	"hep-ph",
	DOI=	"10.1007/JHEP10(2016)016",
	JOURNAL=	"JHEP",
	VOLUME=	"10",
	PAGES=	"016",
	YEAR=	"2016",
}

@ARTICLE{Sirunyan:2019dfx,
	AUTHOR=	"{CMS Collaboration}",
	TITLE=	"Extraction and validation of a new set of {CMS} {\PYTHIA8} tunes from underlying-event measurements",
	EPRINT=	"1903.12179",
	ARCHIVEPREFIX=	"arXiv",
	PRIMARYCLASS=	"hep-ex",
	DOI=	"10.1140/epjc/s10052-019-7499-4",
	JOURNAL=	"Eur. Phys. J. C",
	VOLUME=	"80",
	PAGES=	"4",
	YEAR=	"2020",
}

@ARTICLE{Ball_2017,
	AUTHOR=	"Ball, R. D. and Bertone, V. and Carrazza, S. and Del Debbio, L. and Forte, S. and Groth-Merrild, P. and Guffanti, A. and Hartland, N. P. and Kassabov, Z. and Latorre, J. I. and Nocera, E. R. and Rojo, J. and Rottoli, L. and Slade, E. and Ubiali, M.",
	COLLABORATION=	"NNPDF",
	TITLE=	"Parton distributions from high-precision collider data",
	EPRINT=	"1706.00428",
	ARCHIVEPREFIX=	"arXiv",
	PRIMARYCLASS=	"hep-ph",
	DOI=	"10.1140/epjc/s10052-017-5199-5",
	JOURNAL=	"Eur. Phys. J. C",
	VOLUME=	"77",
	PAGES=	"663",
	YEAR=	"2017",
}

@ARTICLE{GEANT4,
	AUTHOR=	"Agostinelli, S. and others",
	COLLABORATION=	"GEANT4",
	TITLE=	"{\GEANTfour}---a simulation toolkit",
	DOI=	"10.1016/S0168-9002(03)01368-8",
	JOURNAL=	"Nucl. Instrum. Meth. A",
	VOLUME=	"506",
	PAGES=	"250",
	YEAR=	"2003",
}

@ARTICLE{Sirunyan:2018nqx,
	AUTHOR=	"{CMS Collaboration}",
	TITLE=	"Measurement of the inelastic proton-proton cross section at $\sqrt{s}={13\TeV}$",
	EPRINT=	"1802.02613",
	ARCHIVEPREFIX=	"arXiv",
	PRIMARYCLASS=	"hep-ex",
	DOI=	"10.1007/JHEP07(2018)161",
	JOURNAL=	"JHEP",
	VOLUME=	"07",
	PAGES=	"161",
	YEAR=	"2018",
}

@ARTICLE{ftest,
	AUTHOR=	"Fisher, R. A.",
	TITLE=	"On the interpretation of $\chi^2$ from contingency tables, and the calculation of ${P}$",
	DOI=	"10.2307/2340521",
	JOURNAL=	"J. R. Stat. Soc.",
	VOLUME=	"85",
	PAGES=	"87",
	YEAR=	"1922",
}

@ARTICLE{CMS:2022ged,
	AUTHOR=	"{CMS Collaboration}",
	TITLE=	"Measurement of the \ttbar charge asymmetry in events with highly {Lorentz}-boosted top quarks in ${\Pp\Pp}$ collisions at $\sqrt{s}={13\TeV}$",
	EPRINT=	"2208.02751",
	ARCHIVEPREFIX=	"arXiv",
	PRIMARYCLASS=	"hep-ex",
	DOI=	"10.1016/j.physletb.2023.137703",
	JOURNAL=	"Phys. Lett. B",
	VOLUME=	"846",
	PAGES=	"137703",
	YEAR=	"2023",
}

@ARTICLE{Carena:2016,
	AUTHOR=	"Carena, M. and Liu, Z.",
	TITLE=	"Challenges and opportunities for heavy scalar searches in the \ttbar channel at the {LHC}",
	EPRINT=	"1608.07282",
	ARCHIVEPREFIX=	"arXiv",
	PRIMARYCLASS=	"hep-ph",
	DOI=	"10.1007/JHEP11(2016)159",
	JOURNAL=	"JHEP",
	VOLUME=	"11",
	PAGES=	"159",
	YEAR=	"2016",
}

@ARTICLE{Djouadi:2019,
	AUTHOR=	"Djouadi, A. and Ellis, J. and Popov, A. and Quevillon, J.",
	TITLE=	"Interference effects in \ttbar production at the {LHC} as a window on new physics",
	EPRINT=	"1901.03417",
	ARCHIVEPREFIX=	"arXiv",
	PRIMARYCLASS=	"hep-ph",
	DOI=	"10.1007/JHEP03(2019)119",
	JOURNAL=	"JHEP",
	VOLUME=	"03",
	PAGES=	"119",
	YEAR=	"2019",
}

@MISC{chollet2015keras,
	AUTHOR=	"Chollet, F. and others",
	TITLE=	"\textsc{keras}",
	YEAR=	"2015",
	NOTE=	"Software available from \url{https://keras.io}",
}

@ARTICLE{Thaler:2010tr,
	AUTHOR=	"Thaler, J. and Van Tilburg, K.",
	TITLE=	"Identifying boosted objects with ${N}$-subjettiness",
	EPRINT=	"1011.2268",
	ARCHIVEPREFIX=	"arXiv",
	PRIMARYCLASS=	"hep-ph",
	DOI=	"10.1007/JHEP03(2011)015",
	JOURNAL=	"JHEP",
	VOLUME=	"03",
	PAGES=	"015",
	YEAR=	"2011",
}

@ARTICLE{Thaler:2011gf,
	AUTHOR=	"Thaler, J. and Van Tilburg, K.",
	TITLE=	"Maximizing boosted top identification by minimizing ${N}$-subjettiness",
	EPRINT=	"1108.2701",
	ARCHIVEPREFIX=	"arXiv",
	PRIMARYCLASS=	"hep-ph",
	DOI=	"10.1007/JHEP02(2012)093",
	JOURNAL=	"JHEP",
	VOLUME=	"02",
	PAGES=	"093",
	YEAR=	"2012",
}

@ARTICLE{CMS:HIG17027,
	AUTHOR=	"{CMS Collaboration}",
	TITLE=	"Search for heavy {Higgs} bosons decaying to a top quark pair in proton-proton collisions at $\sqrt{s}={13\TeV}$",
	EPRINT=	"1908.01115",
	ARCHIVEPREFIX=	"arXiv",
	PRIMARYCLASS=	"hep-ex",
	DOI=	"10.1007/JHEP04(2020)171",
	JOURNAL=	"JHEP",
	VOLUME=	"04",
	PAGES=	"171",
	YEAR=	"2020",
}

@ARTICLE{Butterworth_2016,
	AUTHOR=	"Butterworth, J. and others",
	TITLE=	"{PDF4LHC} recommendations for {LHC} \mbox{Run 2}",
	EPRINT=	"1510.03865",
	ARCHIVEPREFIX=	"arXiv",
	PRIMARYCLASS=	"hep-ph",
	DOI=	"10.1088/0954-3899/43/2/023001",
	JOURNAL=	"J. Phys. G",
	VOLUME=	"43",
	PAGES=	"023001",
	YEAR=	"2016",
}

@ARTICLE{CMS:LUM-17-003,
	AUTHOR=	"{CMS Collaboration}",
	TITLE=	"Precision luminosity measurement in proton-proton collisions at $\sqrt{s}={13\TeV}$ in 2015 and 2016 at {CMS}",
	EPRINT=	"2104.01927",
	ARCHIVEPREFIX=	"arXiv",
	PRIMARYCLASS=	"hep-ex",
	DOI=	"10.1140/epjc/s10052-021-09538-2",
	JOURNAL=	"Eur. Phys. J. C",
	VOLUME=	"81",
	PAGES=	"800",
	YEAR=	"2021",
}

@TECHREPORT{CMS:LUM-17-004,
	AUTHOR=	"{CMS Collaboration}",
	TITLE=	"{CMS} luminosity measurement for the 2017 data-taking period at $\sqrt{s}={13\TeV}$",
	TYPE=	"CMS Physics Analysis Summary",
	NUMBER=	"CMS-PAS-LUM-17-004",
	YEAR=	"2018",
	URL=	"https://cds.cern.ch/record/2621960",
}

@TECHREPORT{CMS:LUM-18-002,
	AUTHOR=	"{CMS Collaboration}",
	TITLE=	"{CMS} luminosity measurement for the 2018 data-taking period at $\sqrt{s}={13\TeV}$",
	TYPE=	"CMS Physics Analysis Summary",
	NUMBER=	"CMS-PAS-LUM-18-002",
	YEAR=	"2019",
	URL=	"https://cds.cern.ch/record/2676164",
}

@TECHREPORT{CMS-NOTE-2011-005,
	AUTHOR=	"{ATLAS and CMS Collaborations, and LHC Higgs Combination Group}",
	TITLE=	"Procedure for the {LHC} {Higgs} boson search combination in {Summer} 2011",
	NUMBER=	"CMS-NOTE-2011-005, ATL-PHYS-PUB-2011-11",
	YEAR=	"2011",
	URL=	"https://cds.cern.ch/record/1379837",
}

@ARTICLE{CLS1,
	AUTHOR=	"Junk, T.",
	TITLE=	"Confidence level computation for combining searches with small statistics",
	EPRINT=	"hep-ex/9902006",
	ARCHIVEPREFIX=	"arXiv",
	DOI=	"10.1016/S0168-9002(99)00498-2",
	JOURNAL=	"Nucl. Instrum. Meth. A",
	VOLUME=	"434",
	PAGES=	"435",
	YEAR=	"1999",
}

@ARTICLE{CLS2,
	AUTHOR=	"Read, A. L.",
	TITLE=	"Presentation of search results: The {\CLs} technique",
	DOI=	"10.1088/0954-3899/28/10/313",
	JOURNAL=	"J. Phys. G",
	VOLUME=	"28",
	PAGES=	"2693",
	YEAR=	"2002",
}

@ARTICLE{Cowan:2010js,
	AUTHOR=	"Cowan, G. and Cranmer, K. and Gross, E. and Vitells, O.",
	TITLE=	"Asymptotic formulae for likelihood-based tests of new physics",
	EPRINT=	"1007.1727",
	ARCHIVEPREFIX=	"arXiv",
	PRIMARYCLASS=	"physics.data-an",
	DOI=	"10.1140/epjc/s10052-011-1554-0",
	JOURNAL=	"Eur. Phys. J. C",
	VOLUME=	"71",
	PAGES=	"1554",
	YEAR=	"2011",
	NOTE=	"[Erratum: \DOI{10.1140/epjc/s10052-013-2501-z}]",
}

@ARTICLE{CAT-23-001,
	AUTHOR=	"{CMS Collaboration}",
	TITLE=	"The {CMS} statistical analysis and combination tool: \textsc{combine}",
	EPRINT=	"2404.06614",
	ARCHIVEPREFIX=	"arXiv",
	PRIMARYCLASS=	"physics.data-an",
	DOI=	"10.1007/s41781-024-00121-4",
	JOURNAL=	"Comput. Softw. Big Sci.",
	VOLUME=	"8",
	PAGES=	"19",
	YEAR=	"2024",
}

@INPROCEEDINGS{Verkerke:2003ir,
	AUTHOR=	"Verkerke, W. and Kirkby, D.",
	TITLE=	"The \textsc{RooFit} toolkit for data modeling",
	BOOKTITLE=	"{Proc. 13th International Conference on Computing in High Energy and Nuclear Physics (CHEP 2003): La Jolla CA, United States, March 24--28, 2003}",
	EPRINT=	"physics/0306116",
	ARCHIVEPREFIX=	"arXiv",
	PRIMARYCLASS=	"physics.data-an",
	YEAR=	"2003",
	NOTE=	"[eConf C0303241 (2003) MOLT007]",
	URL=	"https://www.slac.stanford.edu/econf/C0303241/proc/papers/MOLT007.PDF",
}

@INPROCEEDINGS{Moneta:2010pm,
	AUTHOR=	"Moneta, L. and Belasco, K. and Cranmer, K. S. and Kreiss, S. and Lazzaro, A. and Piparo, D. and Schott, G. and Verkerke, W. and Wolf, M.",
	TITLE=	"The \textsc{RooStats} project",
	BOOKTITLE=	"{Proc. 13th International Workshop on Advanced Computing and Analysis Techniques in Physics Research (ACAT 2010): Jaipur, India, February 22--27, 2010}",
	EPRINT=	"1009.1003",
	ARCHIVEPREFIX=	"arXiv",
	PRIMARYCLASS=	"physics.data-an",
	DOI=	"10.22323/1.093.0057",
	YEAR=	"2010",
	NOTE=	"[PoS (ACAT2010) 057]",
}
